\documentclass[aps,pre,twocolumn,showpacs,epsfig,epsf,amssymb,showkeys,tightenlines,floatfix]{revtex4-1}
\usepackage{graphicx}
\usepackage{amssymb}
\usepackage{amsmath}
\usepackage{subfigure}
\usepackage{relsize}
\usepackage{booktabs}
\begin{document}

\title{ Effect of the strength of attraction between nanoparticles on Wormlike micelle-nanoparticle system}

\author{Sk. Mubeena}
\email{mubeena@students.iiserpune.ac.in}
\affiliation{
IISER-Pune, Dr. Homi Bhabha Road, Pune-411008, India.\\
}
\date{\today}
\begin{abstract}
       The nanoparticle-Equilibrium polymer (or Wormlike micellar) system shows morphological changes from percolating network-like structures to non-percolating clusters with a change in the minimum approaching distance (EVP-excluded volume parameter) between nanoparticles and the matrix of equilibrium polymers. The shape anisotropy of nanoparticle clusters can be controlled by changing the polymer density. In this paper, the synergistic self-assembly of nanoparticles inside equilibrium polymeric matrix (or Wormlike micellar matrix) is investigated with respect to the change in the strength of attractive interaction between nanoparticles. A shift in the point of morphological transformation of the system to lower values of EVP as a result of a decrease in the strength of the attractive nanoparticle interaction is reported. We show that the absence of the attractive interaction between nanoparticles leads to the low packing of nanoparticle structures, but does not change the morphological behaviour of the system. We also report the formation of the system spanning sheet-like arrangement of nanoparticles which are arranged in alternate layers of matrix polymers and nanoparticles. 
\end{abstract}
\maketitle
\section{Introduction}

       Nanostructures have a wide range of applications in energy devices ~\cite{bisquert2017nanostructured,daubinger2016hierarchical}, opto-electronic devices ~\cite{yi2012semiconductor}, drug delivery  ~\cite{domenech2013polymer,sambarkar2012polymer,ray2006polymer}, cosmetics ~\cite{arraudeau1989composition,tatum1988organoclay}, food ~\cite{sorrentino2007potential,de2009nanocomposites} and novel functional matearials ~\cite{ingrosso2010colloidal,segala2012ruthenium}. To produce nanostructures, recently bottom-up approach is coming up as a cost-effective and easy method in the nanofabrication industry ~\cite{ seul1995domain, tang2006self}. Using polymeric matrices to assemble nanoparticles is one the prominent method in bottom-up approach ~\cite{1,2,3,4} e.g. production of various nanostructures using di-block copolymers matrix ~\cite{hamley2003nanostructure}. However, tailoring nanostructures with a precise control over their size and shape is a challenge in nanofabrication industry.
 
      In this paper, we employ equilibrium polymeric matrix (or Wormlike micellar matrix) to self-assemble nanoparticles into various kind of structures and investigate the effect of the strength of nanoparticle interaction. Nanoparticles with their high surface to volume ratio make it difficult to disperse them in a polymeric matrix. Therefore, nanoparticles are often grafted by polymers or have a surface modification to get a homogenous dispersion ~\cite{bagwe2006surface}. Apart from the dispersion of nanoparticles in a matrix, it is also important to study the effect of interaction between nanoparticles to get a precise control over tailoring nanostructures with desired properties and shape.

      The nanoparticles self-assemble in an equilibrium polymeric matrix to give rise to various kind of structures, viz. mesoporous networks, nanorods and nanosheets ~\cite{2018arXiv180106933M}. With an increase in minimum approaching distance between nanoparticles and polymers (EVP-excluded volume parameter), a morphological transition of nanoparticles from network-like structures to individual clusters has been shown in the previous study ~\cite{2018arXiv180106933M}. The study also indicates that we can control the anisotropy of the nanoparticle clusters by tuning the density of the matrix. In this paper, we report the shift in the values of the EVP required for the structural change of nanoparticles, as a result of the change in the strength of nanoparticle interaction. We also report a decrease in the packing of nanoclusters with a decrease in the strength of attractive interaction between them. Moreover, formation of system spanning nanoparticle sheets are also observed.

\section{Model and method}
\subsection{Modelling Wormlike micelles}

\begin{figure}
\includegraphics[scale=0.2]{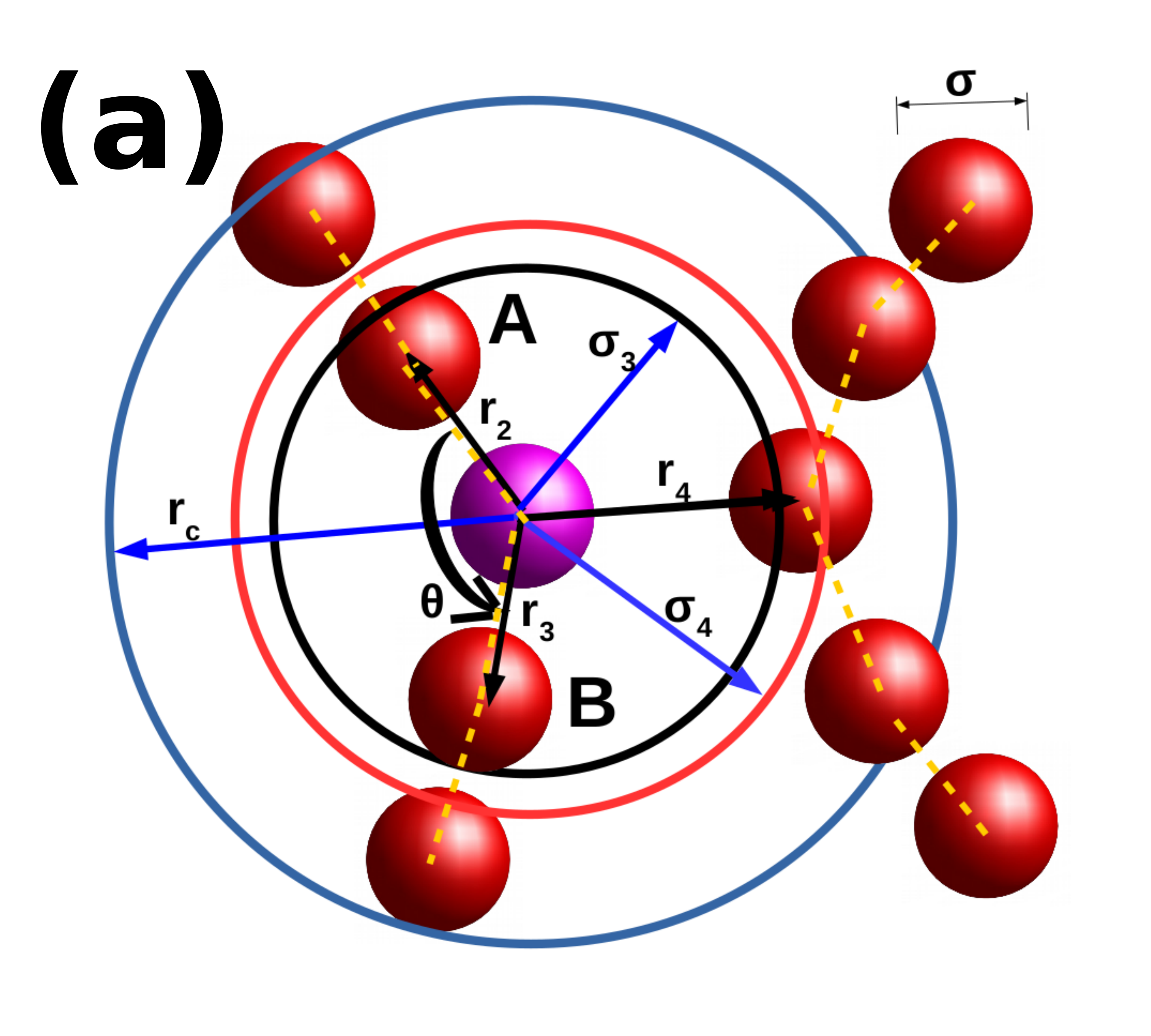}
\includegraphics[scale=0.3]{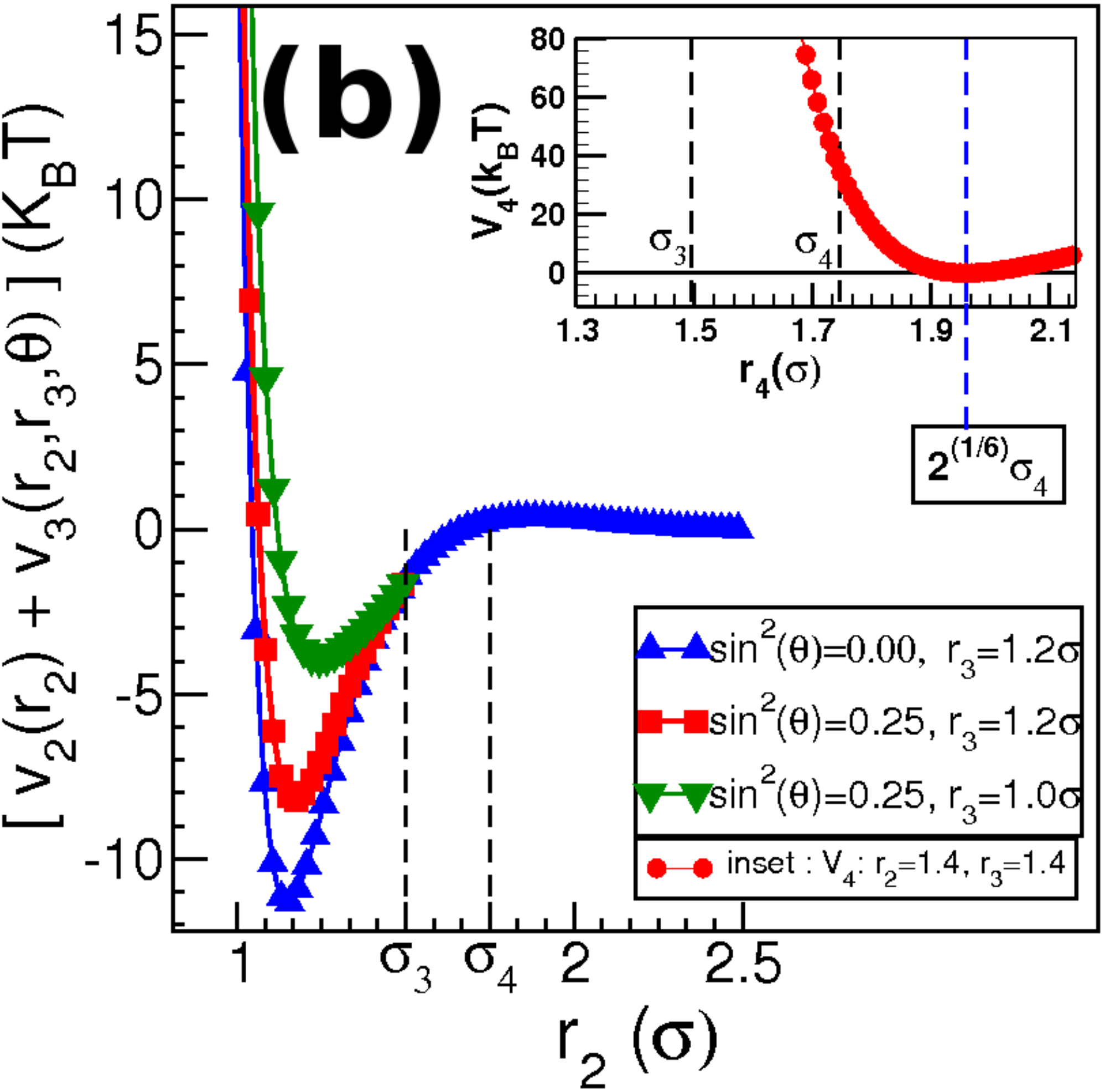}
\caption{(Colour online) The figure shows modelling of the Wormlike micelles. The spheres denote the micellar monomers having size $\sigma$. All the distances are measured with respect to the central monomer shown in pink. These monomers are acted upon by two body potential $V_2$ having a cutoff range of $r_c$. A three-body potential $V_3$ is acting on a triplet with a central monomer (pink) bonded with two other monomers at distances of $r_2$ and $r_3$, forming an angle $\theta$ at the central monomer and having a cutoff range of $\sigma_3$. In addition to these potentials there exists a four-body potential $V_4$ which is a shifted Lennard-Jones potential introduced to prevent branching and having a cutoff distance $2^{1/6}\sigma_4$.}
\label{pot_fig1}
\end{figure}

        The model used in this paper is the same as the model used in ~\cite{2018arXiv180106933M,mubeena2015hierarchical} which is a modified version of the model presented in ~\cite{chatterji2003statistical}. According to this model, the Wormlike micellar chains are coarse-grained as a chain of spherical beads. Each spherical bead (here called as monomer) in the model is assumed to represent a group of amphiphilic molecules at a mesoscopic scale. All the chemical details are ignored here and only the relevant details to describe behaviour at mesoscopic scale are considered. The schematic diagram of the model is shown in Fig.\ref{pot_fig1}(a). The spheres represent the monomers of size $\sigma$, which we set as the unit of length in the system. All the distances are shown with respect to the central monomer (shown in pink). These monomers are allowed to interact with each other using three potentials, a two-body $V_2$, three-body $V_3$ and a four-body potential $V_4$. The behaviour of the three potentials is shown in Fig.\ref{pot_fig1}(b) and the potentials are expressed as follows,

\begin{itemize}
\item{ {\bf{$V_2$: Two body attractive potential}\\}
      For any two monomers at a distance of $r_2$, an attractive Lennard-Jones potential is provided which is modified by an exponential term as shown in Eq. ~\ref{eq1}.
\begin{equation}      
        V_2 = \epsilon [ (\frac{\sigma}{r_2})^{12} - (\frac{\sigma}{r_2})^6 + \epsilon_1 e^{-a r_2/\sigma}]; 
\, \forall  r_2 < r_c.
\label{eq1}
\end{equation}

    Where, $\epsilon=110k_BT$ and the cutoff distance is $r_c=2.5\sigma$ . The exponential term in the above potential creates a maximum at $r_2=1.75\sigma$ which acts as a potential barrier for joining or breaking of monomers from chains. The value of $\epsilon_1$ and $a$ are kept fixed as  $\epsilon_{1}=1.34\epsilon$ and $a= 1.72$. This potential behaviour is shown in Fig.\ref{pot_fig1}(b) where the Y-axis is $V_2+V_3$. When $\theta=0$ then $V_3=0$ (refer below). Therefore, the graph shown by the symbols (blue-triangle) having legends $\sin^2\theta=0$, represents the behaviour of $V_2$. In the graph, $r_2$ is kept fixed at $r_2=\sigma$ (except for the inset figure).
}
\item{ {\bf{$V_3$: Three body potential to add semiflexibility to chians}\\}

     For any monomer that is part of a chain there are two bonded neighbours at a distance of $r_3$ and $r_4$ which subtends an angle $\theta$ at the central monomer (as shown in Fig. \ref{pot_fig1}). The triplet thus formed is then subjected to the following three body potential,
\begin{equation}
V_3 = \epsilon_3 (1 - \frac{r_{2}}{\sigma_3})^2(1 - \frac{r_{3}}{\sigma_3})^2 \sin^2(\theta); 
\, \forall r_{2},r_{3} < \sigma_3. 
\label{eq2}
\end{equation}

Where, the value of $\epsilon_{3}=6075k_{B}T$ and the cutoff distance $\sigma_3$ is kept fixed at $1.5\sigma$. The leading terms inside the two brackets ensure that the potential and force goes smoothly to zero at the cutoff of $\sigma_3$.
}
\item{ {\bf{$V_4$: Four body repulsive potential between chains}\\}
For any monomer with two bonded neighbours at distances $r_2$ and $r_3$, any other monomer at a distance $r_4$ approaching the first monomer to form a branch (refer Fig.\ref{pot_fig1}(a)) will be repelled with the following potential,
\begin{equation}
V_4 = \epsilon_4 (1 - \frac{r_{2}}{\sigma_3})^2(1 - \frac{r_{3}}{\sigma_3})^2 \times V_{LJ}(\sigma_4,r_4) 
\label{eq3}
\end{equation}
The cutoff distance for this potential $\sigma_4$ is chosen such that $ \sigma_3 < \sigma_4 < r_c$ and is fixed at $\sigma_4=1.75\sigma$. The leadinig terms in the brackets are necessary to make the force and potential smoothly approaching zero at the cutoff distance. Since, those terms in the brackets approackes zero as $r_2$ or $r_3$ approaches $\sigma_3$, therefore the value of $\epsilon_4$ is decided to give a very high value $\epsilon_4=2.53 \times 10^5 k_BT$ to ensure enough repulsion between the chains. The behaviour of $V_4$ is shown in the inset of Fig.\ref{pot_fig1}(b). It should be noted that, we refer micellar chains as dispersed if the distance between chains is $> 1.75\sigma$. With distance between chains of monomers is $< 1.75\sigma$, then we refer them as clusters of chains.
}
\end{itemize}

     Using Monte Carlo technique the system is allowed to equilibrate from a randomly initialized state. After around $(5-6)\times 10^5$ iterations, the system evolves to form Wormlike chains of monomers having an exponential distribution of chain length ~\cite{mubeena2015hierarchical}. With an increase in micellar density, an isotropic-to-nematic transition is observed and have been reported in detail in ~\cite{mubeena2015hierarchical}. 

\subsection{Modelling nanoparticles}      

        To investigate the phase behaviour of Wormlike micelle-nanopaticle system, model nanoparticles are added in the model Wormlike micellar system described in the above section. Nanoparticles are modelled by Lennard-Jones attractive particles of size $\sigma_n$ and having a cutoff distance $r_{cn}$ with the interacting potential $V_{2n}$ given by,

\begin{equation}
V_{2n} = \epsilon_n[(\frac{\sigma_{n}}{r_n})^{12} - (\frac{\sigma_{n}}{r_n})^6],      \forall     r_n   <=   r_{cn}
\end{equation}

            The cutoff distance $r_{cn}$ is set at $r_{cn}=2\sigma_n$. These nanoparticles interact with monomers via a repelling potential $V_{4n}$ which is a shifted Lennard-Jones potential given by,

\begin{equation}
V_{4n} = \epsilon_{4n}[(\frac{\sigma_{4n}}{r_{mn}})^{12} - (\frac{\sigma_{4n}}{r_{mn}})^6],  \forall r_{mn} <= 2^{1/6}\sigma_{4n}
\end{equation}

Where, $r_{mn}$ indicates the distance between monomers and nanoparticles with the parameter $\sigma_{4n}$ indicating the centre-to-centre distance between the particles. The value of $\sigma_{4n}$ is used as a parameter. The value of the strength of the repulsive interaction is fixed at $\epsilon_{4n}=30k_BT$.

        In summary, in this model (refer Fig.\ref{pot_fig1}(a)), there are coarse-grained particles (spheres) to form micellar chains of the size $\sigma$ that interacts with each other via a Lennard-Jones potential having a cutoff distance $2.5\sigma$. The micellar chains are semiflexible (potential $V_3$) and repel each other with the potential $V_4$ and the minimum approaching distance $\sigma_4=1.75\sigma$. The system also contains nanoparticles of size $\sigma_n$ which are interacting by a Lennard-Jones attractive potential between themselves (potential $V_{2n}$) with a cutoff distance $2\sigma_n$. These monomers and nanoparticles are repelled by a repulsive potential $V_{4n}$ with the minimum approaching distance $\sigma_{4n}$ and cutoff distance $2^{1/6}\sigma_{4n}$. The value of $\sigma_{4n}$ is used as a parameter.

        Using this model the morphological transformations of the system with a change in $\sigma_{4n}$ and micellar density has been established ~\cite{2018arXiv180106933M} (keeping the nanoparticle size fixed at $\sigma_n=1.5\sigma$). Now in this paper, the system is investigated to explore the effect of the strength of attractive interaction between nanoparticles $\epsilon_n$ on the system behaviour. Therefore, we keep the nanoparticle size fixed at $\sigma_n=1.5\sigma$, while using $\epsilon_n$ as a variable along with $\sigma_{4n}$ and the number density of monomers $\rho_m$. We will generate runs in sets where each set consists of runs with a fixed $\epsilon_n$ but $\sigma_{4n}$ and $\rho_m$ as parameters.

\subsection{Method}

          The model is first applied with Metropolis Monte Carlo (MC) method. However, the method seemed to be insufficient to equilibrate the system with high density. Therefore, the system is first evolved with Metropolis Monte Carlo method with (200-300) nanoparticles within a given number density of monomers $\rho_m$ for $10^5$ iterations. This gives the monomers enough time to develop into chain-like structures in the presence of seeding of nanoparticles. Then, a semi-grand canonical Monte Carlo (GCMC) scheme is applied. According to this scheme, for every 50 Monte Carlo steps, 300 attempts are made to add and remove a nanoparticle randomly. Each successful attempt is penalized with an energy gain or loss of $\pm \mu_n$, where, $\mu_n$ is the chemical potential of the system fixed at $\mu_n=-8k_BT$. All the runs were tested with ten independent runs which show convergence to morphologically similar states and their thermodynamic properties converging to same values. It was shown that, for the model system, runs that started with an unmixed state (both nanoparticles and monomers separated) also tend to form a mixed state for the value of $\mu_n$ chosen. Thus, the possibility of a fully phase separated state as a thermodynamically preferred state is thus negated ~\cite{2018arXiv180106933M}. 

         For all the runs in this paper, we first apply Metropolis Monte Carlo method to allow the growth of equilibrium polymeric chains in the presence of seed of nanoparticles. Then, GCMC scheme is switched on for the rest of the run. For each set of parameters, the system is evolved for $\approx (20-40)\times 10^5$ and the thermodynamic properties are averaged for last $(10-20)\times 10^5$ iterations over ten independent runs. The error bars in the plots shown in this paper are smaller than the symbols and hence not visible here.
 

\section{Results}

         Previous studies ~\cite{mubeena2015hierarchical,2018arXiv180106933M} using the same model (presented in the previous section) has reported the morphological transitions of nanoparticles with the increase in $\sigma_{4n}$. Those results were substantiated by showing the convergence of all ten independent runs to same morphological structures. It is emphasized here that, all these studies are comprised of systems that are initialized with a mixed state of nanoparticles and micelles. It was shown that the nanoparticle clusters formed in the system, vary in their shape anisotropy with a change in matrix polymer density.  This paper takes this investigation further by varying the strength of interaction between nanoparticles $\epsilon_n$ along with $\sigma_{4n}$ and monomer no. density $\rho_m$. All the quantities calculated are averaged over ten independent runs.

\begin{figure}
\includegraphics[scale=0.18]{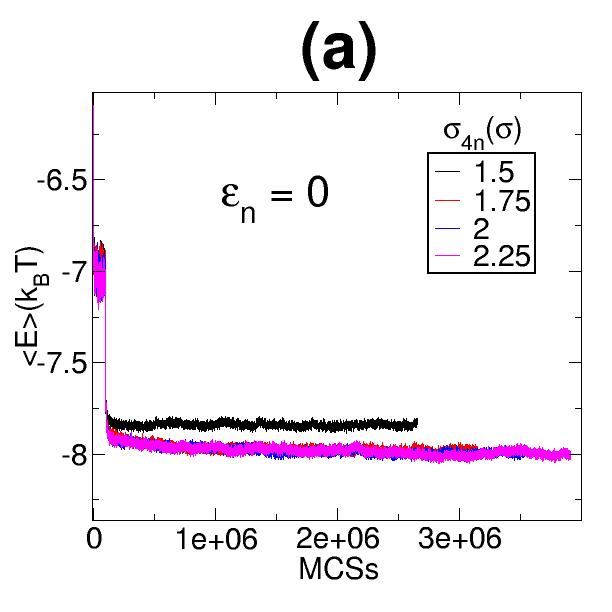}
\includegraphics[scale=0.18]{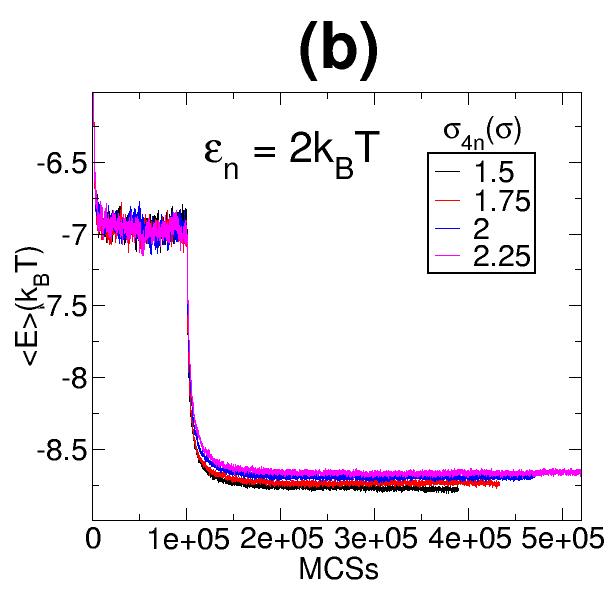}
\caption{(Colour online) The figure shows the evolution of the average energy of the system with Monte Carlo steps for the values of $\epsilon_n$  (a)0 and (b) 2 $k_BT$. Each figure shows graphs for four different values of $\sigma_{4n}=1.5\sigma$, $1.75\sigma$, $2\sigma$ and $2.25\sigma$. All the graphs show a jump in their values at $10^5$ Monte Carlo steps indicating the start of GCMC scheme.}
\label{energies}
\end{figure}

\begin{figure}
\includegraphics[scale=0.2]{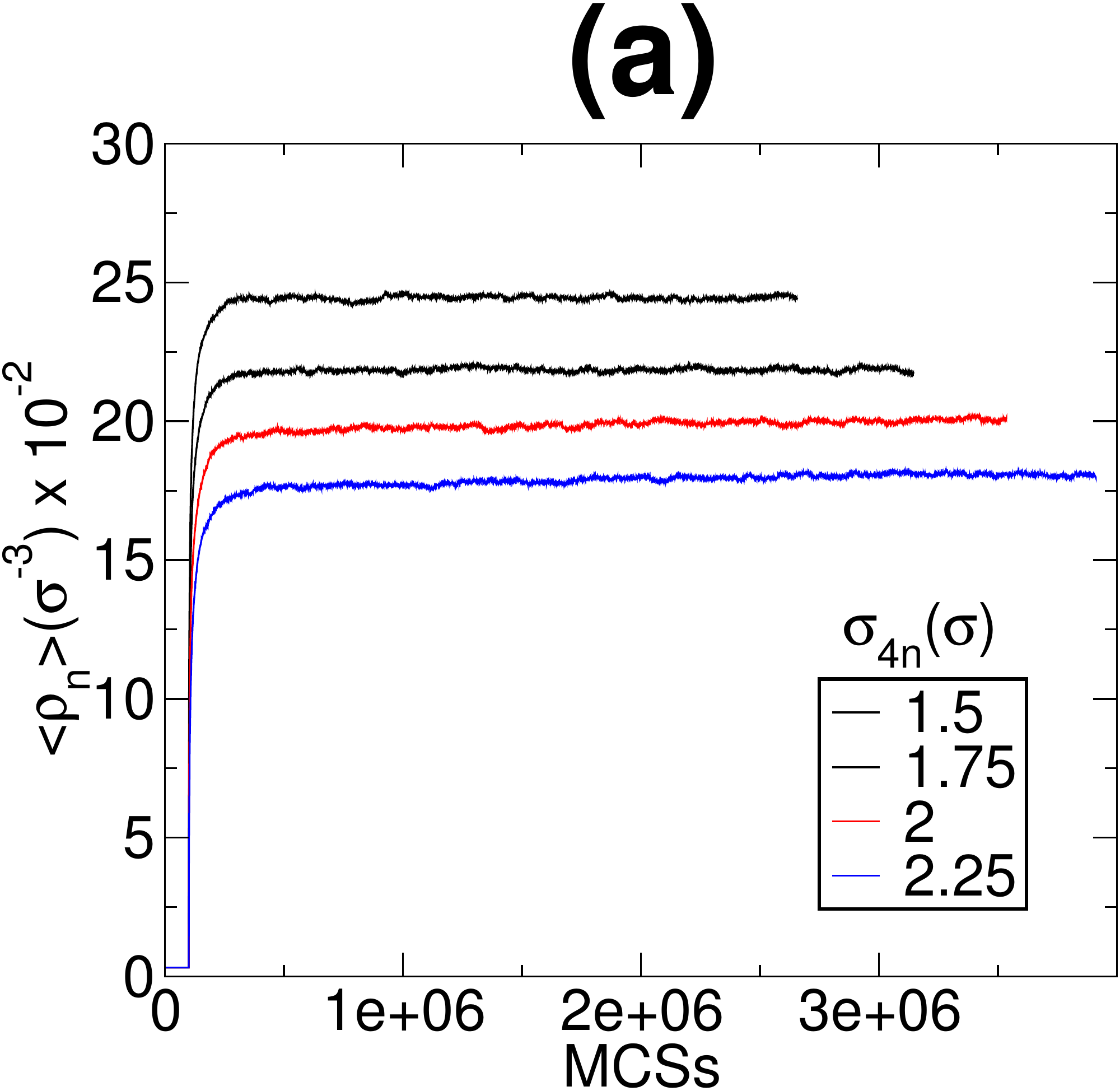}
\includegraphics[scale=0.2]{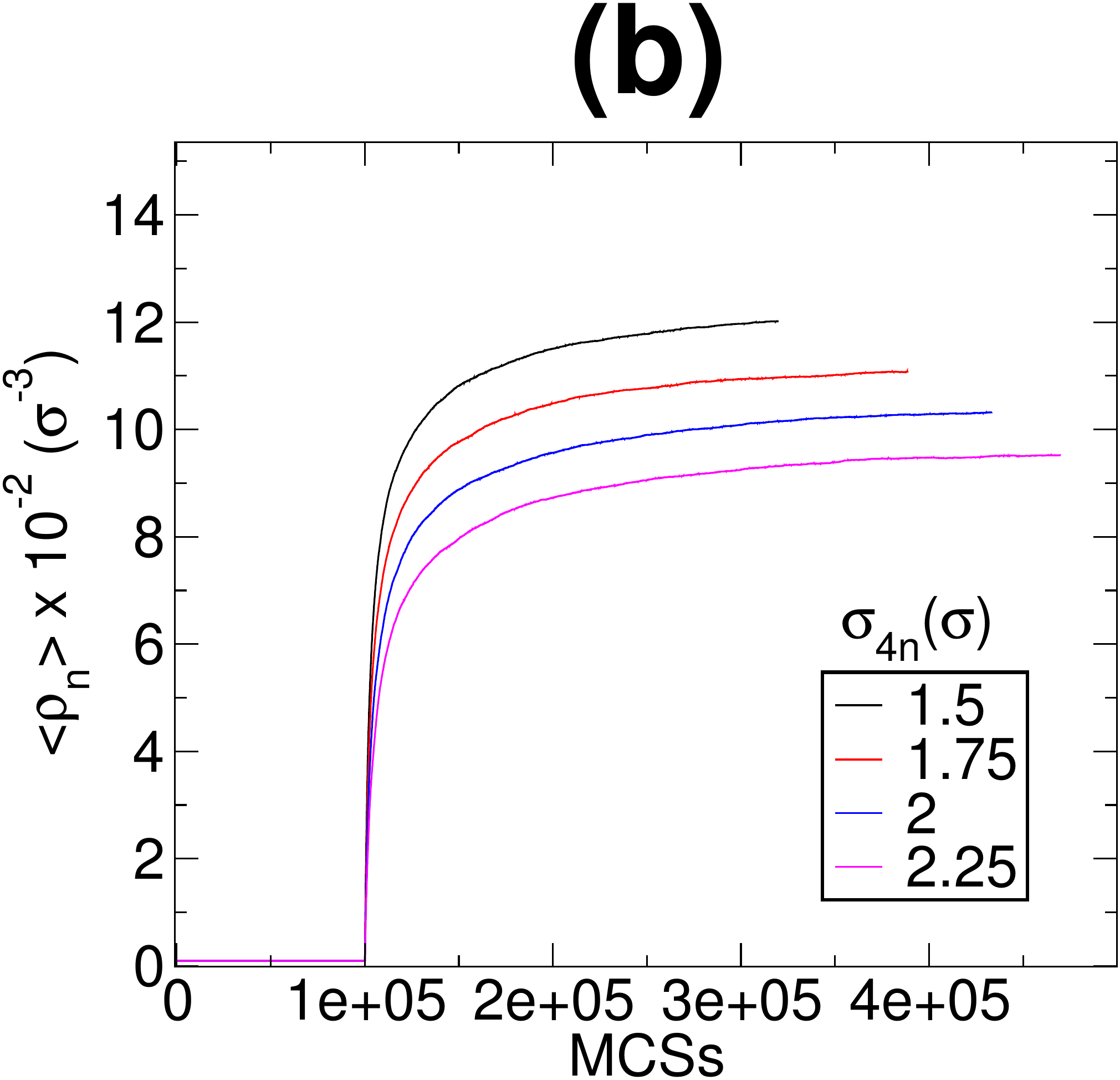}
\caption{(Colour online) The figure shows the evolution of volume fraction of nanoparticles with Monte Carlo steps for values of $\epsilon_n$ (a) 0 and (b) $2k_BT$. Each figure shows the graphs of nanoparticle volume fraction for four different values of $\sigma_{4n}=1.5\sigma$, $1.75\sigma$, $2\sigma$ and $2.25\sigma$. All the graphs show a jump at $10^5$ Monte Carlo steps when the GCMC scheme is switched on.}
\label{numbers}
\end{figure}

            Now, to investigate the effect of strength of interaction between nanoparticles $\epsilon_n$, a set of runs varying in the value of $\rho_m$ and $\sigma_{4n}$ is generated for each value of $\epsilon_n$. For a given value of $\epsilon_n$, the system morphological behaviour is observed and the structural changes are identified. Then these structural changes are compared over different values of $\epsilon_n$ and the change in the value of EVP at which the morphological change occurs are observed. Different values of $\epsilon_n=2k_BT, 5k_BT, 8k_BT$ and $11k_BT$ are considered. Apart from these values, one more case where, there exists no attractive interaction between nanoparticles is also considered. In this case, the nanoparticles are provided with WAC (Week-Anderson-Chandler) potential (similar to the potential expressed in Eq.5). We represent this case by $\epsilon_n=0$ only for convenience. For each value of $\epsilon_n$, a set of runs with four values of number density of monomers $\rho_m=0.037\sigma^{-3}$, $0.074\sigma^{-3}$, $0.093\sigma^{-3}$ and $0.126\sigma^{-3}$ along with varying parameter $\sigma_{4n}$ for each density, are produced. The system is evolved using MC steps for first $10^5$ of iterations and then subjected to GCMC scheme for the rest of the iterations. The system is monitored to ensure that the runs are long enough to produce structures and thermodynamic quantities that are stable over a long run. After around $(2-3)\times 10^5$ iterations, the systems are observed to maintain their morphological states. The behaviour of the average energy of the particles for $\rho_m=0.037\sigma^{-3}$ is shown in Fig. \ref{energies} for two different values of $\epsilon_n$ (a) 0 and (b) $2k_BT$. For each value of $\epsilon_n$, the figure shows graphs for four different values of $\sigma_{4n}=1.5\sigma,1.75\sigma,2\sigma$ and $2.25\sigma$. In both the figures, all the graphs show a jump in their energy values at $10^5$ Monte Carlo Steps (MCSs). These jumps mark the starting of the GCMC scheme where nanoparticles start getting introduced into the system. After around $2\time 10^5$ MCSs, the system morphology is observed to remain same. With the increase in the value of $\epsilon_n$, the system becomes very dense. Therefore, the systems seem to be stuck in some kinetically arrested states for $\epsilon_n > 0$. This can be seen in Fig.2, which shows the evolution of the number of nanoparticles in the simulation box with MCSs for (a) $\epsilon_n=0$ and (b) $\epsilon_n=2k_BT$. After some MCSs, the number of nanoparticle and energy graphs show a very slow increase in its value for $\epsilon_n=2k_BT$ in Fig.\ref{numbers}(b) and Fig.\ref{energies}(b), respectively. Only for $\epsilon_n=0$, the system shows a stable value of energy and the number of nanoparticles as shown in Fig.\ref{energies}(a) and Fig.\ref{numbers}(a). 
Therefore, the systems with higher values of $\epsilon_n$ seem to be in a kinetically arrested state. However, for all the values of $\epsilon_n$, the ten independent runs converge to the same value of energy and morphological structure.

\begin{figure*}
\centering
\includegraphics[scale=0.2]{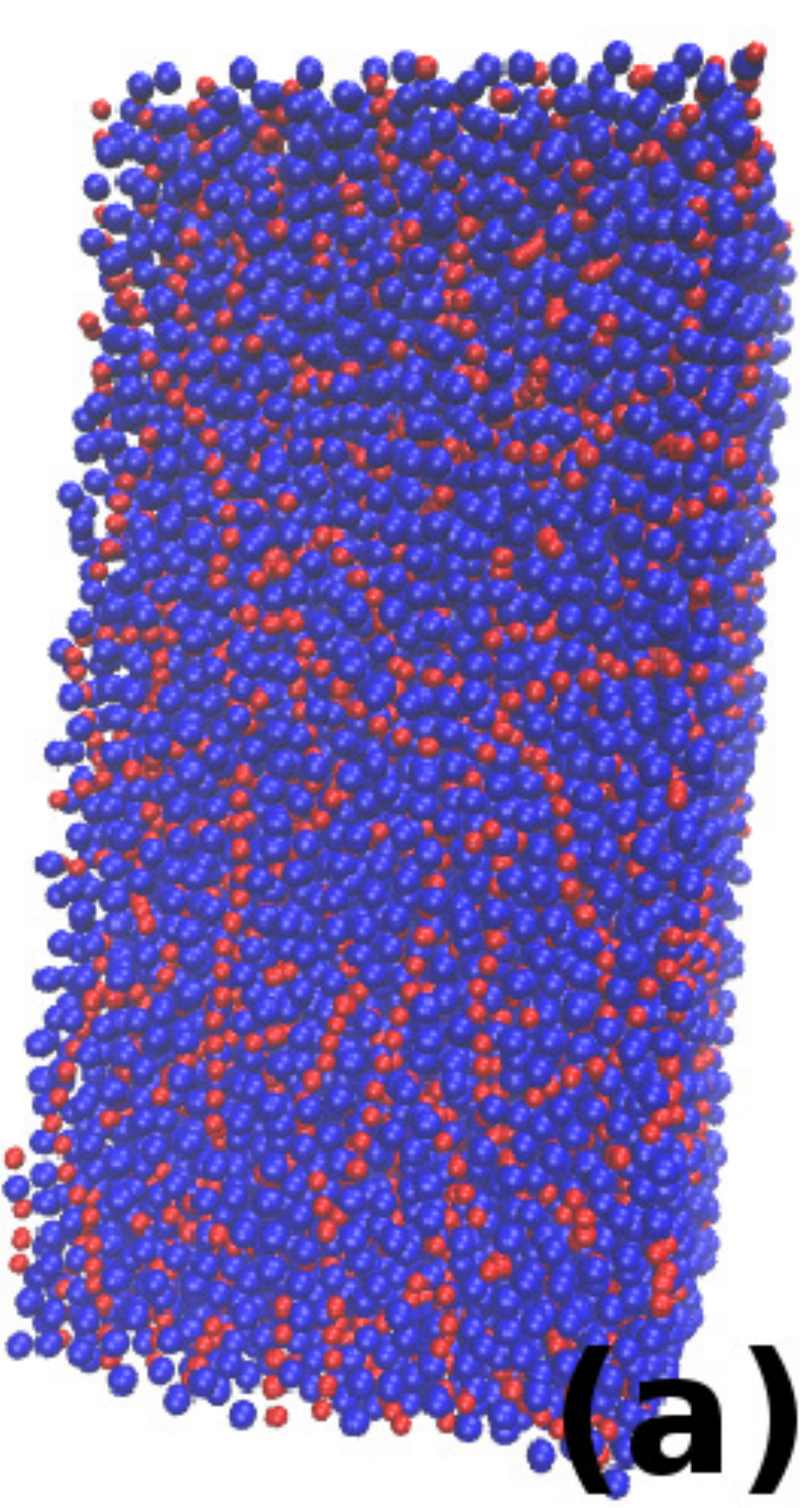}
\hspace{1cm}
\includegraphics[scale=0.2]{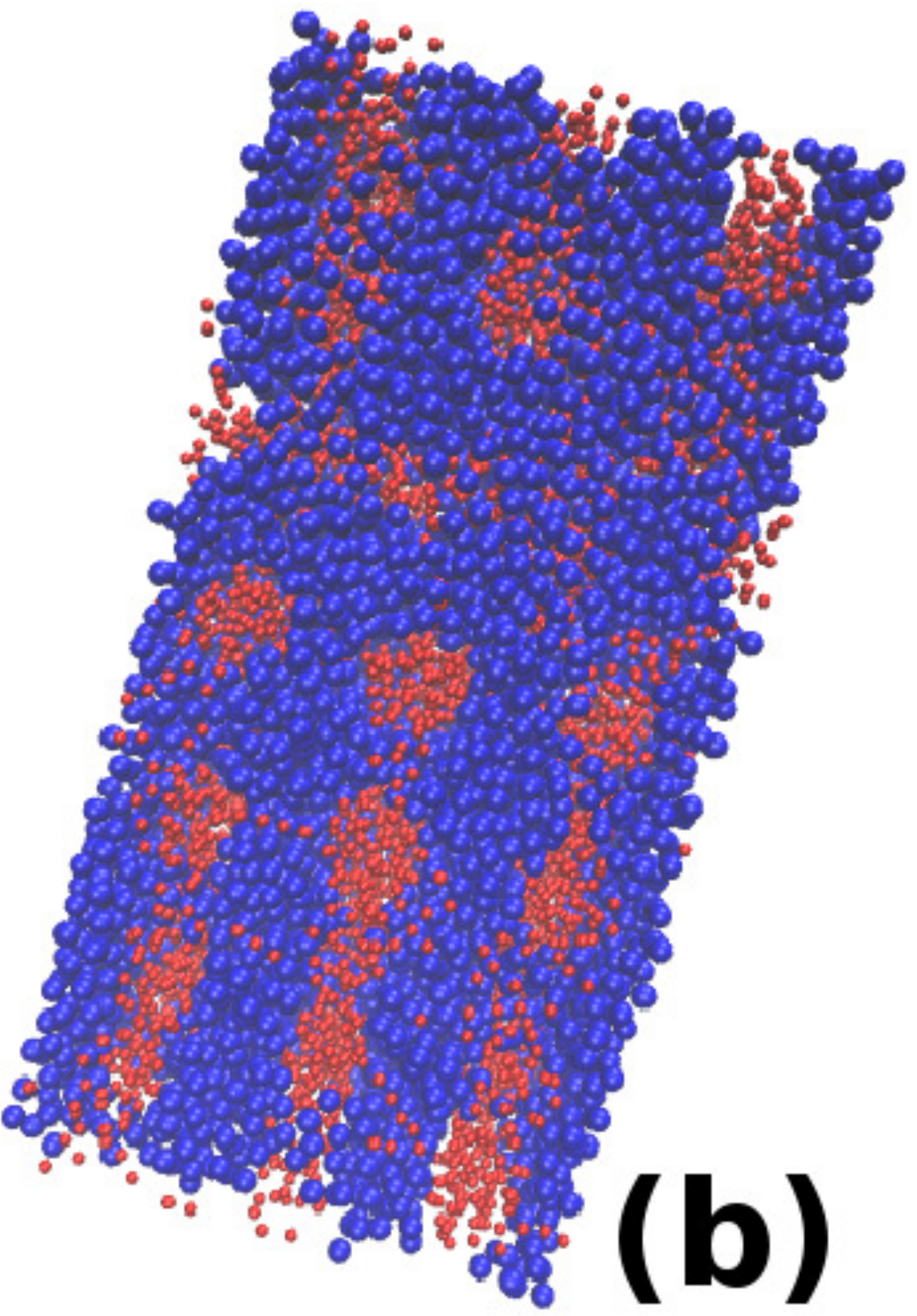}
\hspace{1cm}
\includegraphics[scale=0.2]{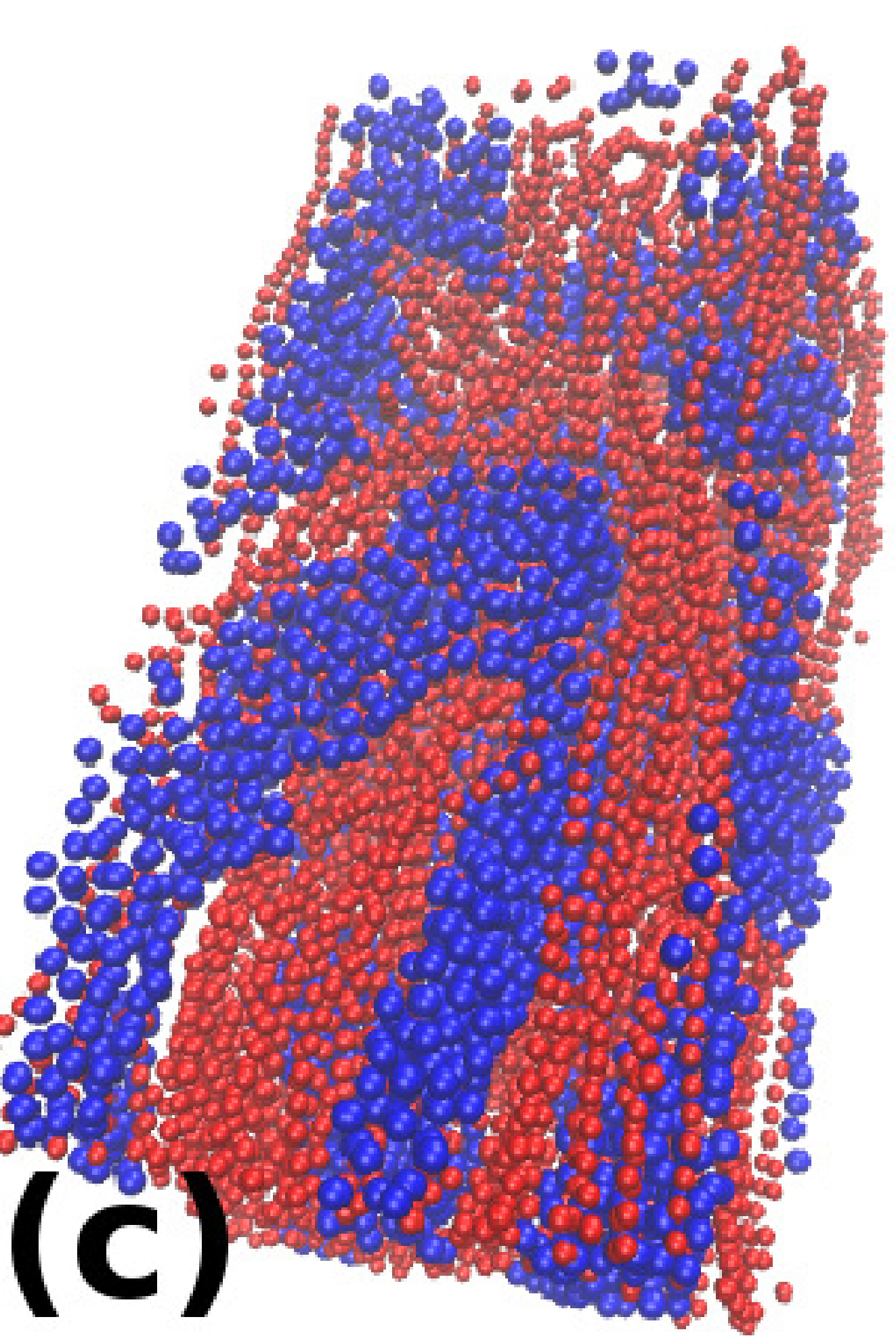}
\hspace{1cm}
\includegraphics[scale=0.2]{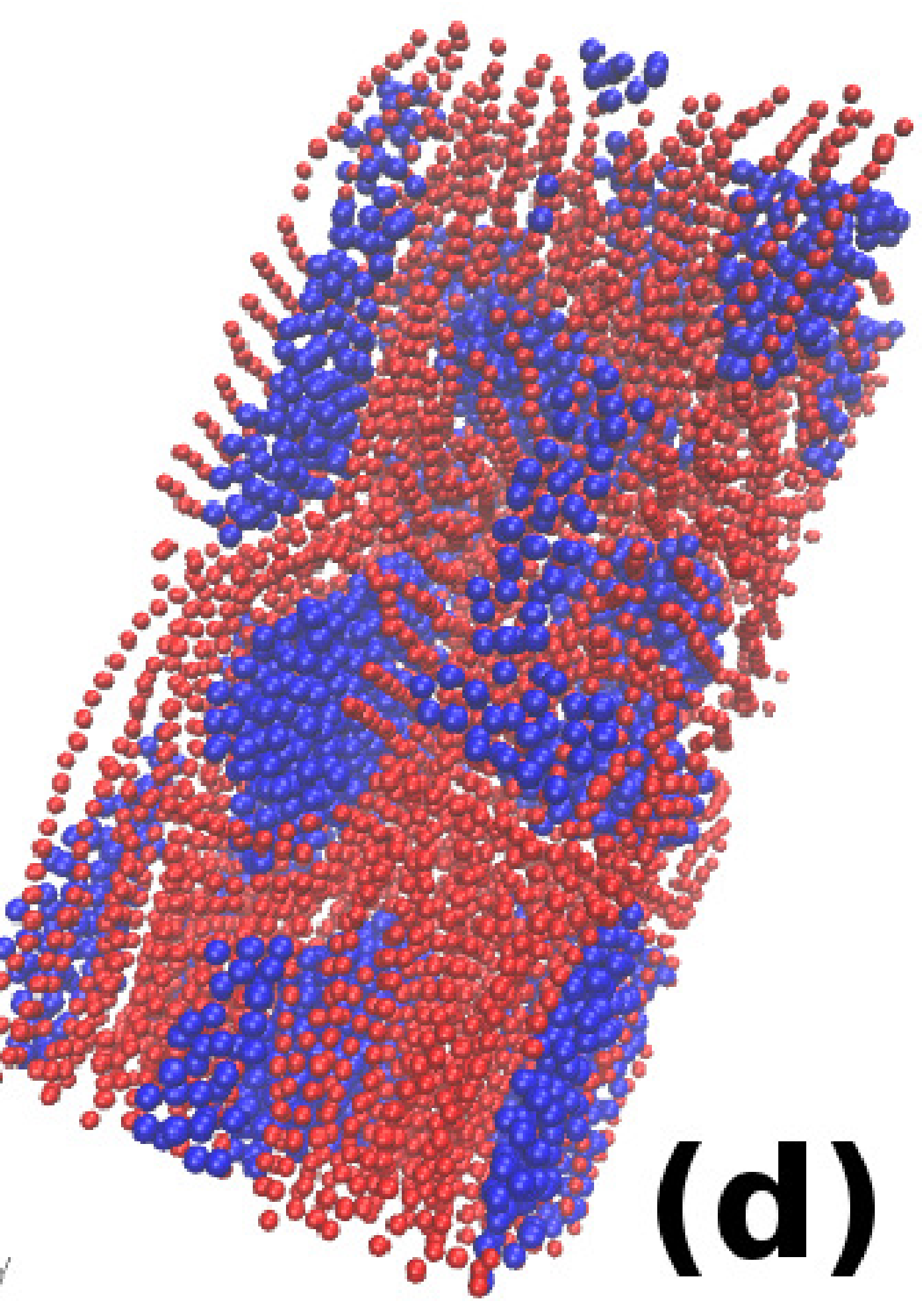} \\
\includegraphics[scale=0.2]{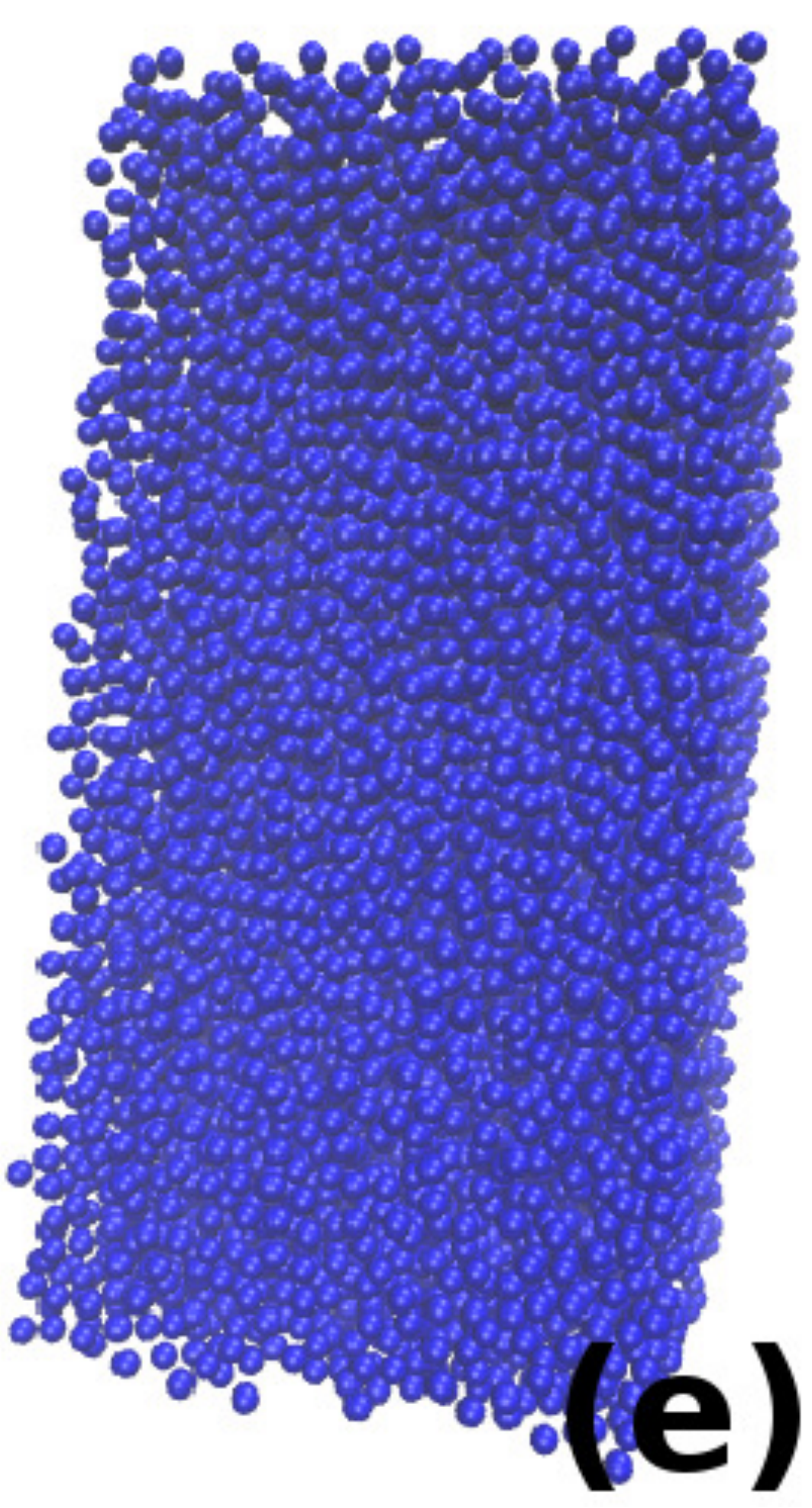}
\hspace{1cm}
\includegraphics[scale=0.2]{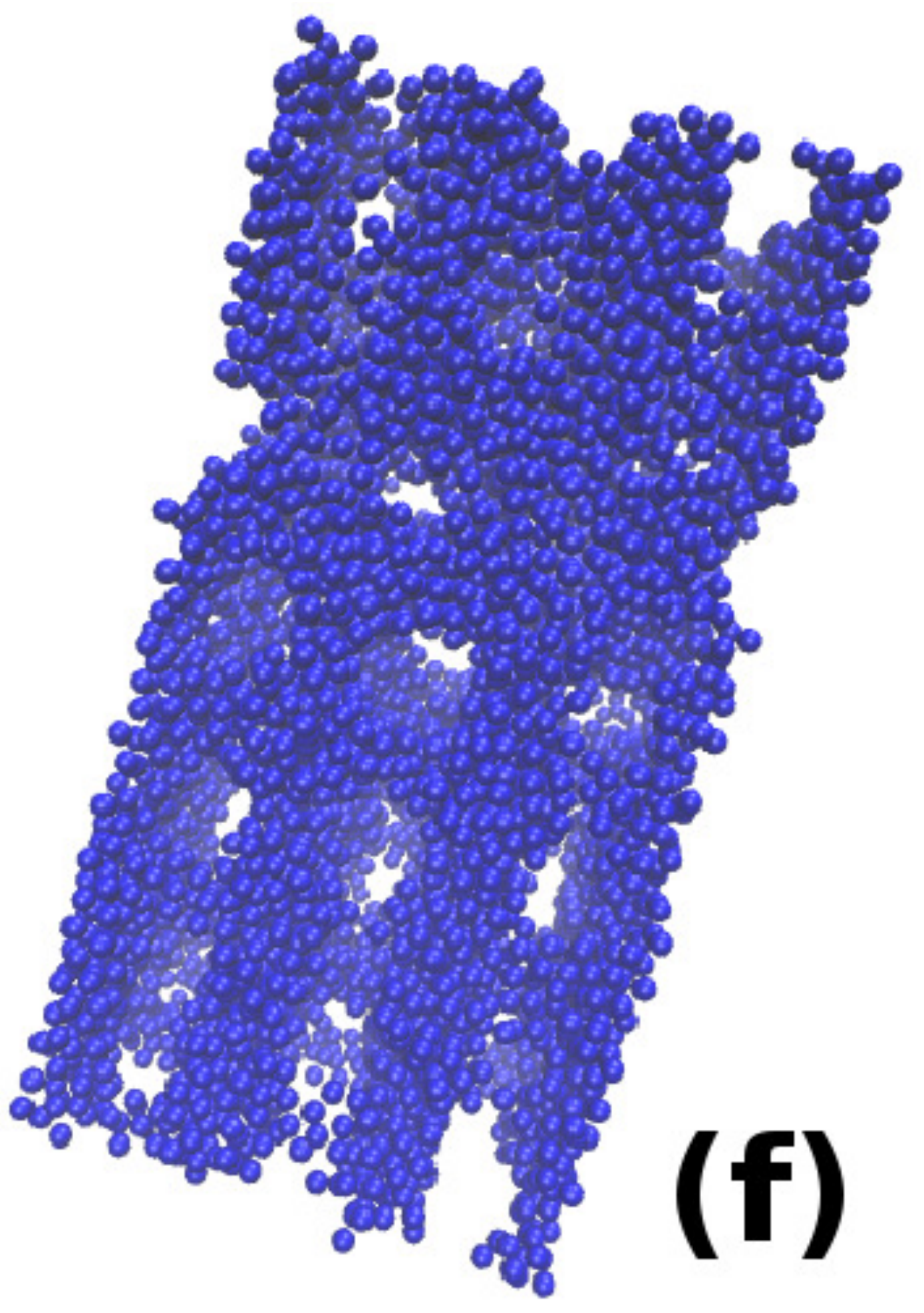}
\hspace{1cm}
\includegraphics[scale=0.2]{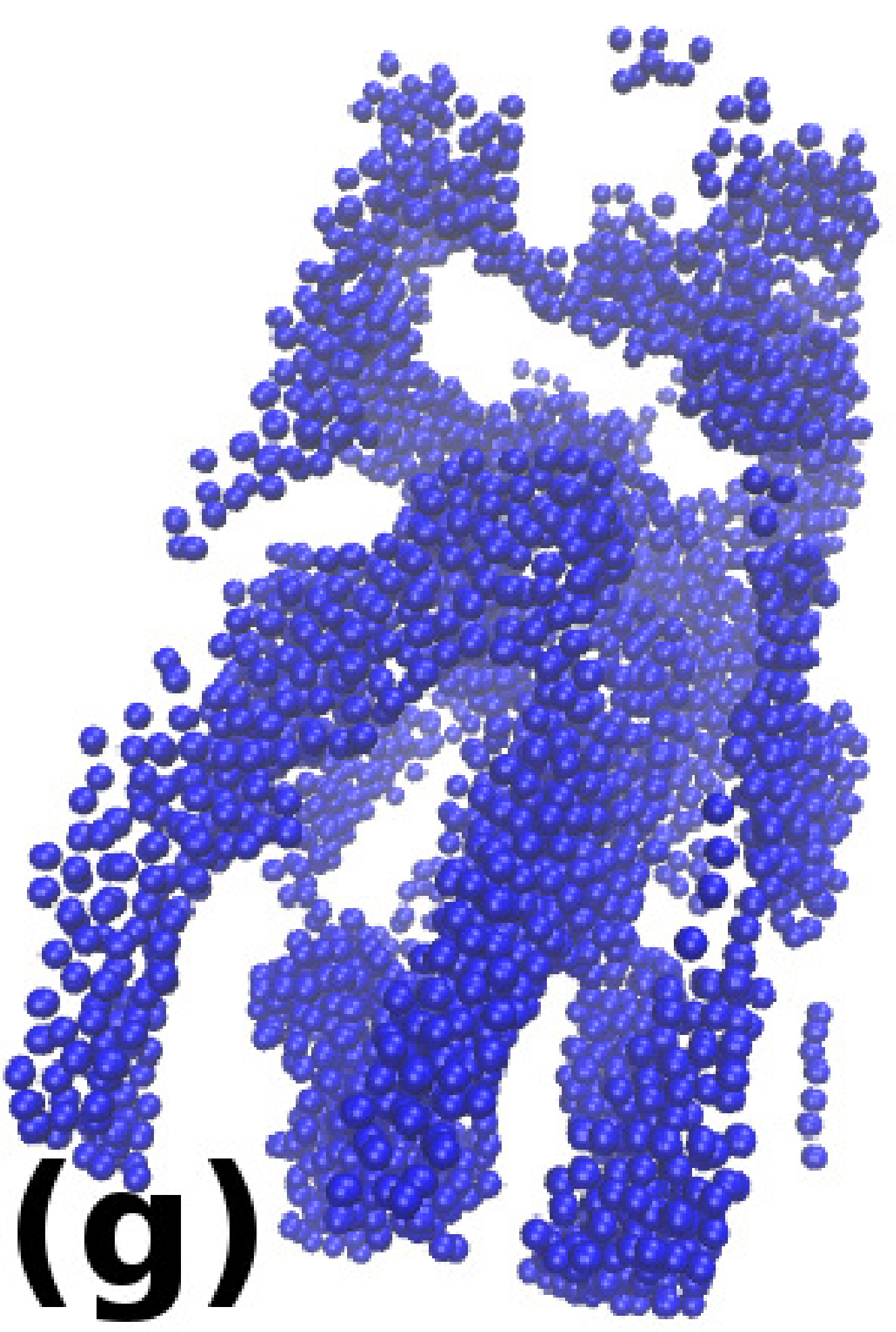}
\hspace{1cm}
\includegraphics[scale=0.2]{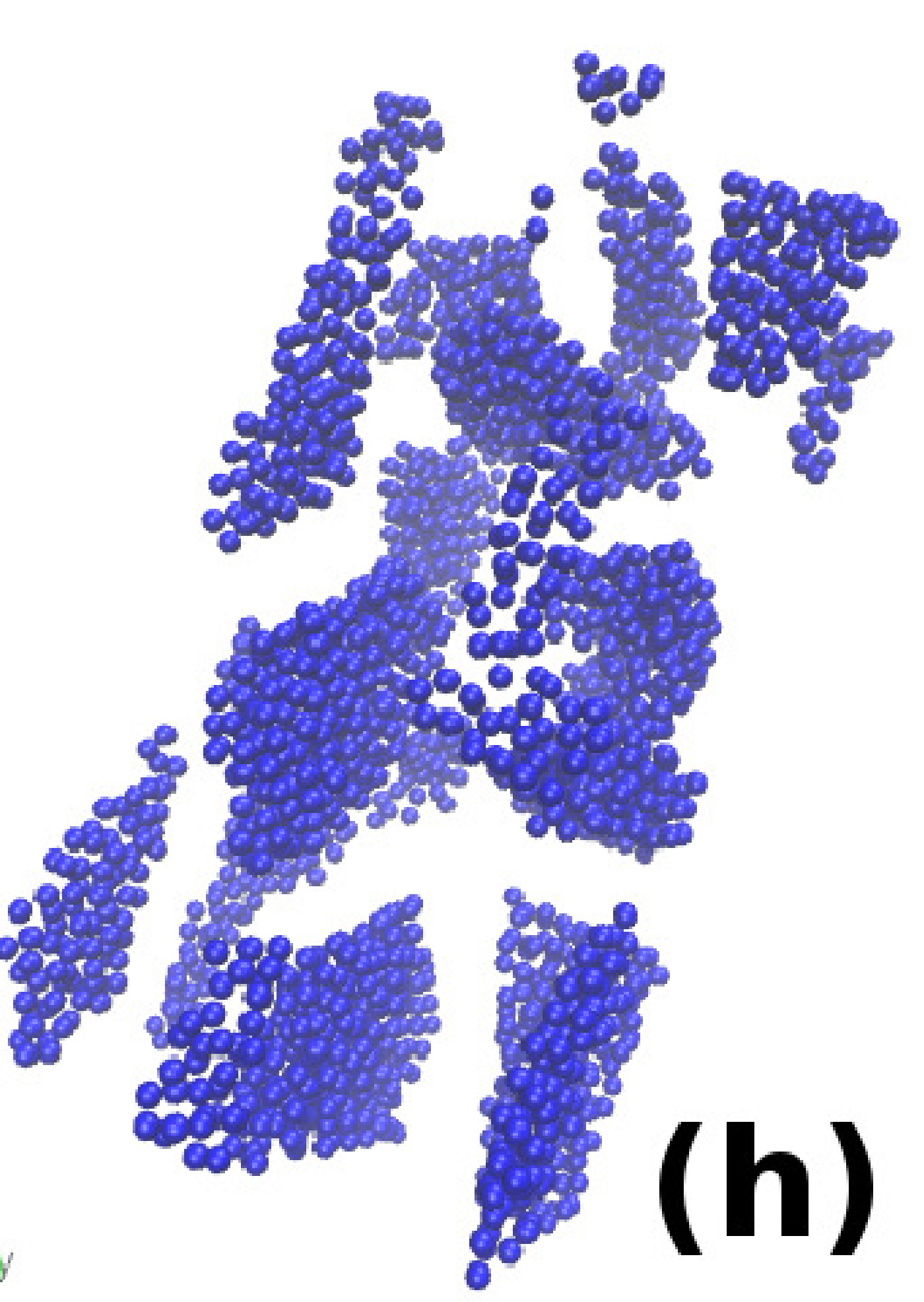}
\caption{(Colour online) The figure shows snapshots for four different values of $\sigma_{4n}$ (a) $1.25\sigma$, (b) $1.75\sigma$, (c) $2.5\sigma$ and (d)$2.75\sigma$ for $\epsilon_n=2k_BT$. The upper row shows both the nanoparticles and monomers, while the lower row shows only nanoparticles.For $\sigma_{4n}=1.25\sigma$, the micellar chains form a dispersed state. For $\sigma_{4n}>1.25\sigma$, the nanoparticles and micellar chains form interpenetrating network-like structures which show a morphological transition for $\sigma_{4n}=2.75\sigma$ forming individual sheets of nanoparticles.}
\label{epsn_2}
\end{figure*}


\begin{figure*}
\centering
\includegraphics[scale=0.2]{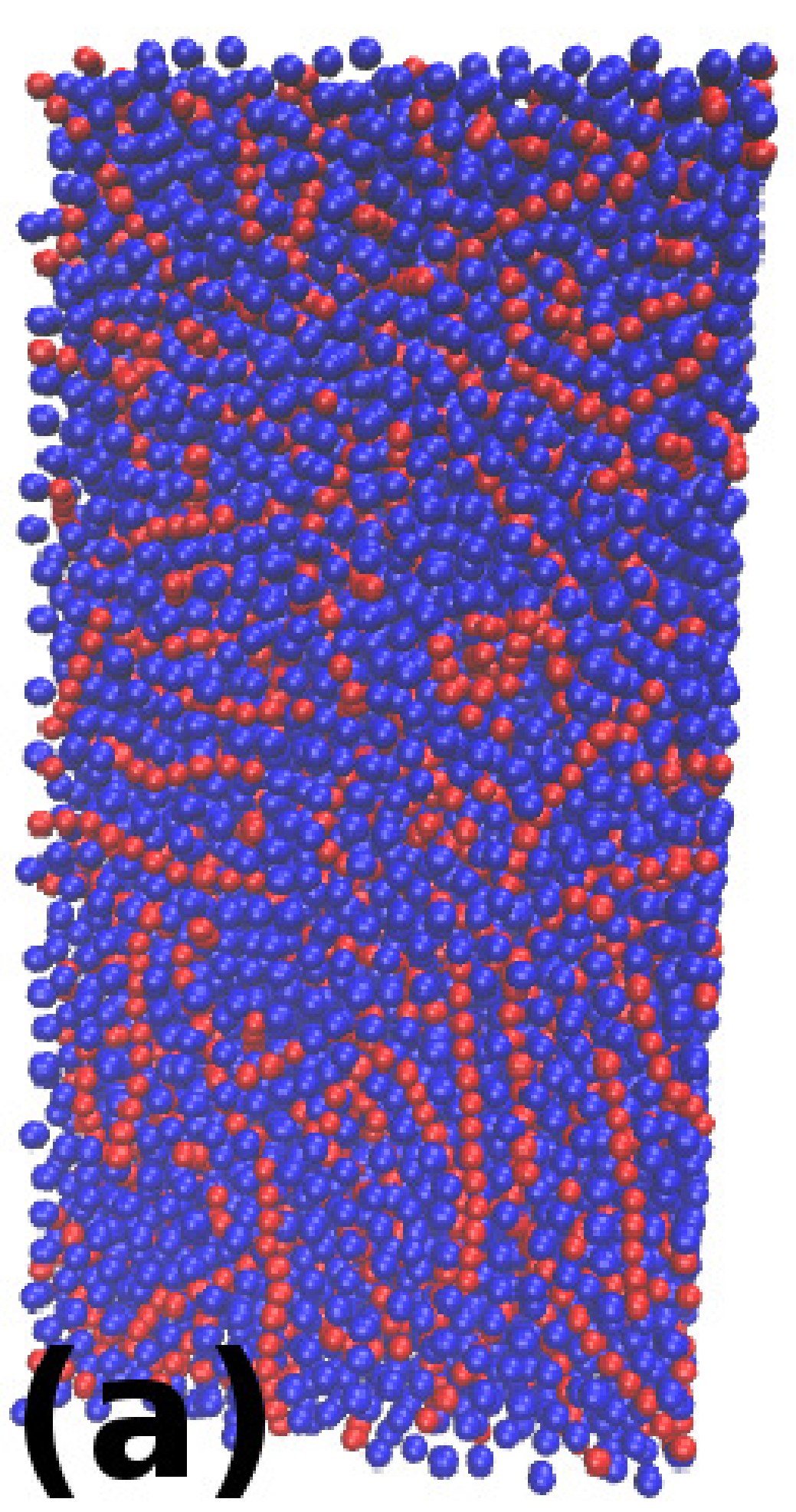}
\hspace{1cm}
\includegraphics[scale=0.2]{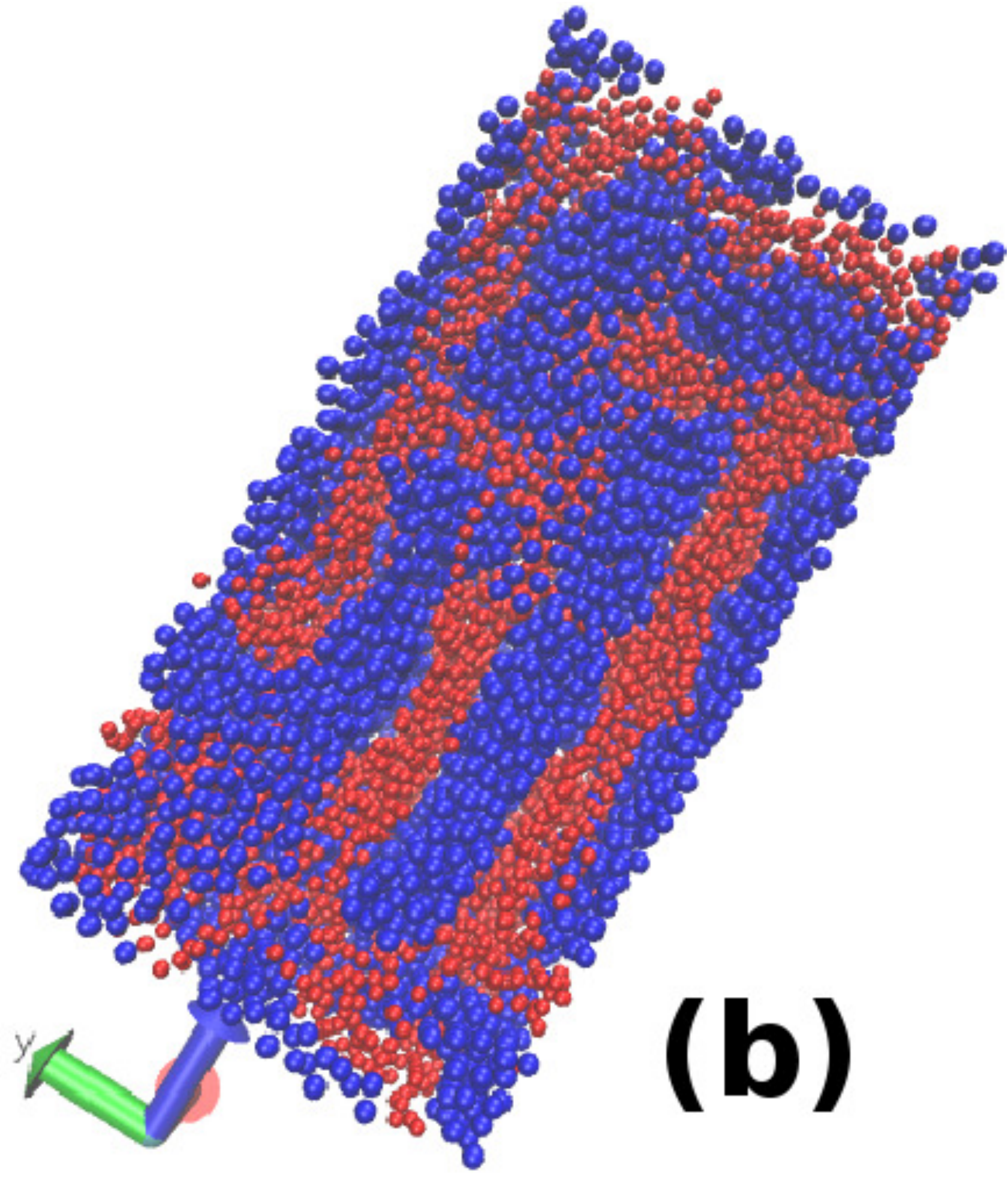}
\hspace{1cm}
\includegraphics[scale=0.2]{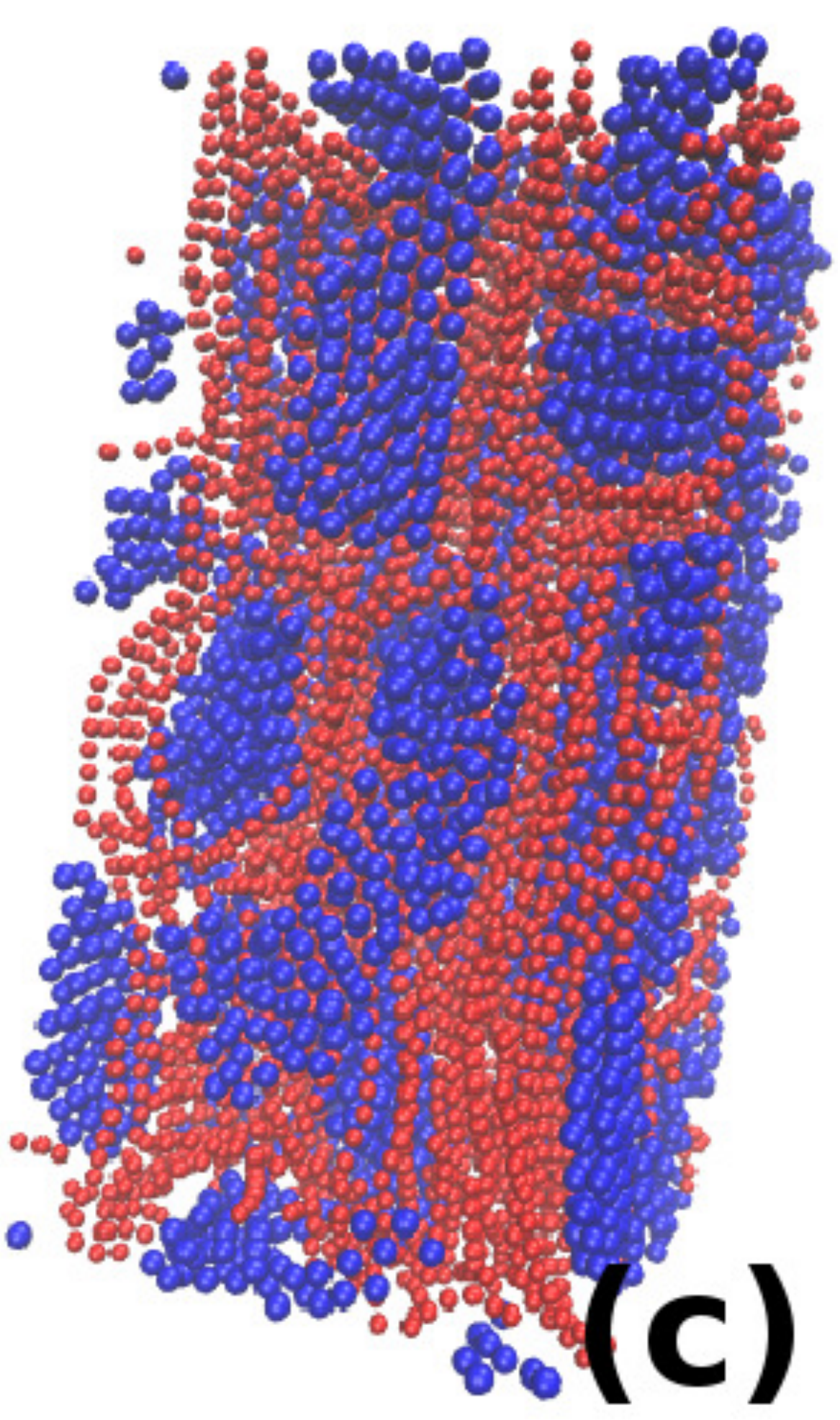}
\hspace{1cm}
\includegraphics[scale=0.2]{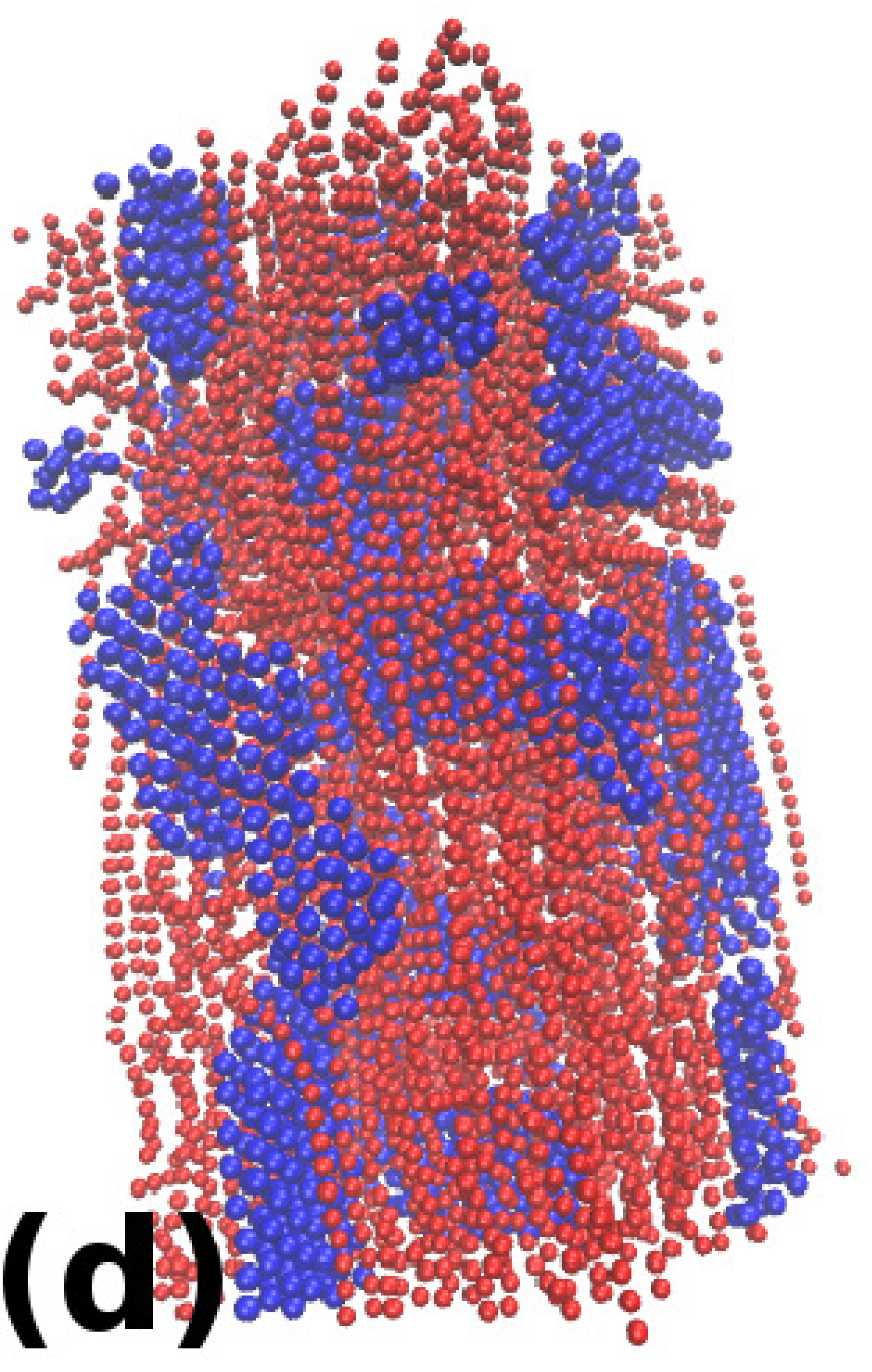} \\
\includegraphics[scale=0.2]{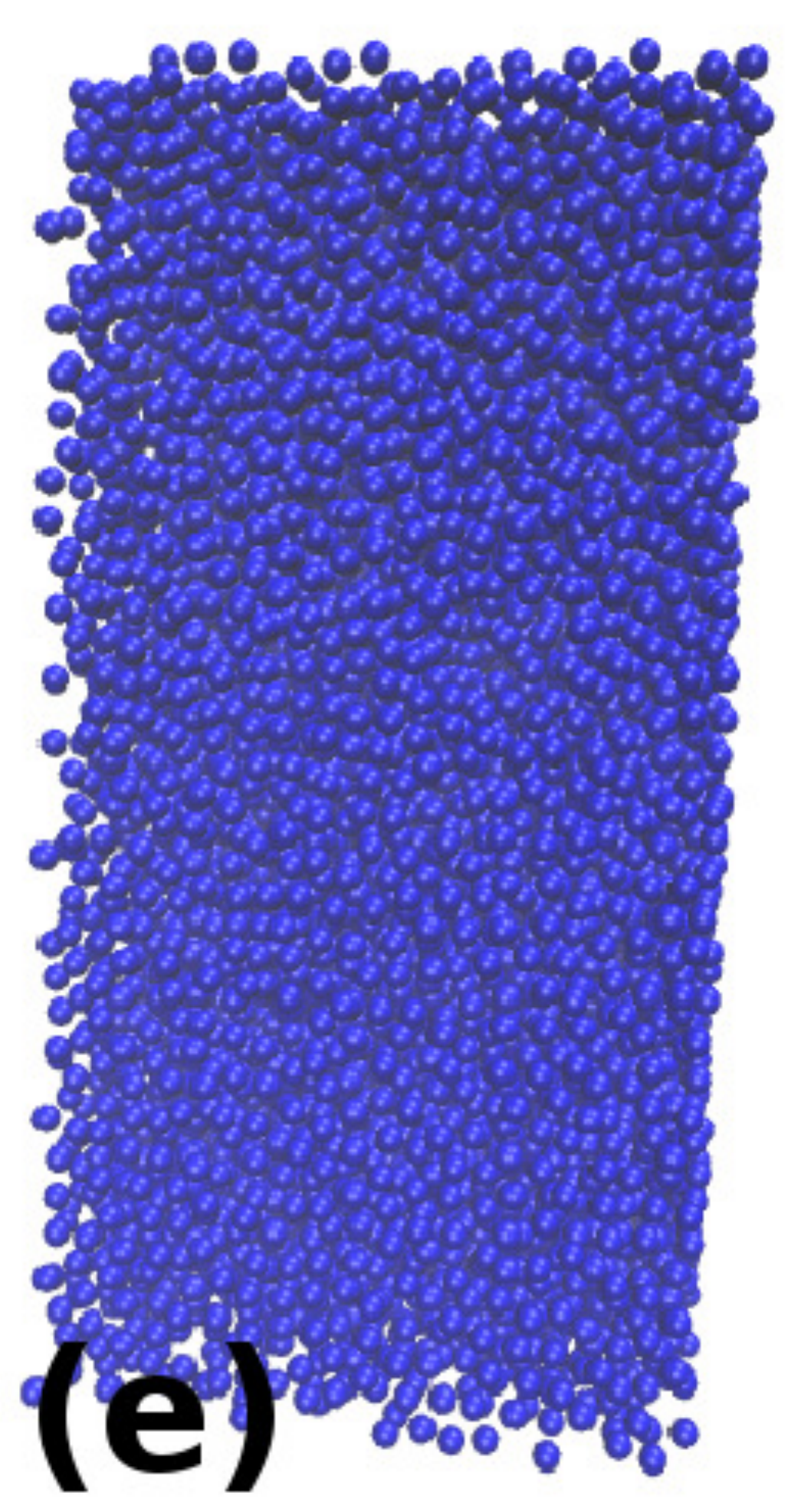}
\hspace{1cm}
\includegraphics[scale=0.2]{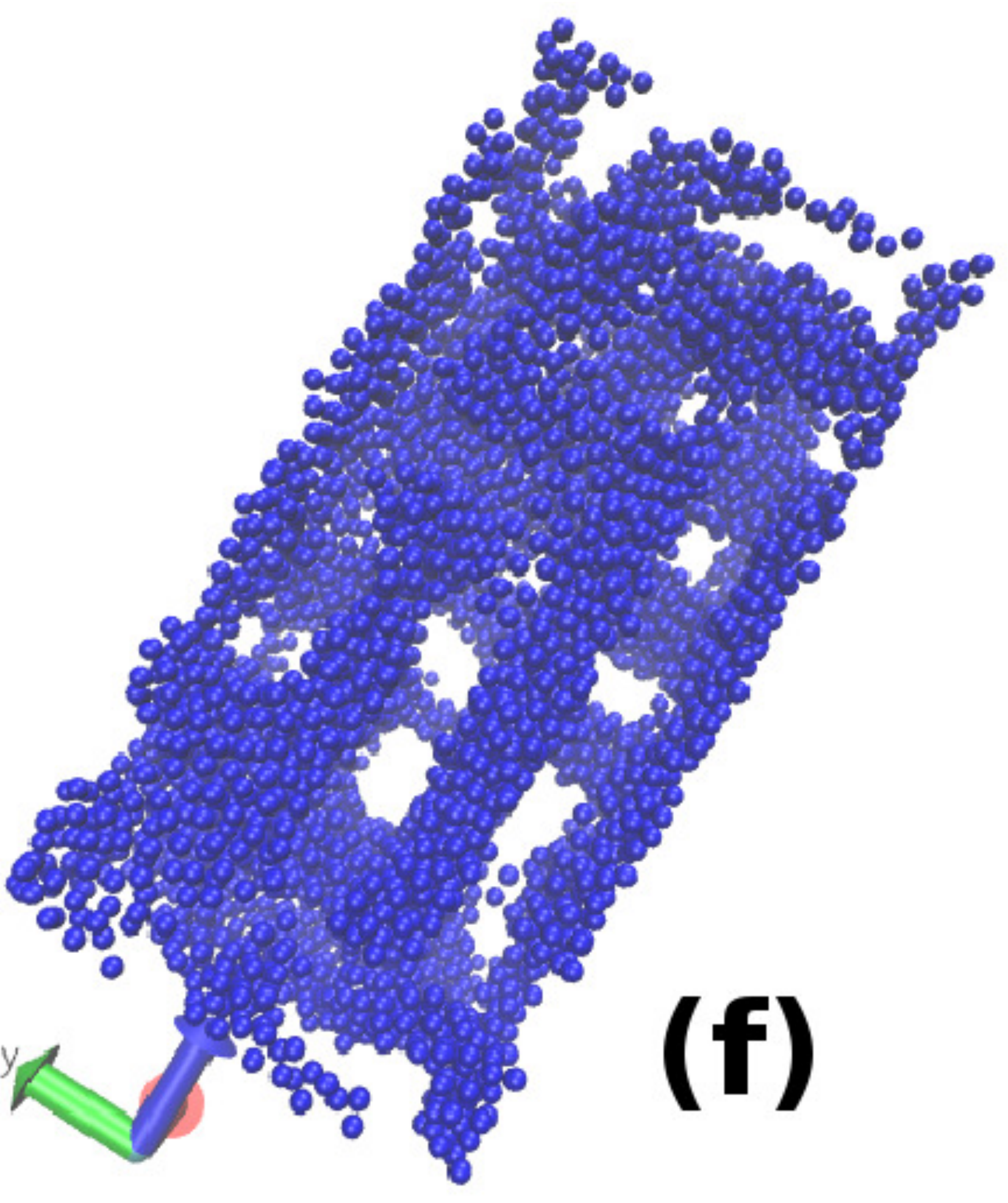}
\hspace{1cm}
\includegraphics[scale=0.2]{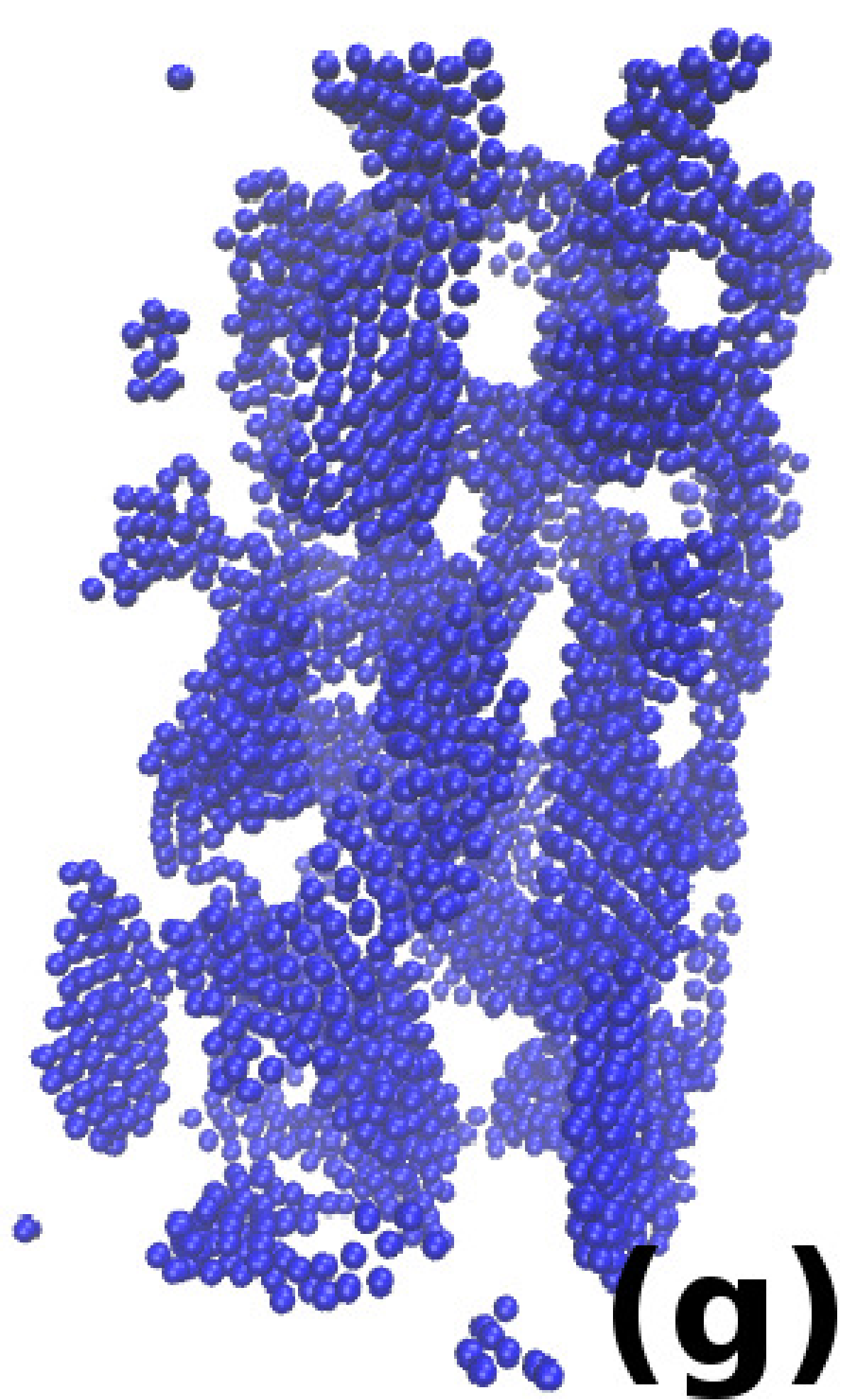}
\hspace{1cm}
\includegraphics[scale=0.2]{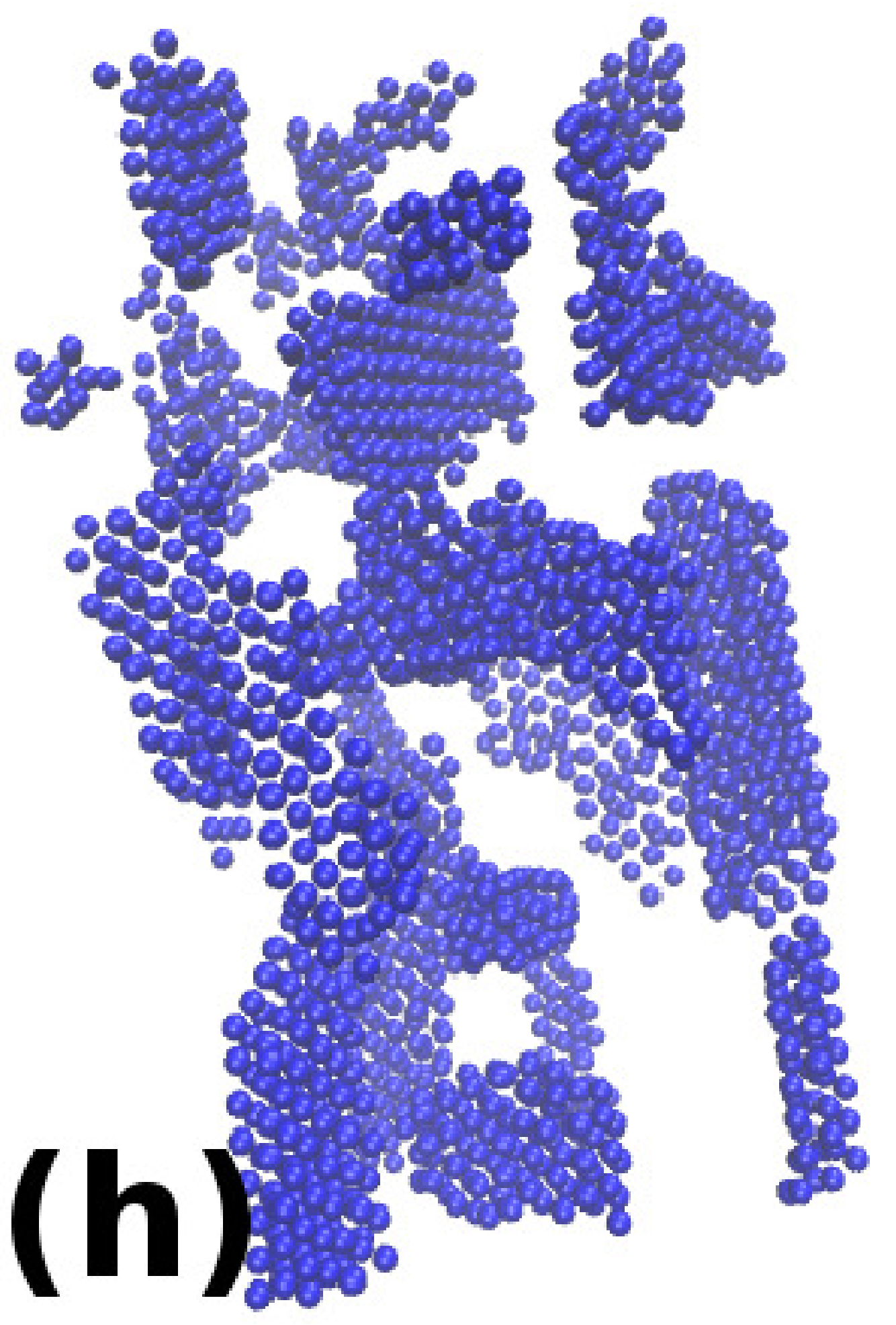}
\caption{(colour online) The figure shows snapshots for  four different values of $\sigma_{4n}$ from (a) to (d) (or from (e) to (h)), $1.25\sigma, 2\sigma, 2.5\sigma$ and $3\sigma$, respectively, for $\epsilon_n=11k_BT$. The upper row shows both the nanoparticles and monomers, while the lower row shows only nanoparticles. The polymeric chains at $\sigma_{4n}=1.25\sigma$ form a dispersed state. For $\sigma_{4n}>1.25\sigma$, the nanoparticles and micellar chains form interpenetrating networks which show a morphological change for $\sigma_{4n}=2.75\sigma$ forming individual sheets of nanoparticles.}
\label{epsn_11}
\end{figure*}

     The observed behaviour is found to be similar for $\rho_m=0.074\sigma^{-3}$, $0.093\sigma^{-3}$ and $0.126\sigma^{-3}$. Therefore, only the snapshots for $\rho_m=0.093\sigma^{-3}$ are used here to illustrate the behaviour for all these densities. Few snapshots for illustration for other densities are also shown at the end. For each value of $\rho_m$ and $\epsilon_n$, the value of $\sigma_{4n}$ is varied. These snapshots are shown in Figs. \ref{epsn_2} and \ref{epsn_11} for values of $\epsilon_n = 2k_BT$ and $11k_BT$ respectively. Each figure shows four different snapshots for different values of $\sigma_{4n}$ increasing from (a) to (d) (or (e) to(h)). The upper row shows both the micelles(red particles) and nanoparticle (blue), while, the lower row shows only nanoparticles. For the size of nanoparticle $\sigma_n=1.5\sigma$ considered here, the minimum value of $\sigma_{4n}$ is $1.25\sigma$ ~\cite{2018arXiv180106933M}. For this value of $\sigma_{4n}=1.25\sigma$, which corresponds to the snapshots in (a) and (e) in both the figures \ref{epsn_2} and \ref{epsn_11}, the micellar chains and nanoparticles form a uniformly mixed state. No two micellar chains are found without nanoparticles in between (i.e no clustering of chains). An increase in the value of $\sigma_{4n}$ from $1.25\sigma$ leads to the formation of clusters of micellar chains. Therefore, in the snapshots shown in (b) in both the figures for different $\epsilon_n$, the micellar chains are found to form a network-like structure. Therefore, nanoparticles in (f) are also forming network-like structures. A further increase in the value of $\sigma_{4n}$, we see nanoparticles forming intermediate states where the network of nanoparticles gradually breaks (figures (c) and (g)). For some higher value of $\sigma_{4n}$ (depending on $\epsilon_n$), the networks show a transition to individual clusters of nanoparticles as shown in figure (h) in both the figures.

Similar morphological changes are observed for other values of $\rho_m$, but with a difference of the anisotropy of nanoparticle clusters. For $\rho_m=0.093\sigma^{-3}$, the nanoparticles form sheet-like structures while, the anisotropy increases with increase in $\rho_m$. Therefore, for $\rho_m=0.126\sigma^{-3}$ rod-like nanoparticle clusters are observed. However, irrespective of the value of $\epsilon_n$, similar morphological transitions are observed for all $\epsilon_n$. This shows that changes in the morphology of the structures are due to change in $\sigma_{4n}$ only. Moreover, the sheet-like morphology of the structures formed in figure (h) in both the figures \ref{epsn_2} and \ref{epsn_11} verifies the previous result that the micellar density governs the morphology of nanoparticle structures. For $\epsilon_n=0$, though, without attractive interaction between nanoparticles a nanostructure cannot be formed, it only shows the arrangement of nanoparticles as a result of the arrangement of micellar chains. Therefore, the interaction between nanoparticles for $\epsilon_n=0$ is the interaction mediated by the micellar matrix. Thus, the value of $\epsilon_n$ does not seem to be affecting the morhological behaviour of the system. However, the change in $\epsilon_n$ shows to effect two things. The value of $\sigma_{4n}$ at which the morphological change occurs and the packing of nanoparticles.

       Examining the figures \ref{epsn_2} and \ref{epsn_11}, we see that the value of $\sigma_{4n}$ at which the nanoparticles undergo a transition from network-like morhology to individual clusters of nanoparticles, gets shifted to a higher value of $\sigma_{4n}$ with increase in $\epsilon_n$. For $\epsilon_n=0$, the breaking of network into individual clusters occurs at $\sigma_{4n}=2.5\sigma$ ~\cite{supporting_material}, but this happens at $\sigma_{4n}=2.75\sigma$ and $3\sigma$ for the value of $\epsilon_n=2 k_BT (or 5k_BT)$ and $11k_BT$ respectively (figures \ref{epsn_2}(h) and \ref{epsn_11}(h)). This is because nanoparticle density increases with increase in $\epsilon_n$, but decreases with increase in the value of $\sigma_{4n}$. Therefore to reach the low nanoparticle density required to produce individual clusters, systems with higher value of $\epsilon_n$ need to get higher values of $\sigma_{4n}$. Though, the value of EVP at which nanoparticles morphology changes from network to non-percolating clusters gets shifted, but the change in micellar chains structure from dispersed state to formation of clusters, with change from $\sigma_{4n}=1.25\sigma$ to $1.5\sigma$ does not change with change in $\epsilon_n$.

\begin{figure}
\centering
\includegraphics[scale=0.2]{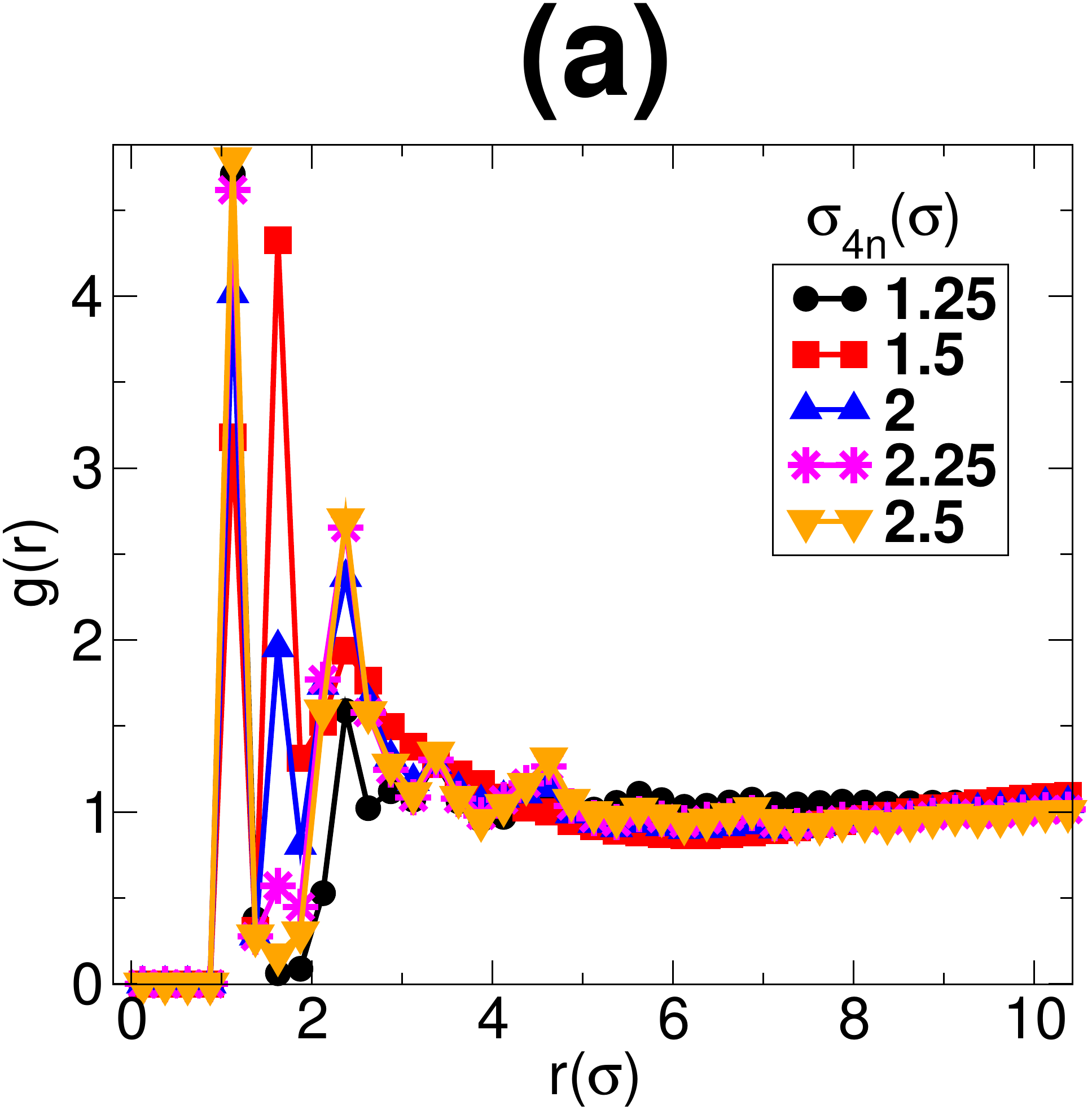}
\includegraphics[scale=0.2]{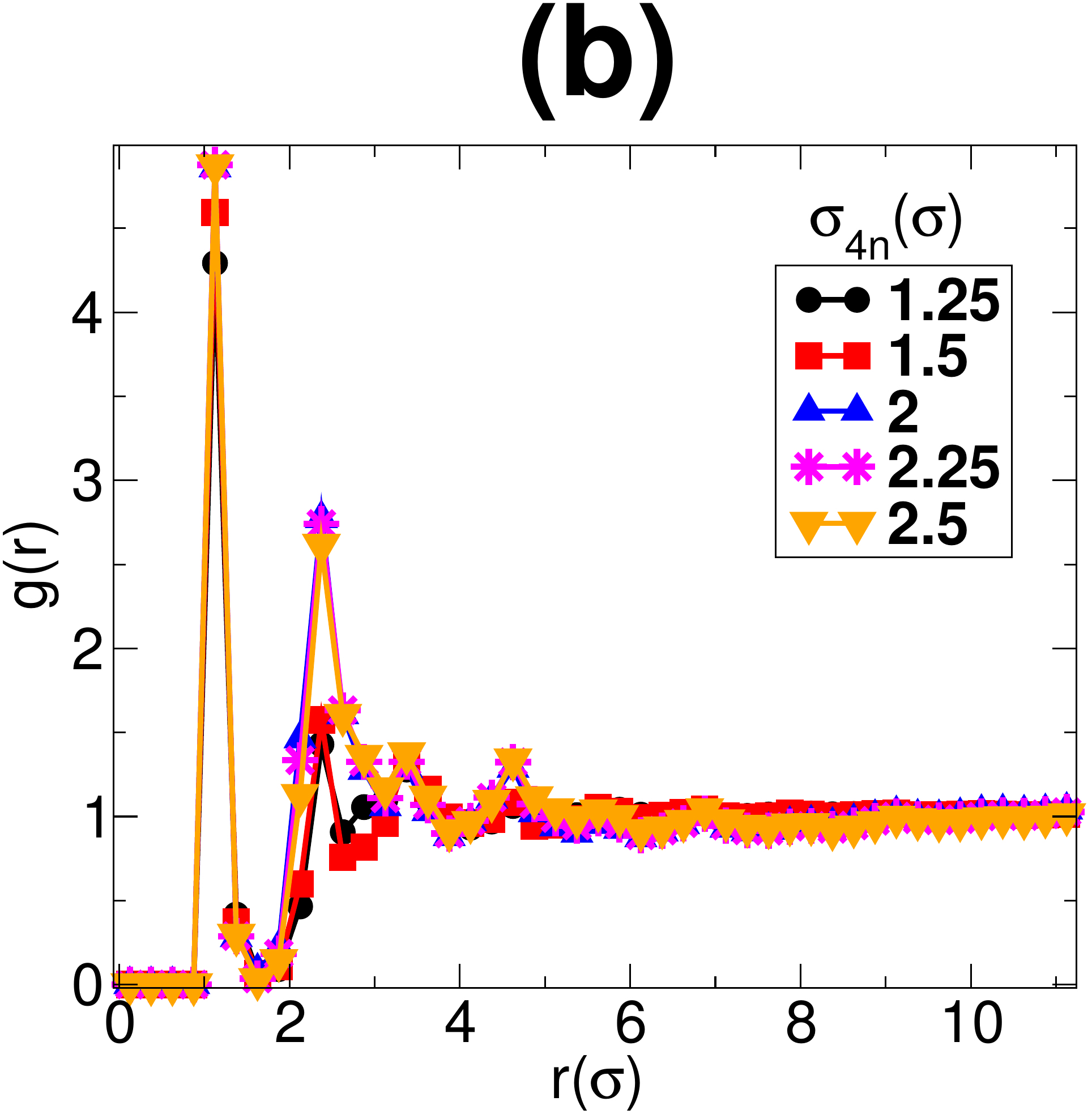}
\caption{(Colour online) The figure shows the pair correlation function of monomers for values of $\epsilon_n$ (a) $8k_BT$ and (b) 0. Each figure shows five different graphs for values of $\sigma_{4n}=1.25\sigma$, $1.5\sigma$, $2\sigma$, $2.25\sigma$ and $2.5\sigma$. A peaks around $1.75\sigma$ indicates clustering of micellar chains. For $\epsilon_n=8k_BT$, there appear peaks around $1.75\sigma$, which decreases with increase in $\sigma_{4n}$. However, there are no peaks around $1.75\sigma$ for $\epsilon_n=0$.}
\label{corr_mono}
\end{figure}

      An increase in the value of $\sigma_{4n}$ demands an increase in the excluded volume of the system. Therefore, with increase in $\sigma_{4n}$ from $1.25\sigma$ to $1.5\sigma$, the system reorganizes itself to lower its excluded volume due to both $V_4$ and $V_{4n}$. This reorganization evokes a competition between the excluded volume due to $V_4$ and the excluded volume due to $V_{4n}$. When the value of $\sigma_{4n}$ increases from $1.25\sigma$ to $1.5\sigma$, then the micellar chains reorganizes to form clusters in order to decrease $V_{4n}$, but increasing $V_4$. This is because the number (or energy) of nanoparticles is higher than monomers (this we will see in the following sections). When the value of $\sigma_{4n}$ increases to a high value such that decreasing the distance between chains is more costly in terms of energy, then the nanoparticle density is decreased and the nanoparticle network starts breaking. When the nanoparticle network breaks to an extent that micellar chains get enough volume to be away from each other's repulsive interaction ($V_4$) range ($2^(1/6)\sigma_4$), then the total excluded volume of the system decreases. This change in the distance between micellar chains can be confirmed by plotting the pair correlation function. This is shown in Fig.~\ref{corr_mono}(a).
      
     The figure~\ref{corr_mono}(a) shows plots of monomer pair correlation function for $\epsilon_n=2k_BT$. The behaviour of the pair correlation function for monomers is similar for all $\epsilon_n>0$. The Fig.\ref{corr_mono}(a) shows graphs for five different values of $\sigma_{4n}=1.25\sigma$, $1.5\sigma$, $2\sigma$, $2.25\sigma$ and $2.5\sigma$. For monomer pair correlation function, a first peak is expected at $\approx \sigma$ and its multiples indicating the monomers which are part of a chain. Peaks are also expected to occur at $1.75\sigma$ and its multiples if the chains are within the range of repulsive potential $V_4$. In the figure, the pair correlation function for $\sigma_{4n}=1.25\sigma$ does not show any peak around $1.75\sigma$. This shows that the chains are dispersed in between nanoparticles as shown in the snapshots in Figs. \ref{epsn_2}(a) and \ref{epsn_11}(a). With the increase in $\sigma_{4n}$ from $1.25\sigma$, there appears peaks around $1.75\sigma$ and its multiples. This shows the formation of clusters of micellar chains that join to form a network as shown in Figs.\ref{epsn_2}(b) and \ref{epsn_11}(b). However, this peak decreases in its height with the increase in $\sigma_{4n}$. This decrease is due to the decrease in the density of nanoparticles or breaking of the nanoparticle network that decreases the excluded volume between micellar chains. For $\sigma_{4n}=2.75\sigma$, the network breaks to an extent that micellar chains get enough volume to be out of the range of the repulsive interaction ($V_4$) from each other.

 If nanoparticle energy is lower (-ve) than micellar chains, then an increase in $\sigma_{4n}$ will lead to more clustering of micellar chains such that the distance between micellar chains decreases while the distance between micellar chains and nanoparticles increases (micellar chains are "pushed" by nanoparticles). However, if nanoparticle energy is higher, then nanoparticle density decreases without any decrease in the distance between micellar chains. This competition between nanoparticles and micellar chains energies can be clearly observed by examining the pair correlation functions for different values of $\epsilon_n$. The Figure ~\ref{corr_mono}(b) represents the monomer pair correlation function for $\epsilon_n=0$. Comparing the two figures Fig.\ref{corr_mono}(a) and Fig.\ref{corr_mono}(b), we can clearly see that there are no peaks around $1.75\sigma$ for $\epsilon_n=0$. This clearly shows that energy of nanoparticles in case of $\epsilon_n=0$ is not competitive with monomers and hence, micellar chains do not form clusters. Therefore, for $\epsilon_n=0$, with the increase in $\sigma_{4n}$, the number of nanoparticles decreases (or the nanoparticle network breaks) without decreasing the distance between micellar chains.

\begin{figure}
\includegraphics[scale=0.3]{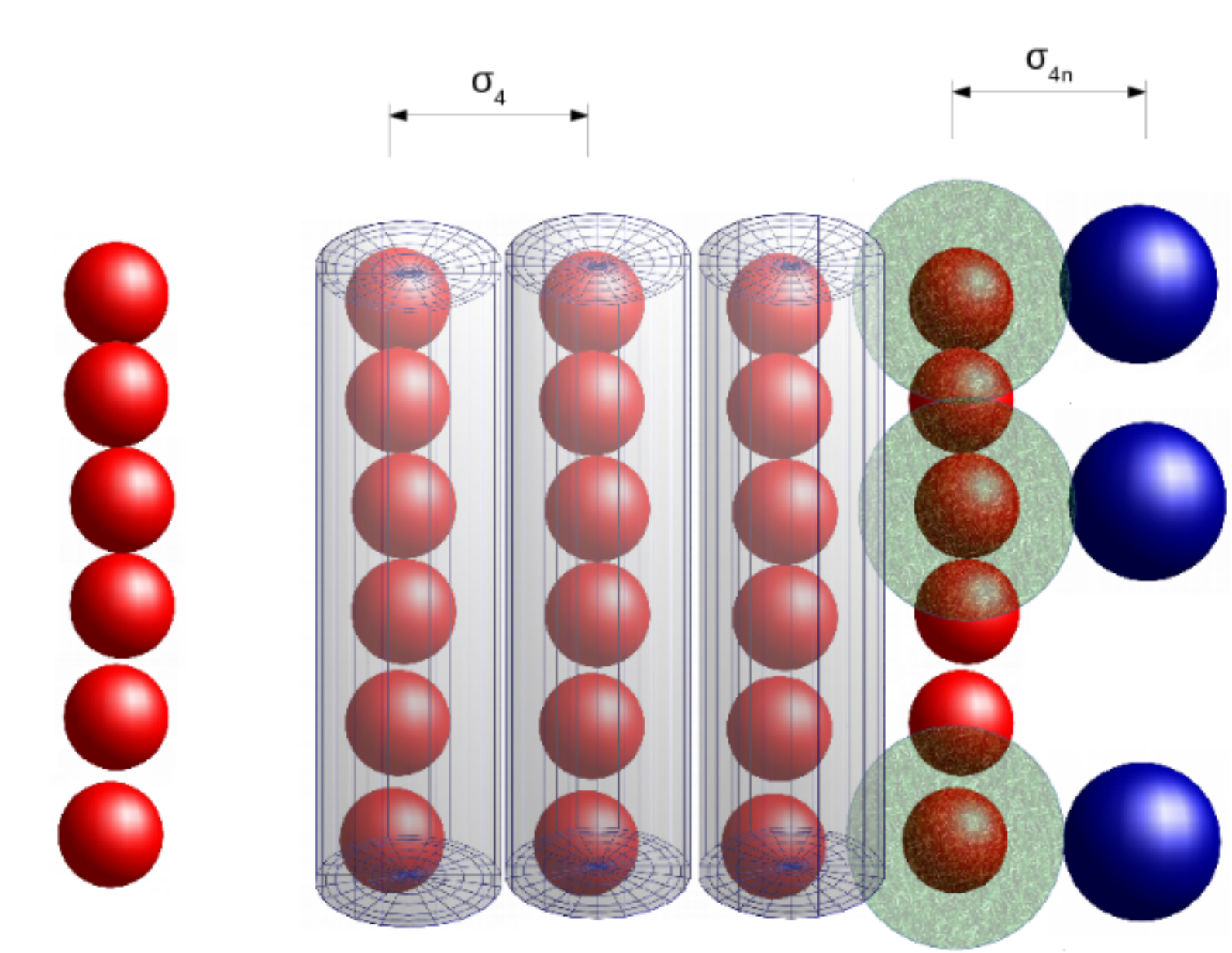}
\caption{(Colour online) The figure explains the calculation of the effective volume of micelles. if any two micellar chains are at a distance $r <= \sigma_4$ from each other, they are considered as cylinders of radius $\sigma_4/2$ shown as a shaded region(red). When a nanoparticle is at a distance $r <= \sigma_{4n}$ from a monomer, then the monomer is assumed as a sphere of radius $\sigma_{4n}-\sigma_n/2$. }
\label{eff_vol}
\end{figure}

    Thus we see that with the increase in $\sigma_{4n}$ the total excluded volume of the system increases as a result of which the system reorganizes itself. Therefore, it is realized that the behaviour of the system can be explained well if we take into account of the excluded volume in the system. To take into account of the excluded volume, the volume of the matrix polymeric chains are described along with the excluded volume which we call as the effective volume of micelles (or monomers). The effective volume of monomers ${V_m}^{eff}$ is defined as the total excluded volume due to repulsive interactions between chains of monomers $V_4$ and in between the monomers and nanoparticles $V_{4n}$ in addition to the volume of monomers. The scheme to calculate the effective volume of micelles (or EPs) is shown in Fig.\ref{eff_vol}. The figure shows that any two micellar chains at a distance $r < \sigma_4$ ($\sigma_4$, the cutoff distance for $V_4$) are considered as cylinders of diameter $\sigma_4$, while any monomer at a distance $r < \sigma_{4n}$ ($\sigma_{4n}$, cutoff distance for $V_{4n}$) is considered as a sphere of radius $\sigma_{4n}-\sigma_n/2$.
        
         To calculate the effective volume of matrix polymers, a suitable algorithm is used to first sort out monomers which are part of a single chain. Then all the chains involved in the repulsive interaction $V_4$ with other chains or repelling a nanoparticle with $V_{4n}$ are found. Then using the scheme explained in Fig.\ref{eff_vol}, the effective volume of chains is calculated. This effective volume not only depends on the value of $\sigma_{4n}$ but also on the arrangement of the constituent particles that determines the number pairs of particles repelling each other. This, in turn, depends on the density of nanoparticles. Using the scheme shown in Fig.\ref{eff_vol}, the effective monomer volume fraction may get slightly over-estimated, but that is insignificant and does not affect the results.

\begin{figure}
\includegraphics[scale=0.2]{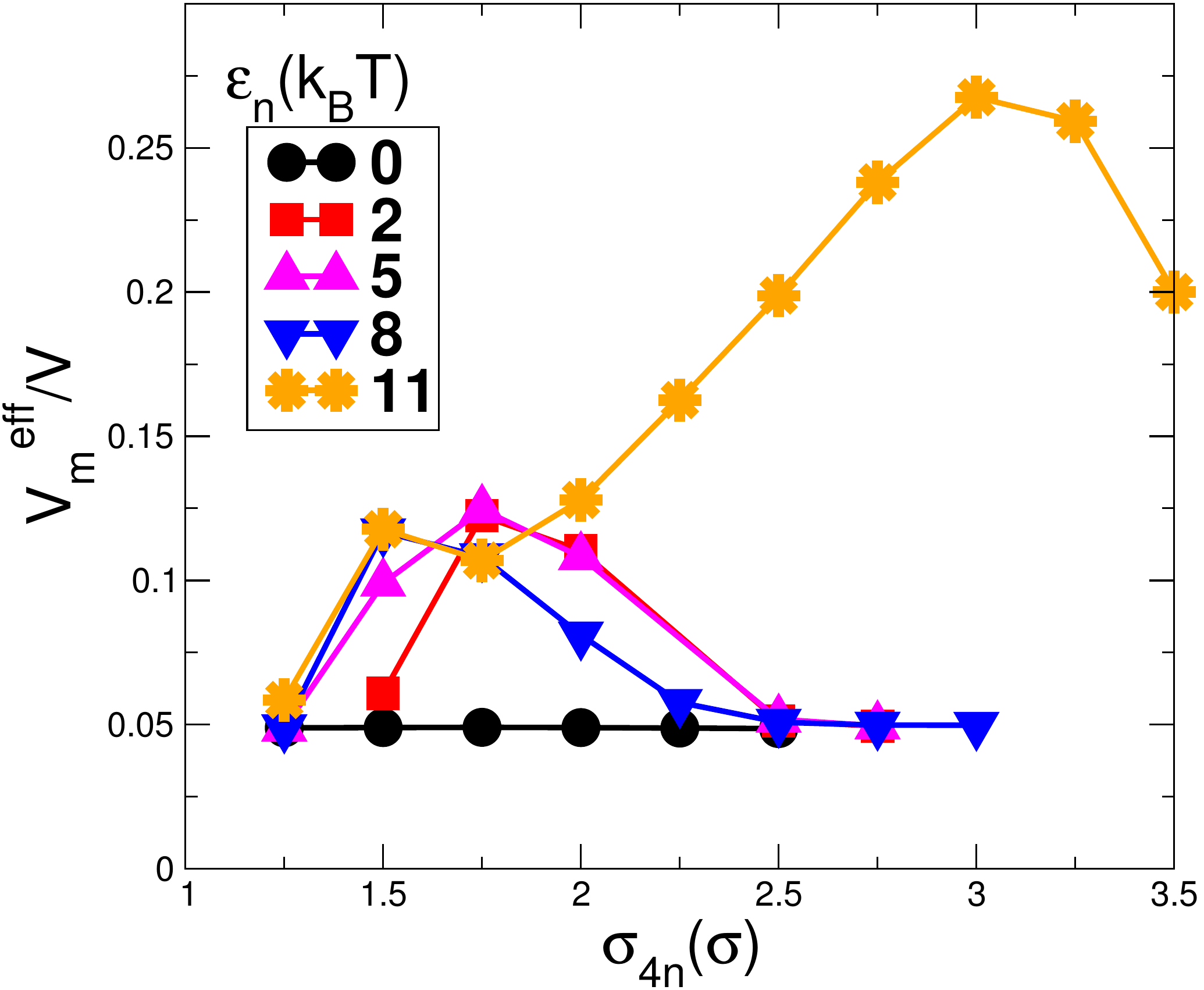}
\includegraphics[scale=0.2]{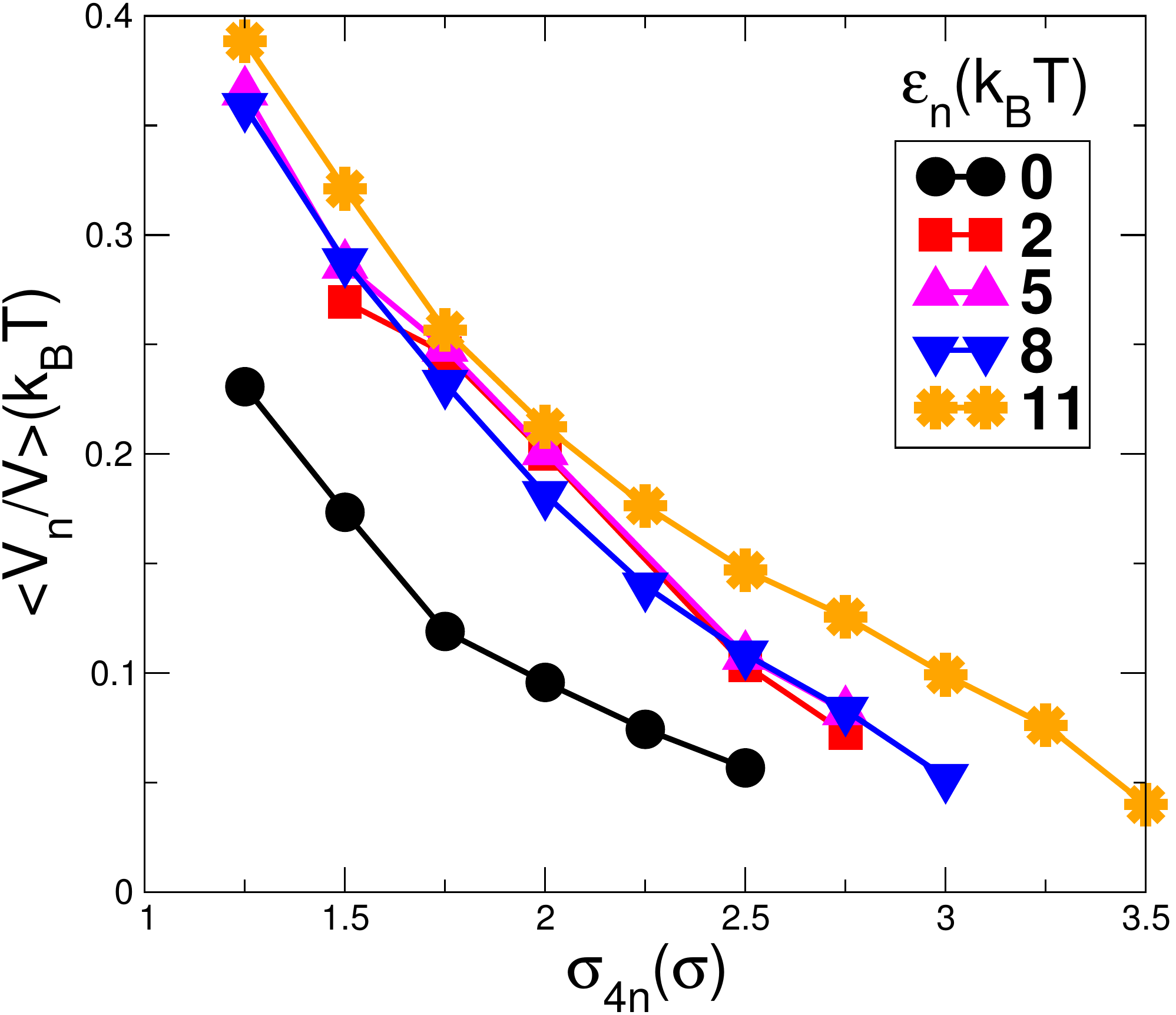}
\caption{(Colour online) The figure shows (a) the effective volume fraction of monomers and (b) volume fraction of nanoparticles. The effective volume of micelles shows an increase in its value with increase in $\sigma_{4n}$ marking the change from a dispersed state to the formation of clusters of chains. With further increase in $\sigma_{4n}$, it decreases for all values of $\epsilon_n$ except for $\epsilon_n=11k_BT$ indicating the presence of a high competition between the clusters of nanoparticles and monomer chains. The changes in $\epsilon_n=0$ are of the very low order of magnitude and appear to be constant in this graph. The nanoparticle volume fraction shows a decreasing behaviour with an increase in the value of $\sigma_{4n}$ for all the values of $\epsilon_n$. For $\epsilon_n=0$, the values of nanoparticle volume fractions are lower than the values for other  $\epsilon_n$. }
\label{vol_frac}
\end{figure}

          The behaviour of the effective volume fraction of monomers is shown in Fig.~\ref{vol_frac}(a) and Fig.~\ref{vol_frac}(b) shows the volume fraction of nanoparticles. Each figure shows graphs for different values of $\epsilon_n$.  It should be noted that for a given value of $\sigma_{4n}$, a high value of the effective volume of monomers indicates the presence of a large number of pairs of particles having repulsive interactions between monomer chains or between monomers and nanoparticles. In Fig~\ref{vol_frac}(a), the effective volume shows an increase in its value with increase in $\sigma_{4n}$ from $1.25\sigma$, representing the change from a dispersed state of chains to its clusters. Then it shows a decrease in its value for further increase in $\sigma_{4n}$ for all the values of $\epsilon_n$ except $\epsilon_n=11k_BT$. A decrease in the value of ${V_m}^{eff}$ shows the breaking of nanoparticle network to the extent that micellar chains get enough volume to be away from each other's repulsive interaction range.  We see that only for $\epsilon_n=11k_BT$, the value of effective volume keeps on increasing with increase in $\sigma_{4n}$ and then show a decrease at a higher value of $\sigma_{4n}=3\sigma$. The changes in ${V_m}^{eff}$ are insignificant for $\epsilon_n=0$ and it shows a nearly constant low value of ${V_m}^{eff}$. Observing the graphs for nanoparticle volume fraction in Fig.\ref{vol_frac}(b) one can see that the nanoparticle volume fraction not only decreases with increase in $\sigma_{4n}$ but also decreases slightly for a decrease in $\epsilon_n$. The plots show nearly similar values for all $\epsilon_n > 0$ but, show distinctly lower values for the case of $\epsilon_n=0$. As explained earlier, the increase in the excluded volume is because of the competition between the clusters of nanoparticles and monomer chains. Therefore, the behaviour of the ${V_m}^{eff}$ can be confirmed by plotting the average energies of the nanoparticles and monomers.

\begin{figure}
\includegraphics[scale=0.2]{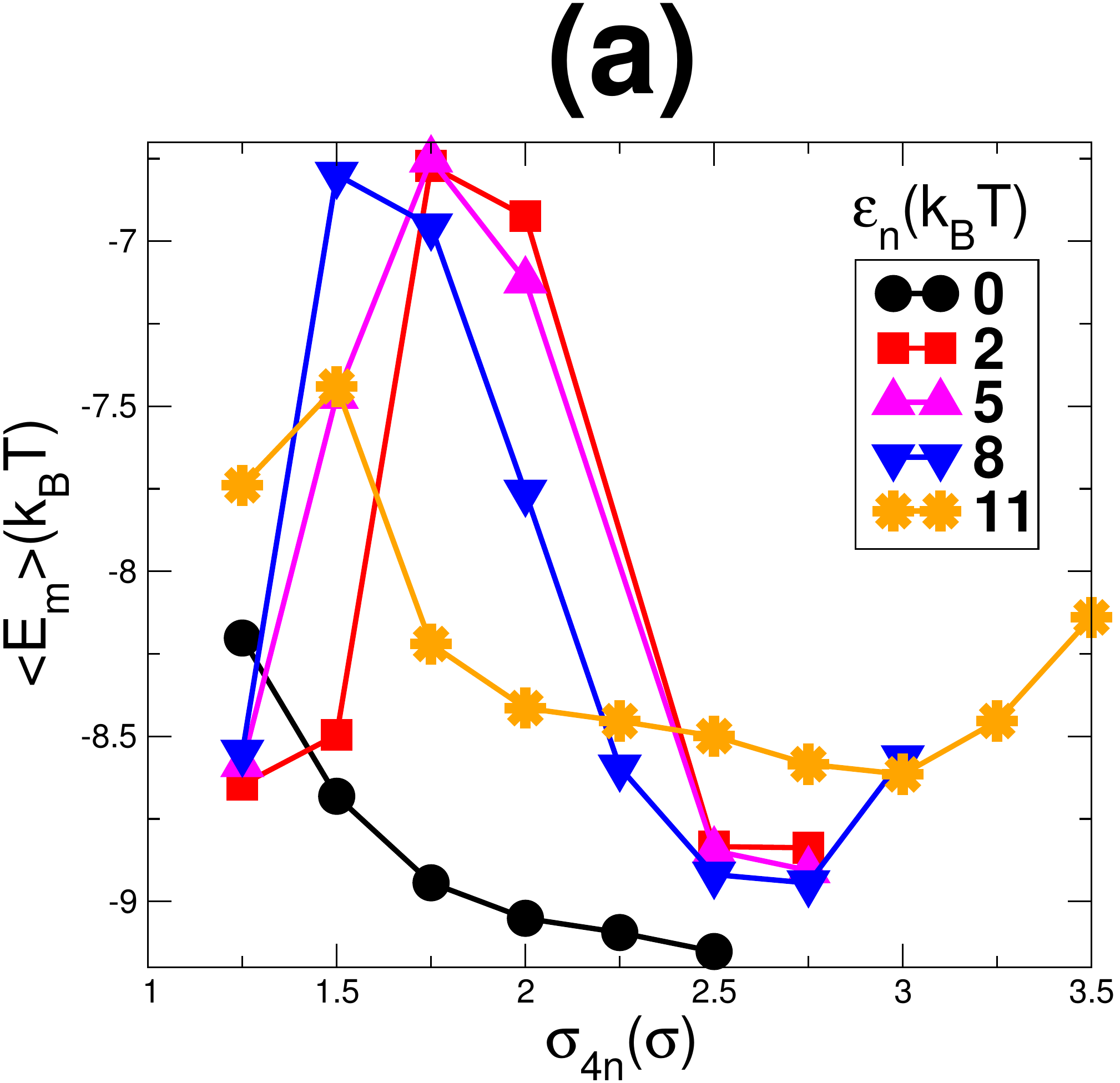}
\includegraphics[scale=0.2]{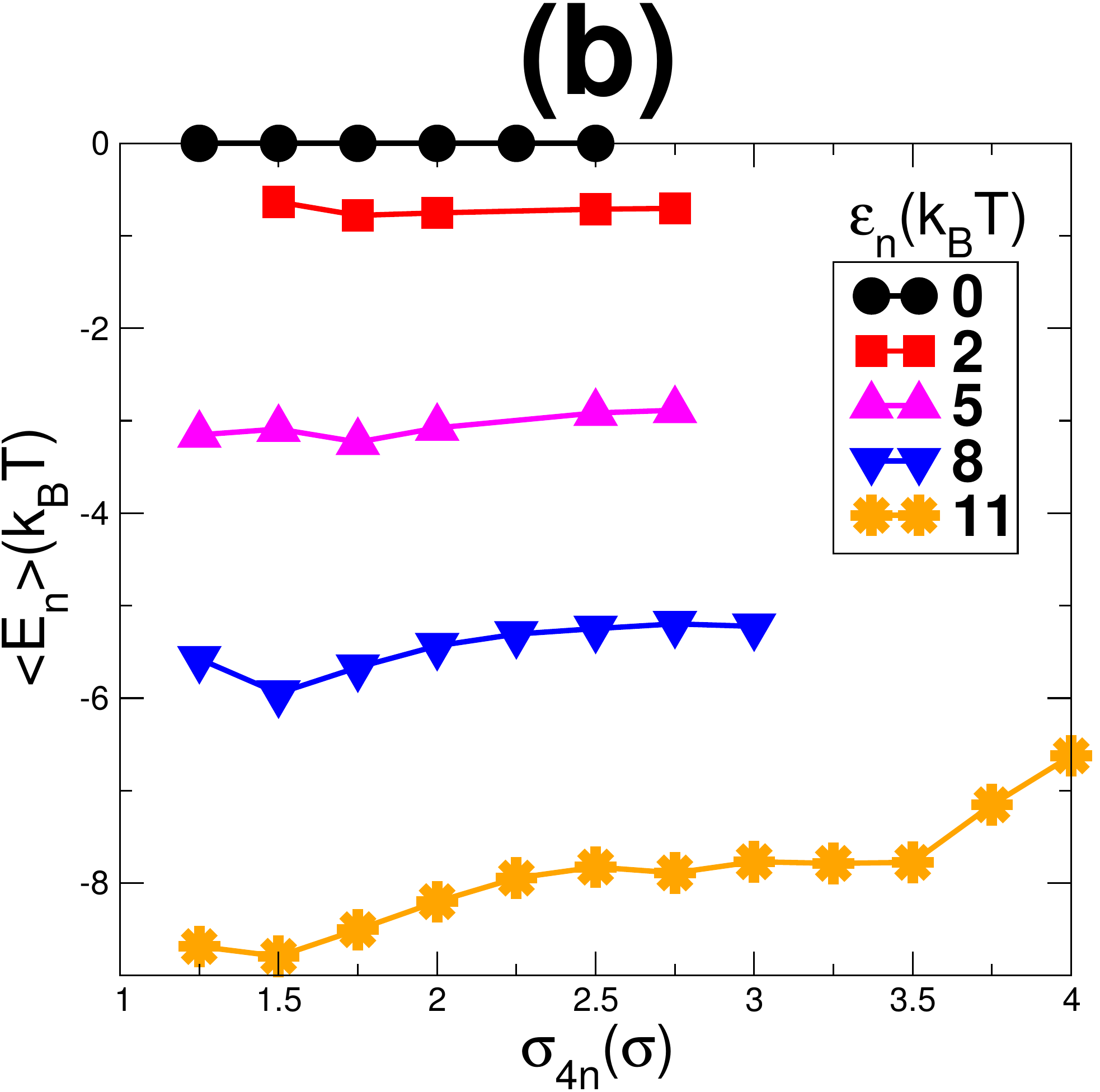}
\includegraphics[scale=0.2]{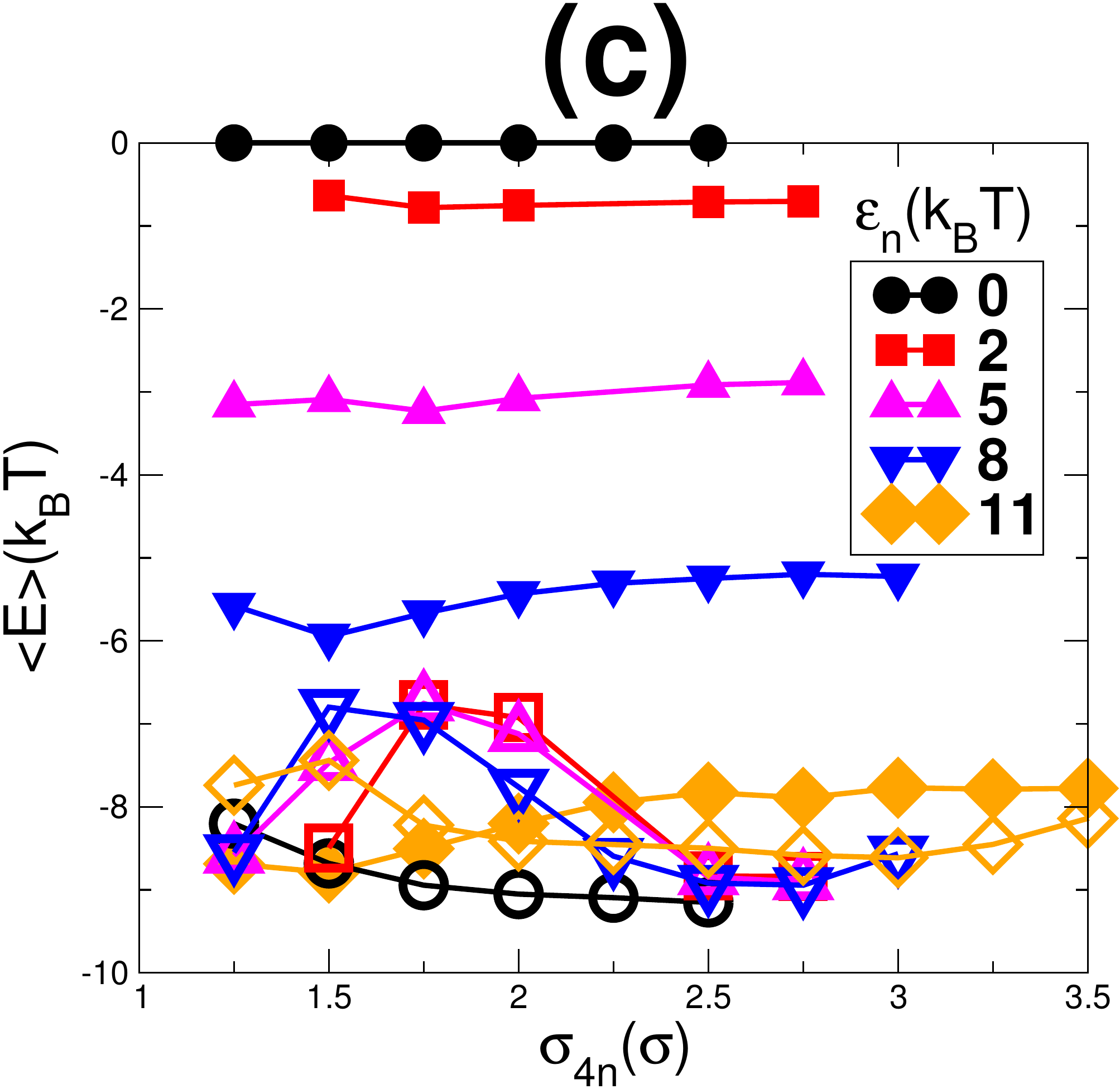}
\caption{(colour online) The figure shows the average energy of (a) monomers, (b) nanoparticles and (c) both (monomers - empty symbols and nanoparticles - filled symbols). Comparing the two energies, it can be seen that only for $\epsilon_n=11k_BT$, the energies of monomers and nanoparticles are competitive. For other values of $\epsilon_n$, the gap between the energies is higher as shown in (c).}
\label{avr_en}
\end{figure}
     The plots of average energies of monomers and nanoparticles are shown in Figs.~\ref{avr_en}(a) and \ref{avr_en}(b) respectively. Figure \ref{avr_en}(c) shows the plots for both the energies. Each figure represents graphs for different values of $\epsilon_n$. Except for $\epsilon_n=0$, all other values of $\epsilon_n$ indicates the two transformation points in the morphology of the system by showing a non-monotonic behaviour. With the increase in $\sigma_{4n}$ from $1.25\sigma$ to a higher value, the monomer energy shows an increase in its value while the nanoparticle energy shows a decrease in its value. The increase in energy of monomers is due to the increased repulsive interaction between chains due to clustering of monomer chains. This point corresponds to the transformation from a dispersed state of chains to network-like structures. With further increase in $\sigma_{4n}$ the nanoparticle network breaks (volume fraction of nanoparticle decreases). Due to the breaking of the network, the available volume for monomer chains increase and hence their distance from each other increases, which results in a decrease in their repulsive interaction. Hence, the monomer energy shows a decrease (more -ve) in its value. For a higher value of $\sigma_{4n}$ (depending on $\epsilon_n$), the energy of monomers again show an increase. This increase in energy is due to a decrease in the effective volume of monomers because of the breaking of nanoparticle network. This leads to a lower chain length of monomers hence increasing their energy ~\cite{2018arXiv180106933M}. Comparing the energies of monomers and nanoparticles, it can be seen that only the values of energies of monomers and nanoparticles for $\epsilon_n=11k_BT$ are relatively comparable. For lower values of $\epsilon_n$, nanoparticle energy is higher compared to monomers as shown in Fig.\ref{avr_en}(c) and the gap between them increases with the decrease in $\epsilon_n$. This confirms the observed behaviour of the effective volume of monomers for different values of $\epsilon_n$ shown in Fig.\ref{vol_frac}(a).

\begin{figure}
\centering
\includegraphics[scale=0.3]{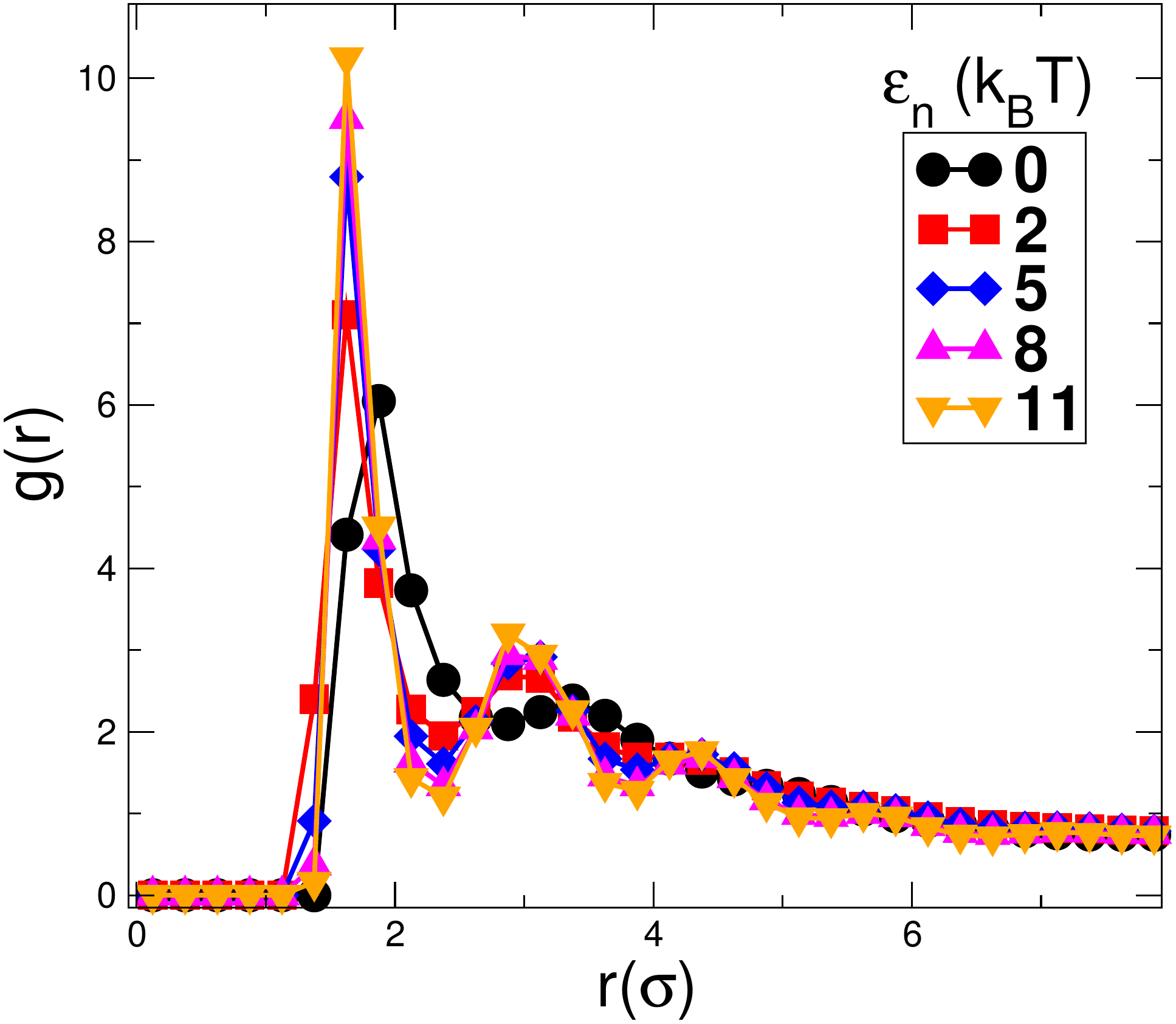}
\caption{(colour online) The figure shows the pair correlation function for nanoparticles with $\sigma_{4n}=2.5\sigma$ and for different values of $\epsilon_n$. The height of the peaks decreases with the decrease in $\epsilon_n$. The lowest and broader peak for $\epsilon_n=0$ shows that the nanoparticle packing, in this case, is lowest.}
\label{corr_nano}
\end{figure}
        
      Apart from the shift in the morphological transformation point (the value of EVP) with the change in $\epsilon_n$, one more change with the change in $\epsilon_n$ can be easily noticed. With the decrease in the value of $\epsilon_n$, the nanoparticle packing decreases. To get an insight into the arrangement of nanoparticles, the pair correlation function g(r) for nanoparticles is plotted in Fig.\ref{corr_nano}. It shows the pair correlation function for nanoparticles with $\sigma_{4n}=2.5\sigma$ and different values of $\epsilon_n=0$, $2k_BT$, $5k_BT$ and $11k_BT$ as indicated by the symbols. With the decrease in $\epsilon_n$, the height of the peak decreases. For $\epsilon_n=0$, the peaks are broader and have a relatively short range of correlation. Hence, the packing of the nanoparticles is lowest in case of $\epsilon_n=0$.

\begin{figure}
\centering
\includegraphics[scale=0.2]{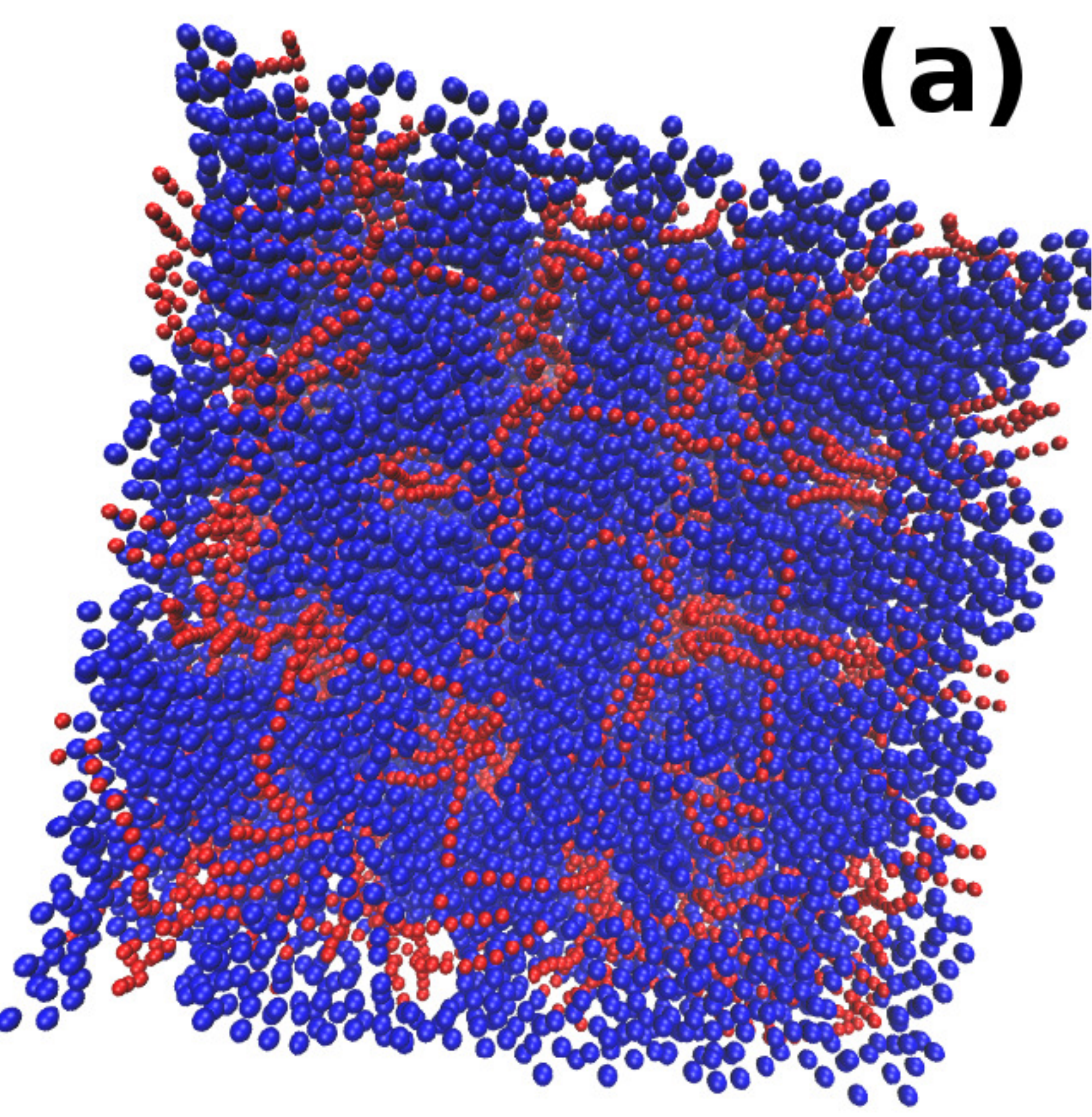}
\includegraphics[scale=0.2]{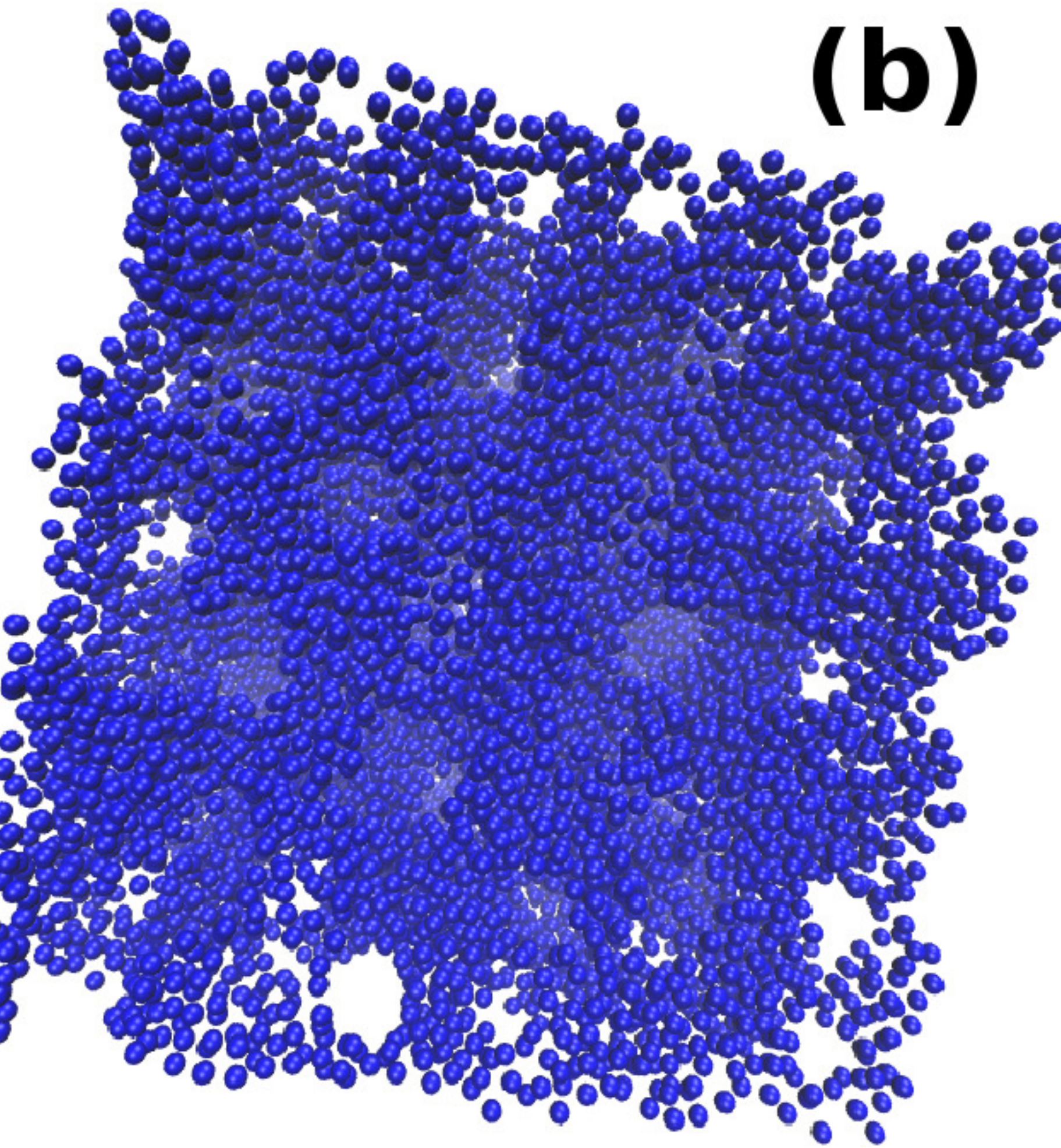}
\caption{(Colour online) The figure shows the snapshots for $\rho_m=0.037\sigma^{-3}, \epsilon_n=0, \sigma_{4n}=2.75\sigma$ reproduced in a larger box size of $60\times 60\times 60\sigma^{3}$. Figure (a) shows both the nanoparticles (blue) and monomers (red) while, figure in (b) shows only nanoparticles. The structure obtained for the larger box size are similar to smaller box size shown in ~\cite{supporting_material}.}
\label{2000_box}
\end{figure}

        As shown earlier ~\cite{2018arXiv180106933M}, the same kind of behaviour is shown by all the densities of micelles, $\rho_m=0.074\sigma^{-3}$, $0.093\sigma^{-3}$ and $0.126\sigma^{-3}$ except for $\rho_m=0.037\sigma^{-3}$. In case of $\rho_m=0.037\sigma^{-3}$, a change in $\sigma_{4n}$ from $1.25\sigma$ to $1.5\sigma$ leads to the clustering of micellar chains which joins to form a network-like structure similar to other densities. However, no change from network to individual clusters of nanoparticles is observed for this micellar density. For all the values of $\sigma_{4n}>1.25\sigma$, the system shows the formation of a network of nanoparticle clusters and micellar chains with no further structural change observed for any value of $\sigma_{4n}$ considered here. When the system is subjected to the change in the value of $\epsilon_n$, there was no change in the structural morphology observed for $\sigma_{4n}>1.25\sigma$ as well. For all the values of $\sigma_{4n}>1.25\sigma$ considered and all the values of $\epsilon_n$ considered, the system shows only network-like structures. A comparison of the systems with $\rho_m=0.037\sigma^{-3}$ for different values of $\epsilon_n$ is shown in ~\cite{supporting_material}. The systems for $\rho_m=0.037\sigma^{-3}$ are also reproduced in a bigger box size of $60\times 60\times 60\sigma^3$. No change in the structures is observed. One of the snapshots for $\sigma_{4n}=2.75\sigma$ and $\epsilon_n=0$ is shown in figure \ref{2000_box} (a) and (b). The snapshot in (a) shows the nanoparticles (blue) and micelles (red) both while, only nanoparticles from the snapshot in (a) are shown in (b). The snapshots show the network of nanoparticle clusters interpenetrating with the network of micellar chains.


\begin{figure}
\centering
\includegraphics[scale=0.2]{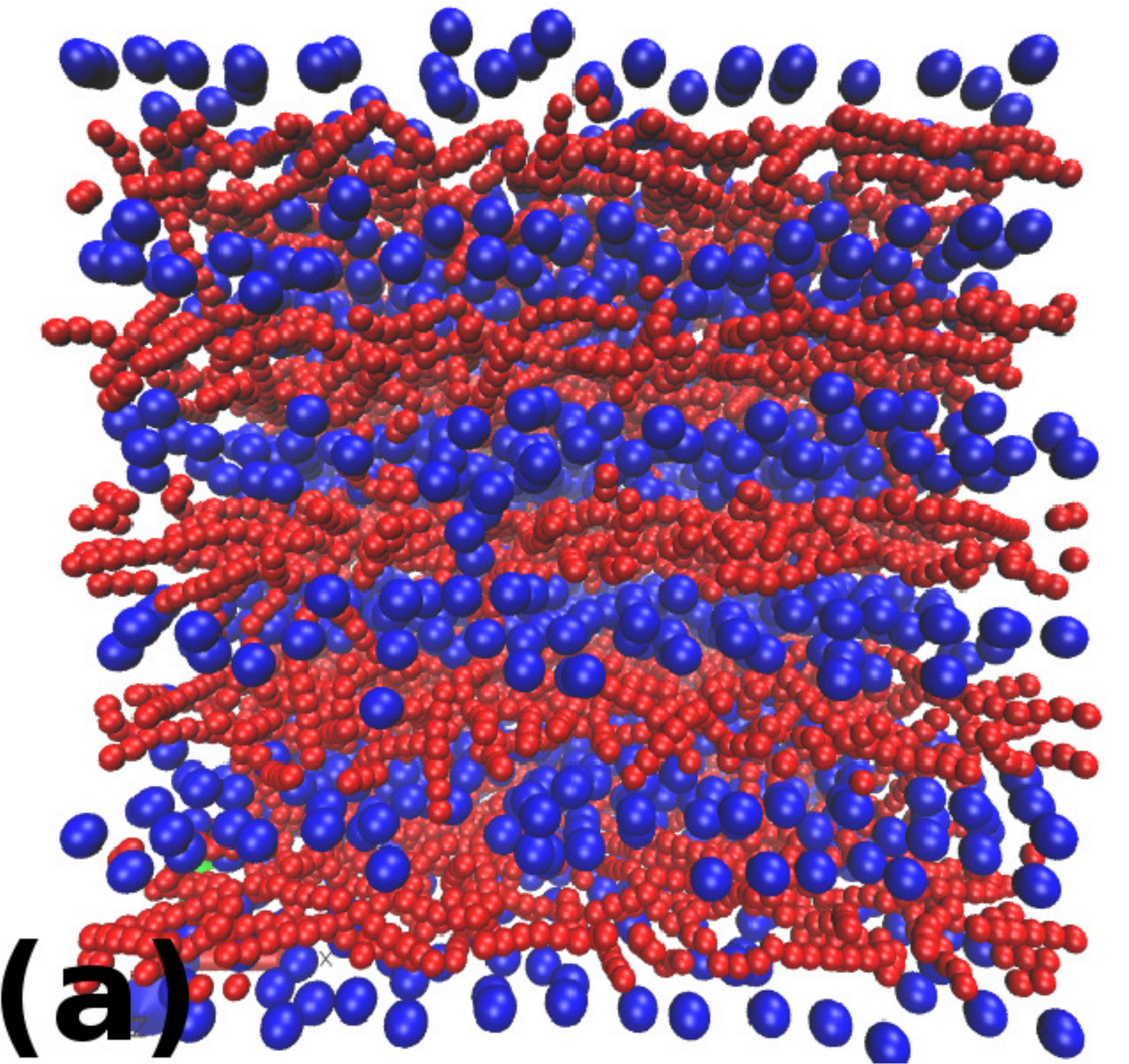}
\includegraphics[scale=0.2]{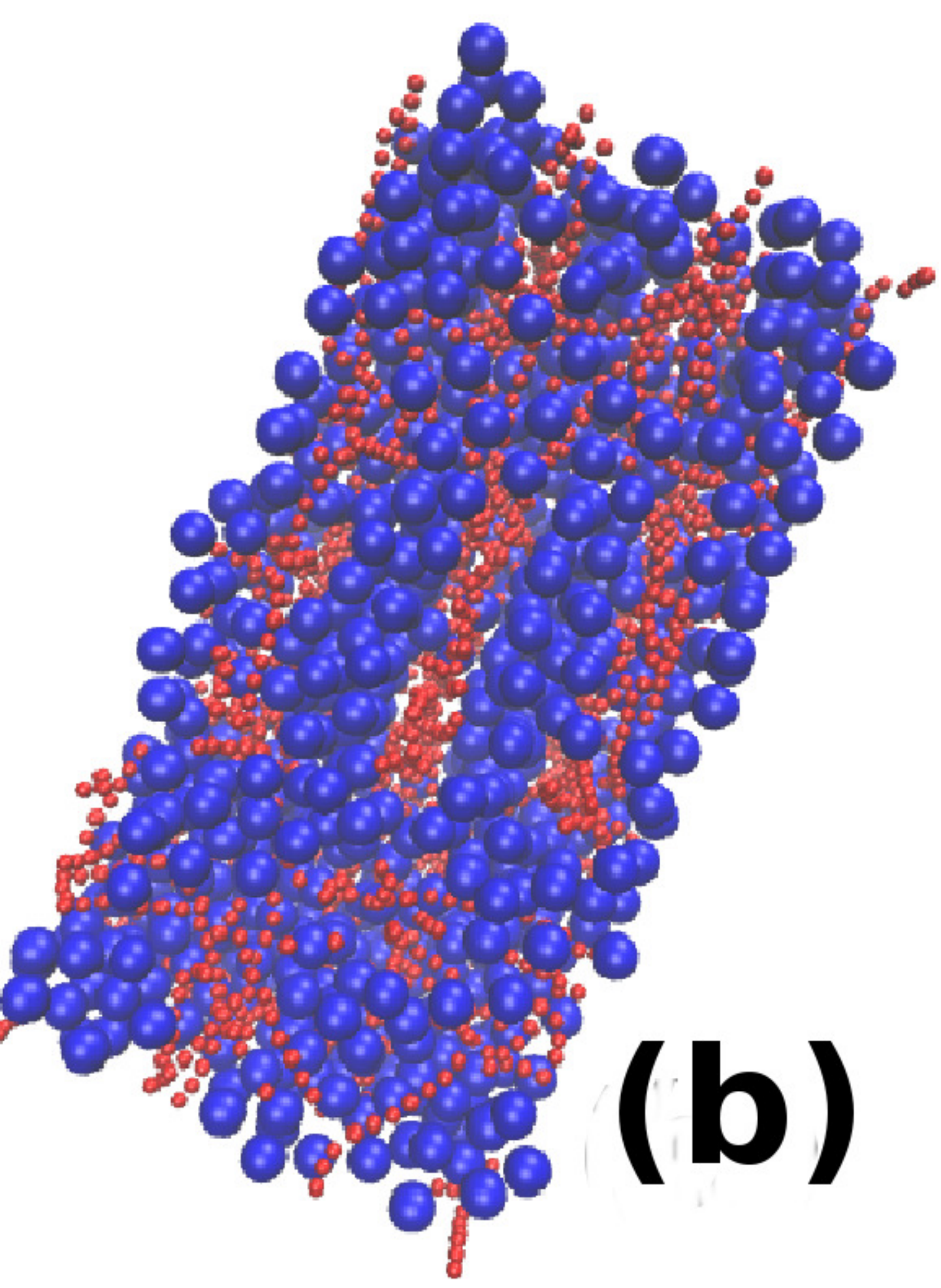} \\
\includegraphics[scale=0.2]{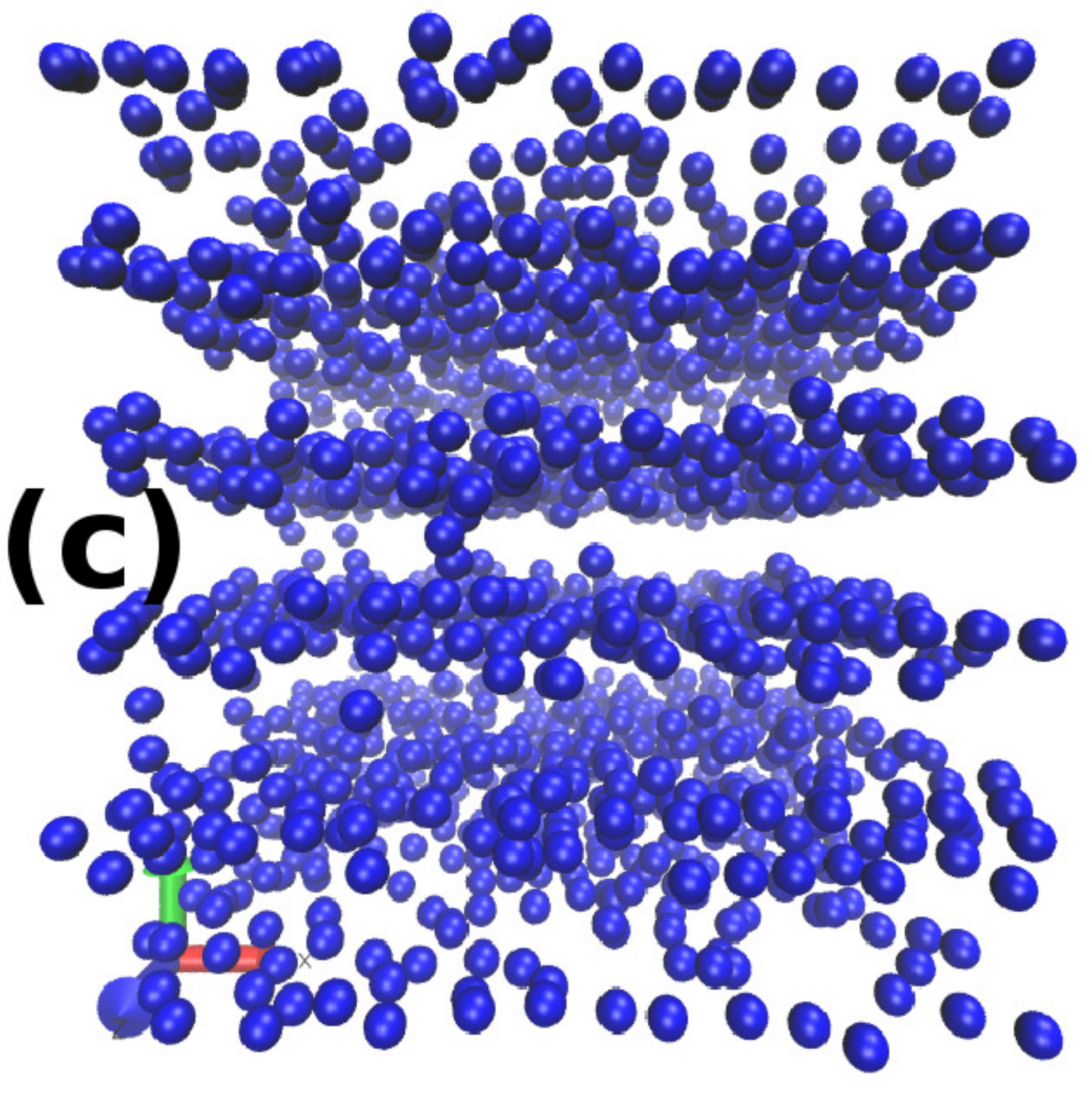}
\includegraphics[scale=0.2]{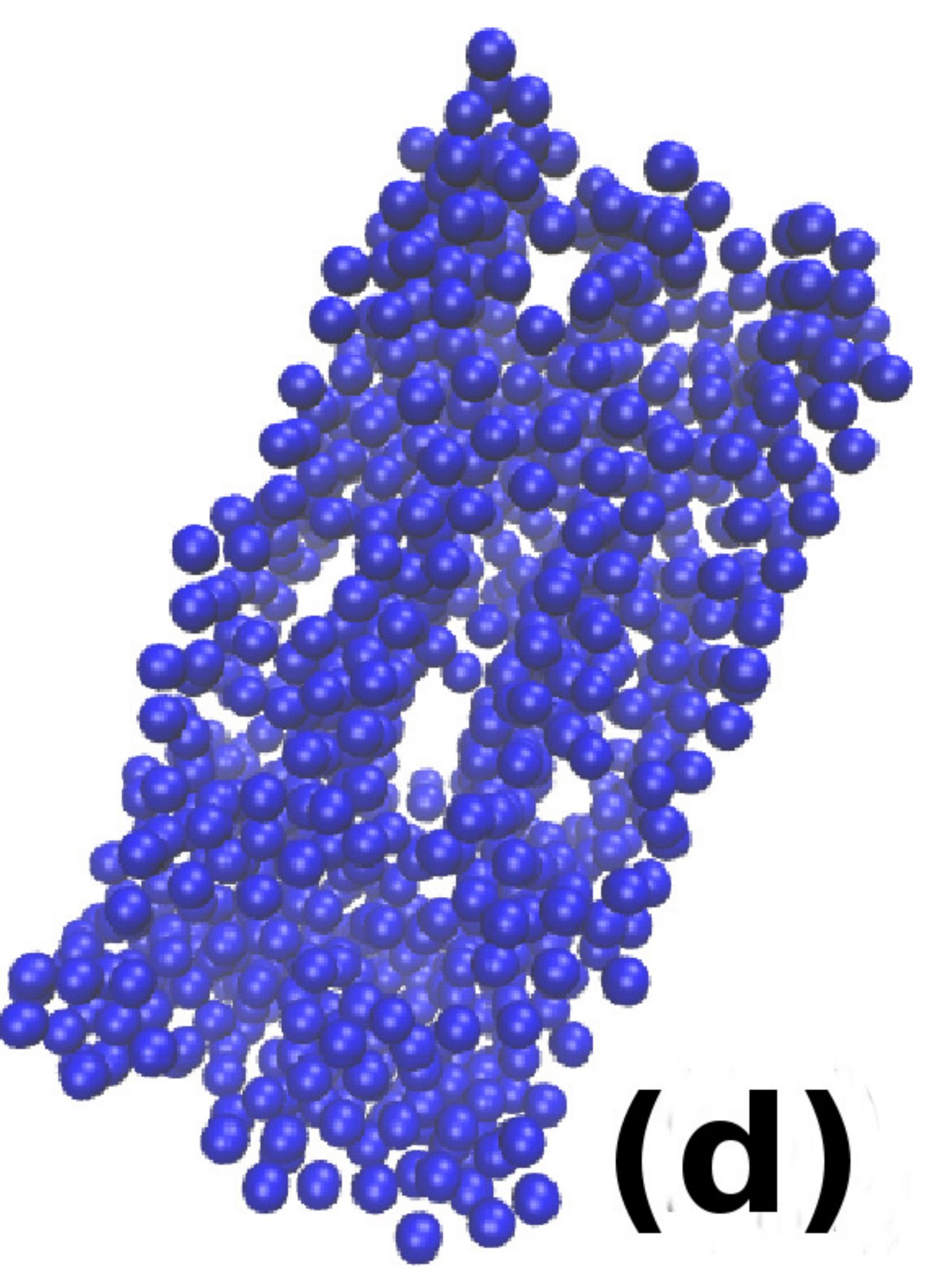}
\caption{(Colour online) Figure shows the snapshots for $\rho_m=0.037\sigma^{-3}, \sigma_n=3\sigma, \sigma_{4n}=3.25\sigma$ for two different values of $\epsilon_n$   (a) 0 and (b) $11k_BT$. The upper row shows both the nanoparticles and monomers while the lower row only shows the nanoparticles. The point of the structural change for nanoparticles gets shifted to lower values of $\sigma_{4n}$ with a decrease in $\epsilon_n$. Therefore, the left snapshots show a system forming system spanning sheet of nanoparticles while the right figure is still in the regime of percolating network-like structure in spite of having the same values of $\sigma_{4n}$.}
\label{size}
\end{figure}

\begin{figure}
\includegraphics[scale=0.2]{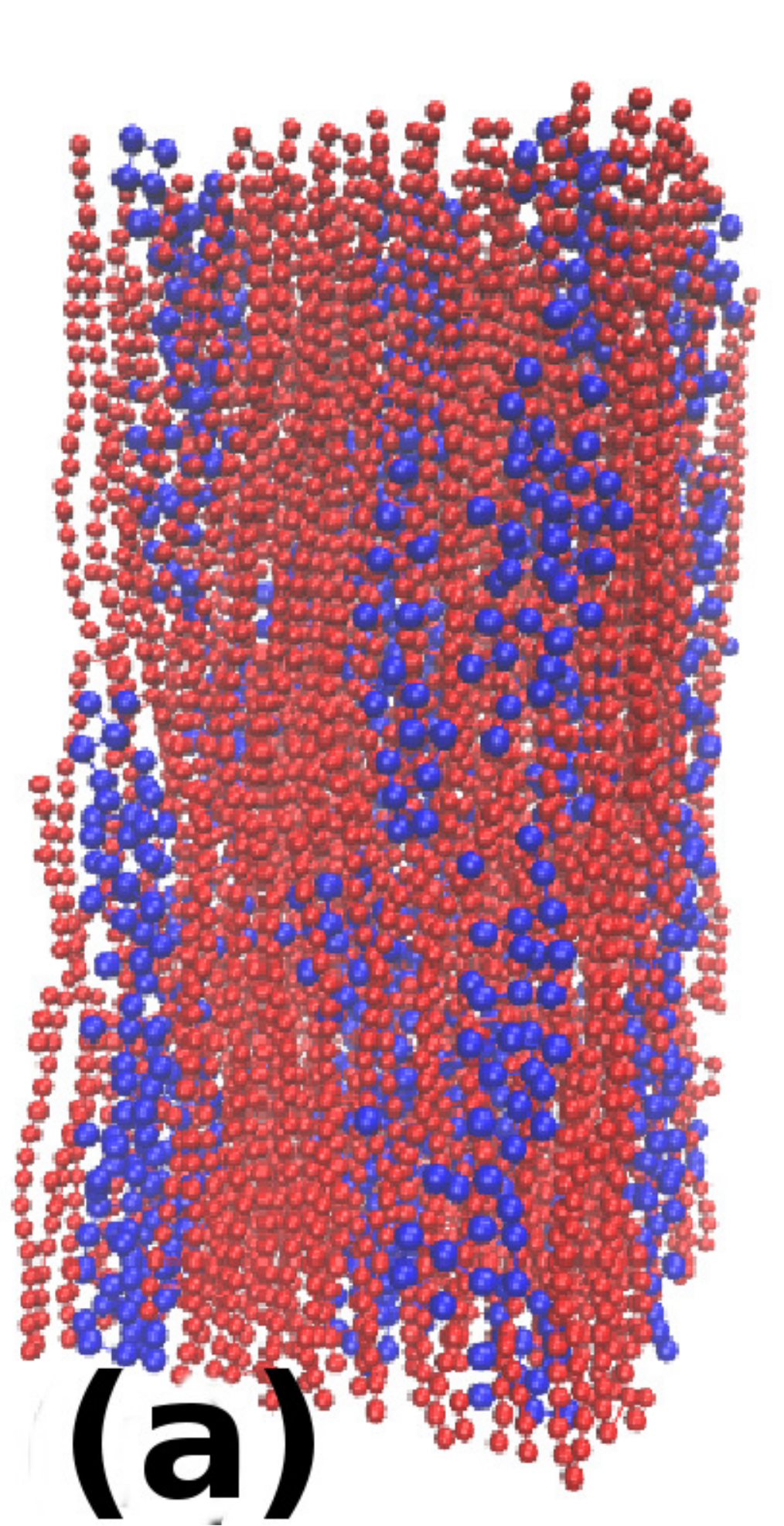} 
\includegraphics[scale=0.2]{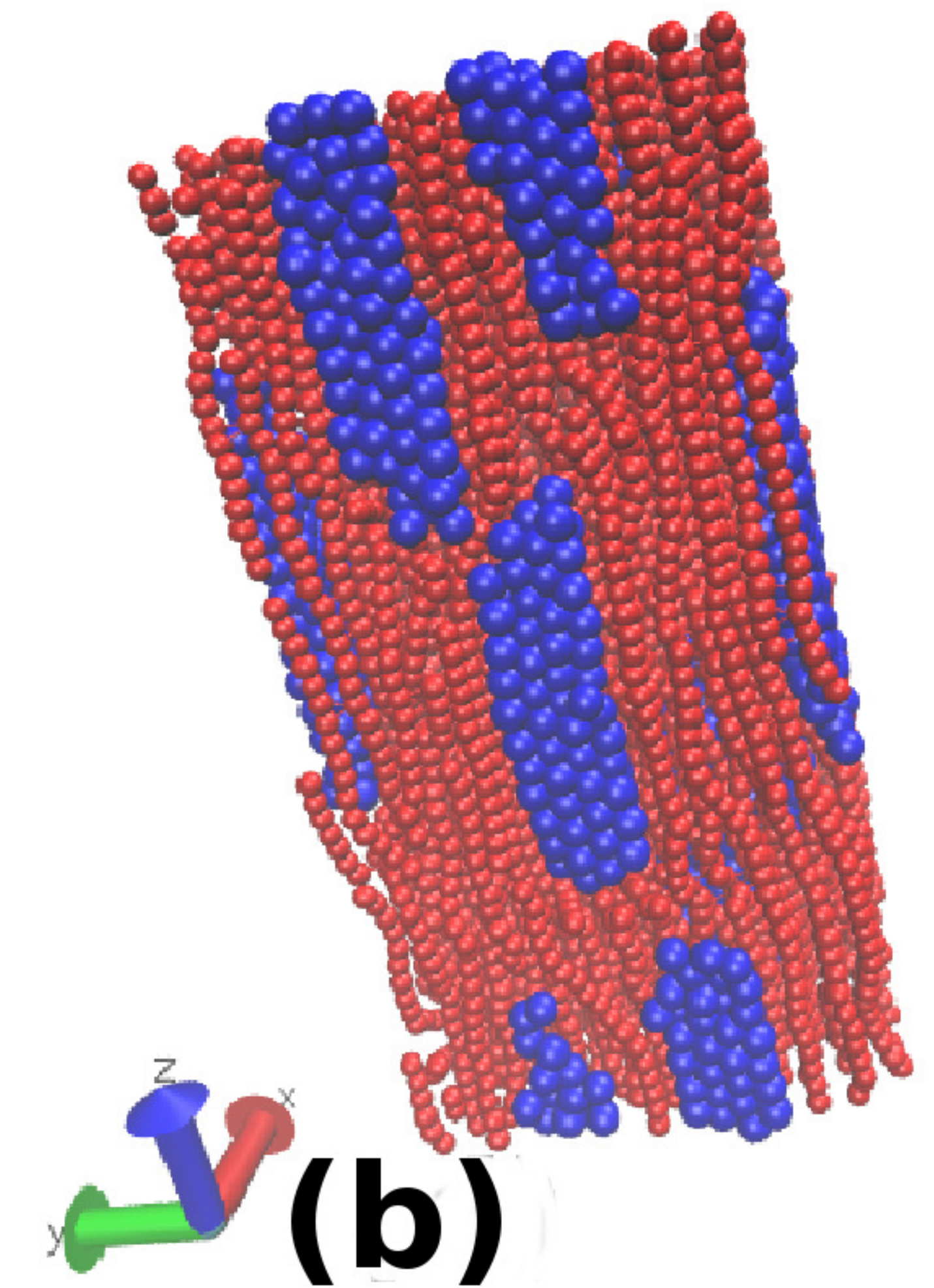} \\
\includegraphics[scale=0.2]{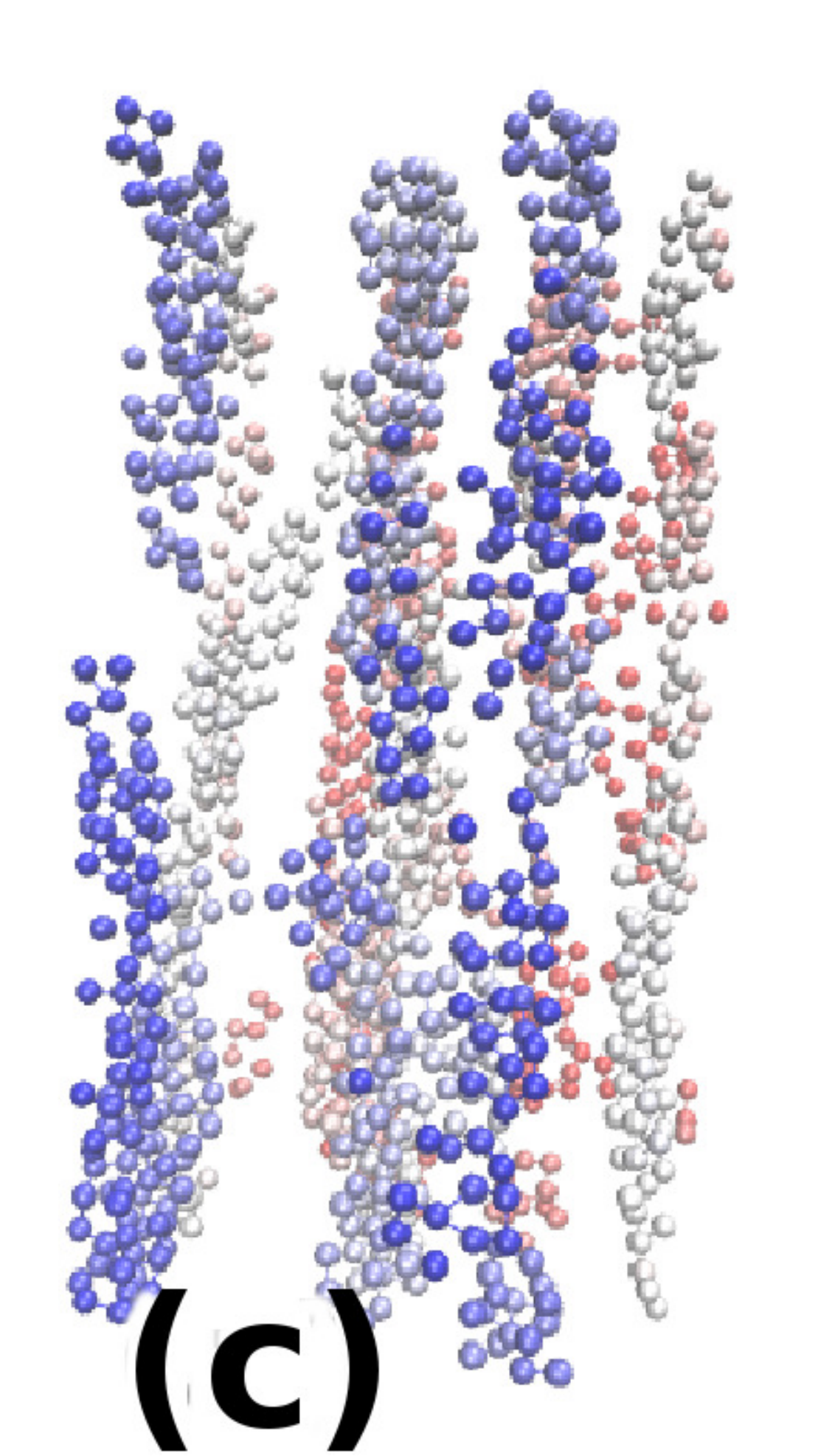} 
\includegraphics[scale=0.18]{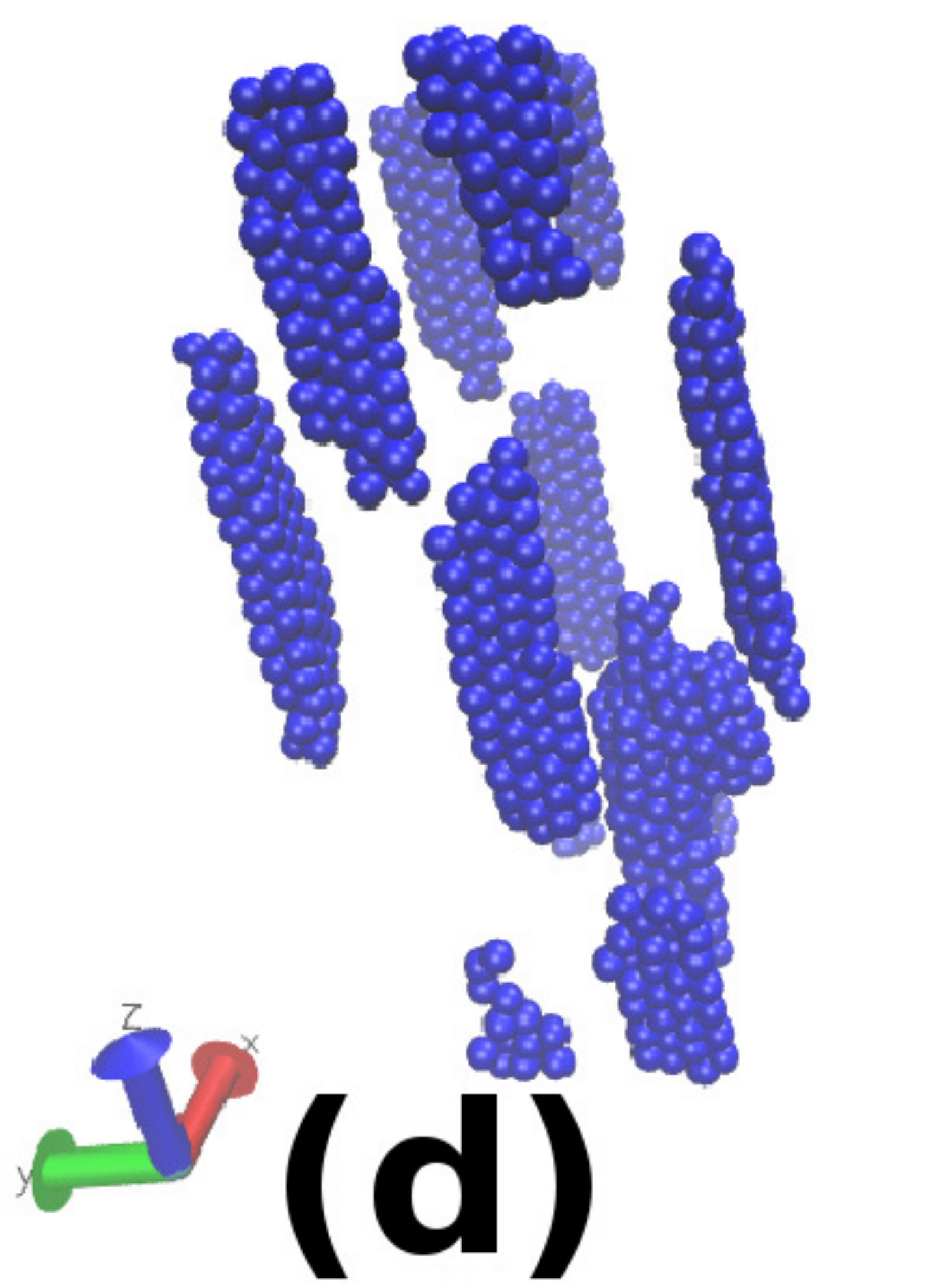} 
\caption{(Colour online) The figure shows the snapshots for the $\rho_m=0.126\sigma^{-3}$ and $\sigma_{4n}=2\sigma$ but (a) $\epsilon_n$=0  and (b) $\epsilon_n=11k_BT$. The upper row only shows both the nanoparticles(blue) and micelles(red) while the lower row only shows nanoparticles. Only figure (c) has a gradient in its colour along one of the shorter axes of the box. Comparision of both the figure shows that lower value of $\epsilon_n$ leads to low volume fraction and low packing of nanoparticles. Therefore the rods formed in figure (c) do not show a well-packed structure compared to the right figure and are also thinner.}
\label{dens}
\end{figure}

     Throughout the paper, the value of the size of nanoparticles is kept constant $\sigma_n=1.5\sigma$. For the range of values of $\sigma_{4n}$ considered, no change from network to individual clusters for nanoparticles is observed for the value of $\rho_m=0.037\sigma^{-3}$. However, for nanoparticle size $\sigma_n=3\sigma$ the transition occur at the value of $\sigma_{4n}=3.25\sigma$ for the value of $\epsilon_n=0$. While keeping the other parameters same, for $\epsilon_n=11k_BT$, no transition is observed. This is another example of the shift in the value of EVP for the system morphological change with the change in $\epsilon_n$. This is shown in figure \ref{size}. The figure shows snapshots for $\sigma_n=3\sigma$ and $\sigma_{4n}=3.25\sigma$ for two different values of $\epsilon_n$ ((a) and (c)) 0 and  ((b) and (d)) $11k_BT$. The figures in upper row show both the nanoparticles (blue) and monomer chains (red) while in the lower row only nanoparticles are shown. The snapshot in (a) shows system spanning sheets of nanoparticles (arranged in alternate layers of nanoparticles and micellar chains) while the snapshot in (b) shows a network-like structure of nanoparticles. Apart from that, the arrangement or packing of nanoparticles is relatively low in (a) compared to (b). 
        A similar example is shown in figure \ref{dens}. It shows two snapshots for (a) $\epsilon_n=0$ and (b) $\epsilon_n=11k_BT$ keeping the values of $\rho_m=0.126\sigma^{-3}$ and $\sigma_{4n}=2.25\sigma$ and $\sigma_n=1.5\sigma$ same for both. The upper row shows both nanoparticles (blue) and monomers (red) while only nanoparticles are shown in the lower row. Only for the figure (c), there exists a gradient in the colour varying from red to blue along one of the shorter axis of the box. The system with this micellar no. density $\rho_m=0.126\sigma^{-3}$ is shown to be producing rod-like morphology of nanoparticle clusters ~\cite{2018arXiv180106933M}. Here, the snapshots are shown for the value of $\sigma_{4n}$ past the point of transformation from network to individual clusters. Hence the systems are showing rodlike structures of nanoparticles. But the rods in case of $\epsilon_n=0$ (in (c))  can be seen as thinner compared to the rods in (d) for $\epsilon_n=11k_BT$. Moreover, one can clearly see the difference in the packing of nanoparticles. The nanoparticles are well packed in (d) compared to (c).

\section{Summary:}

    A detailed investigation on the effect of the strength of interaction between nanoparticles on the structural behaviour of the Wormlike micelles-nanoparticles system is carried out. It is shown that with the decrease in the value of $\epsilon_n$, the point (value of EVP) of the transition from network to individual clusters of nanoparticles gets shifted to the lower value of $\sigma_{4n}$. It is also shown that this shift in transition point is due to a decrease in the nanoparticle volume fraction with the decrease in $\epsilon_n$. For the case of $\epsilon_n=0, \sigma_{n}=3\sigma$ and $\sigma_{4n}=3.25\sigma$, system spanning sheet-like arrangement of nanoparticles is reported. The investigation also shows that a decrease in the value of $\epsilon_n$ leads to a decrease in the packing of nanoparticles.

\bibliographystyle{apsrev4-1}
\bibliography{paper_today}

\section{Supporting Material}
\begin{figure*}
\centering
\includegraphics[scale=0.2]{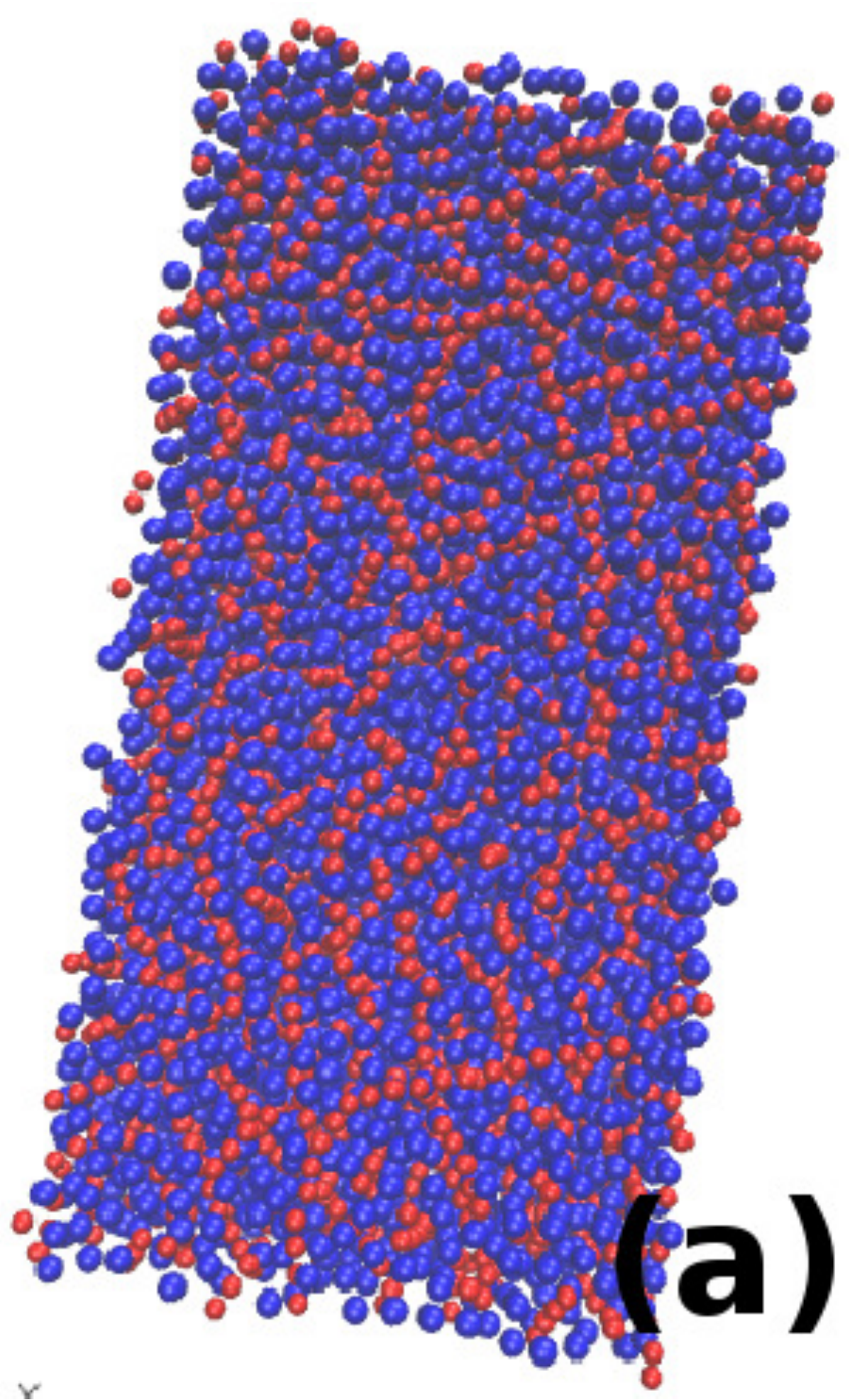}
\hspace{1cm}
\includegraphics[scale=0.2]{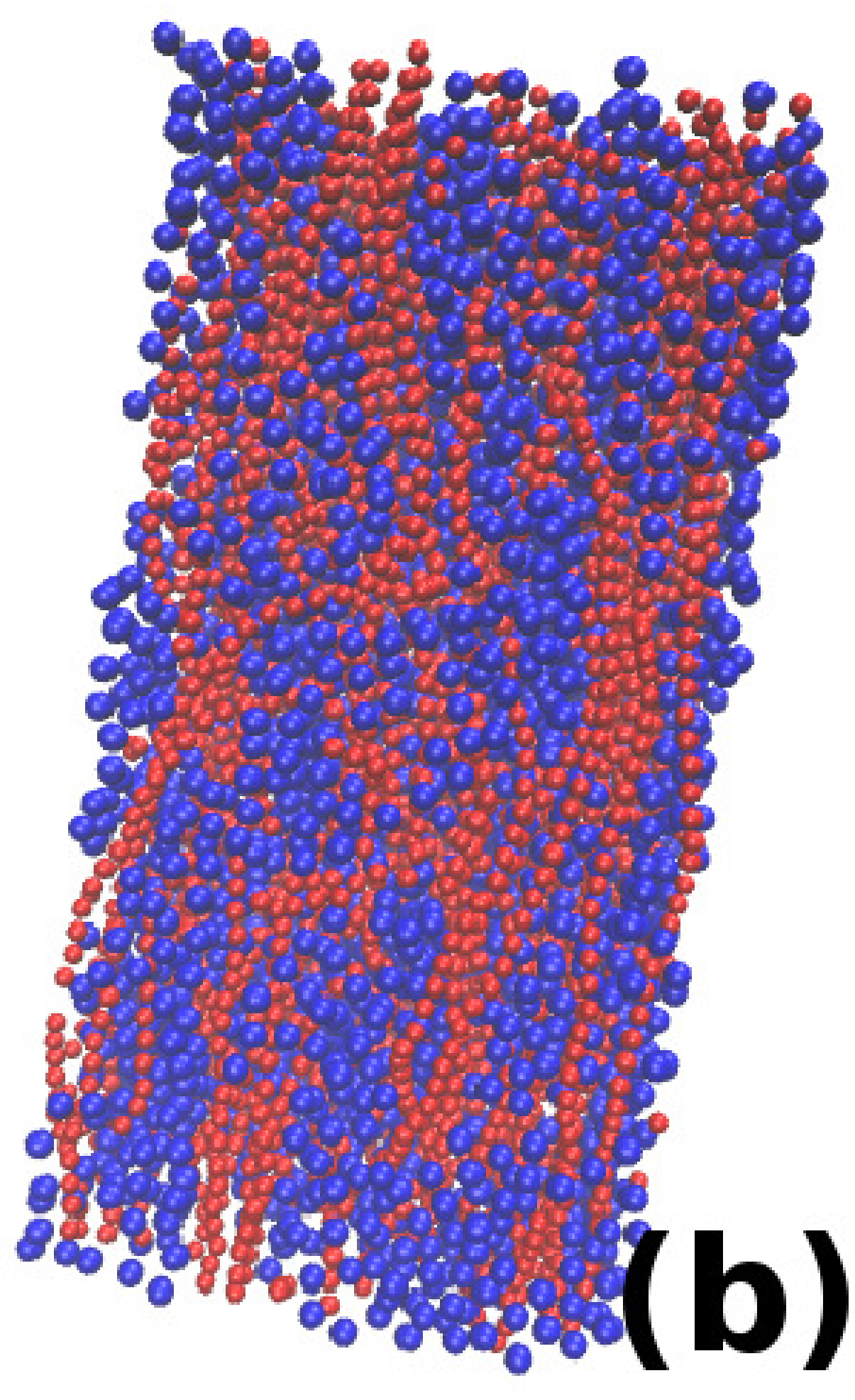}
\hspace{1cm}
\includegraphics[scale=0.2]{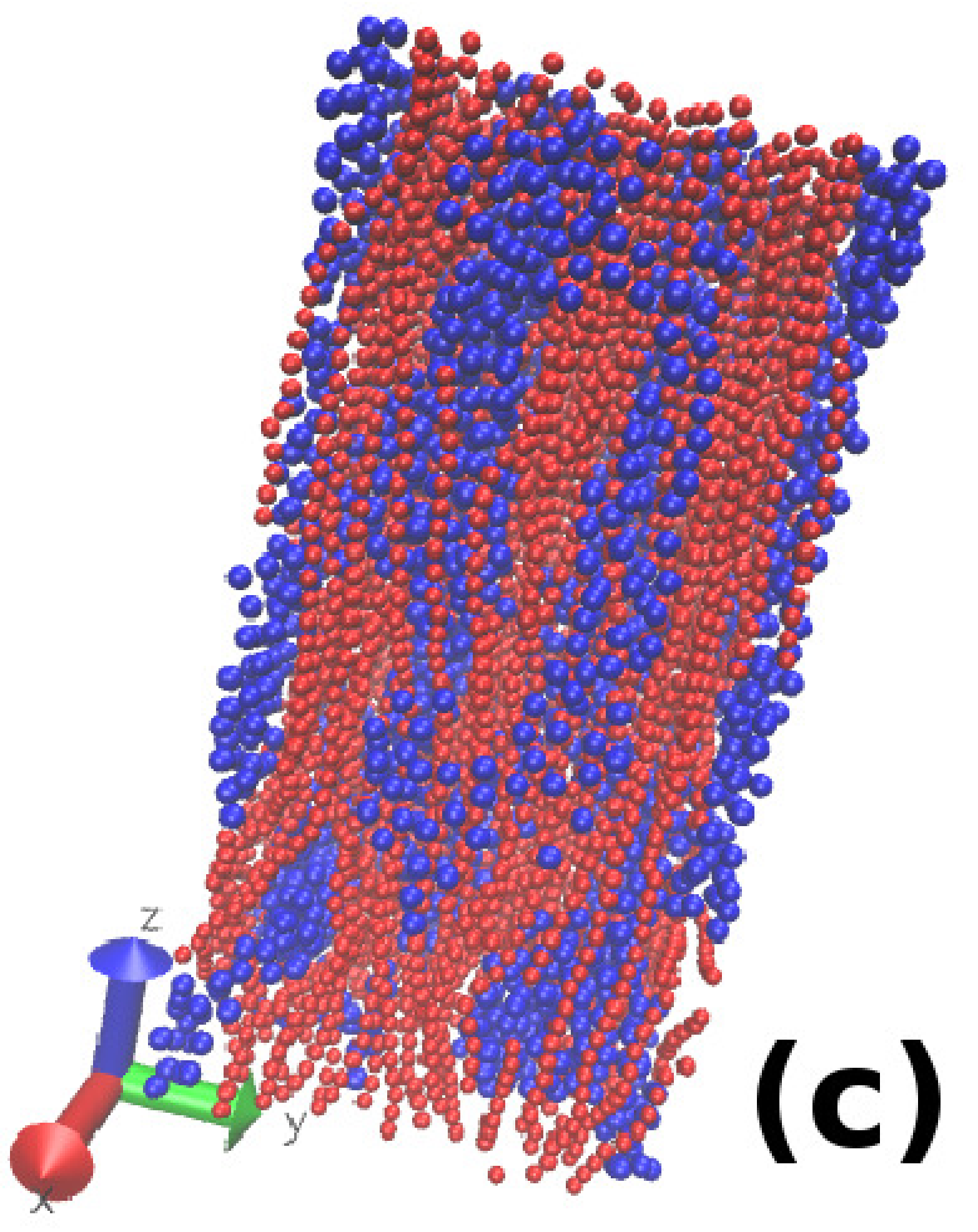}
\hspace{1cm}
\includegraphics[scale=0.2]{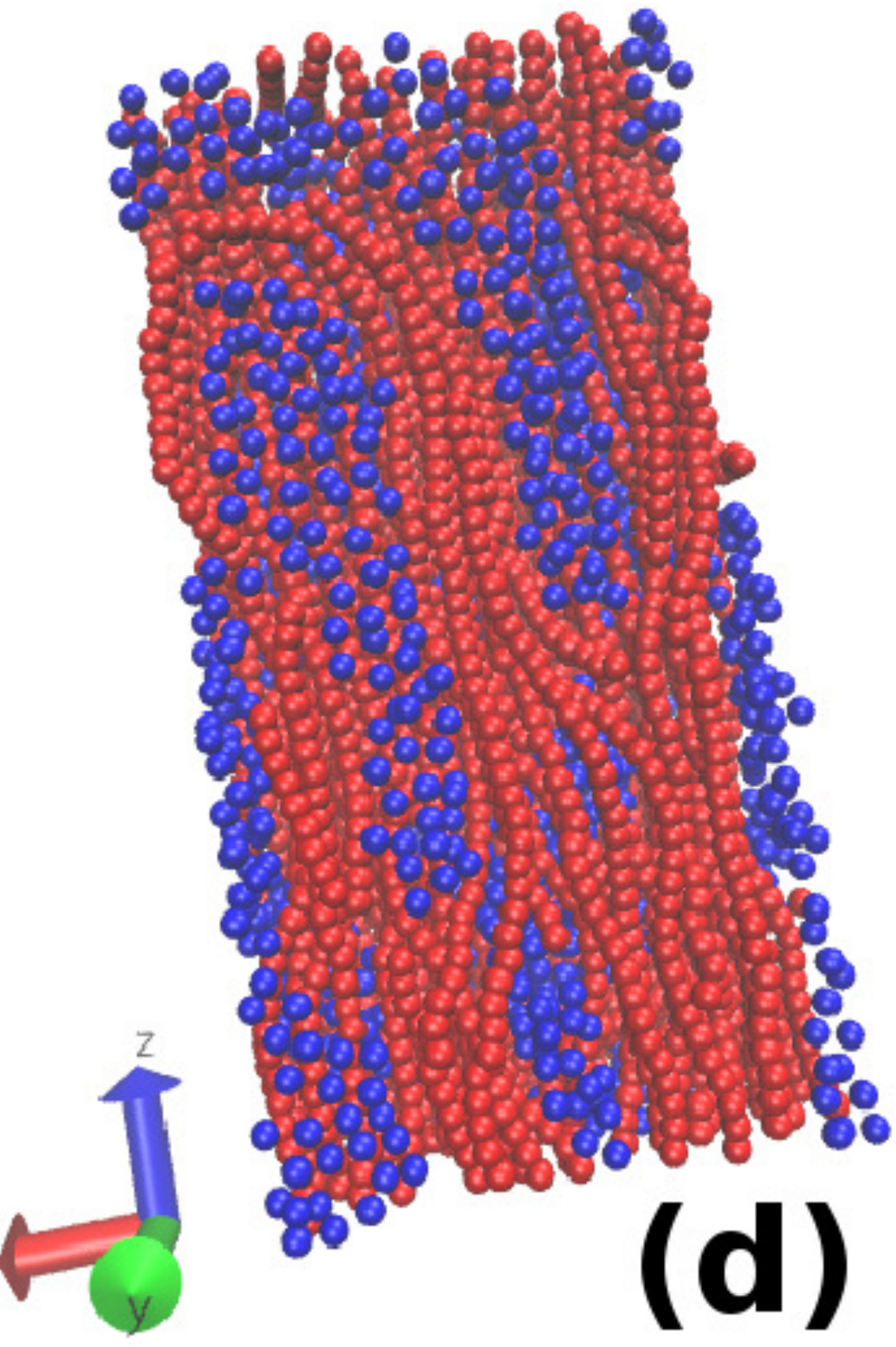} \\
\includegraphics[scale=0.2]{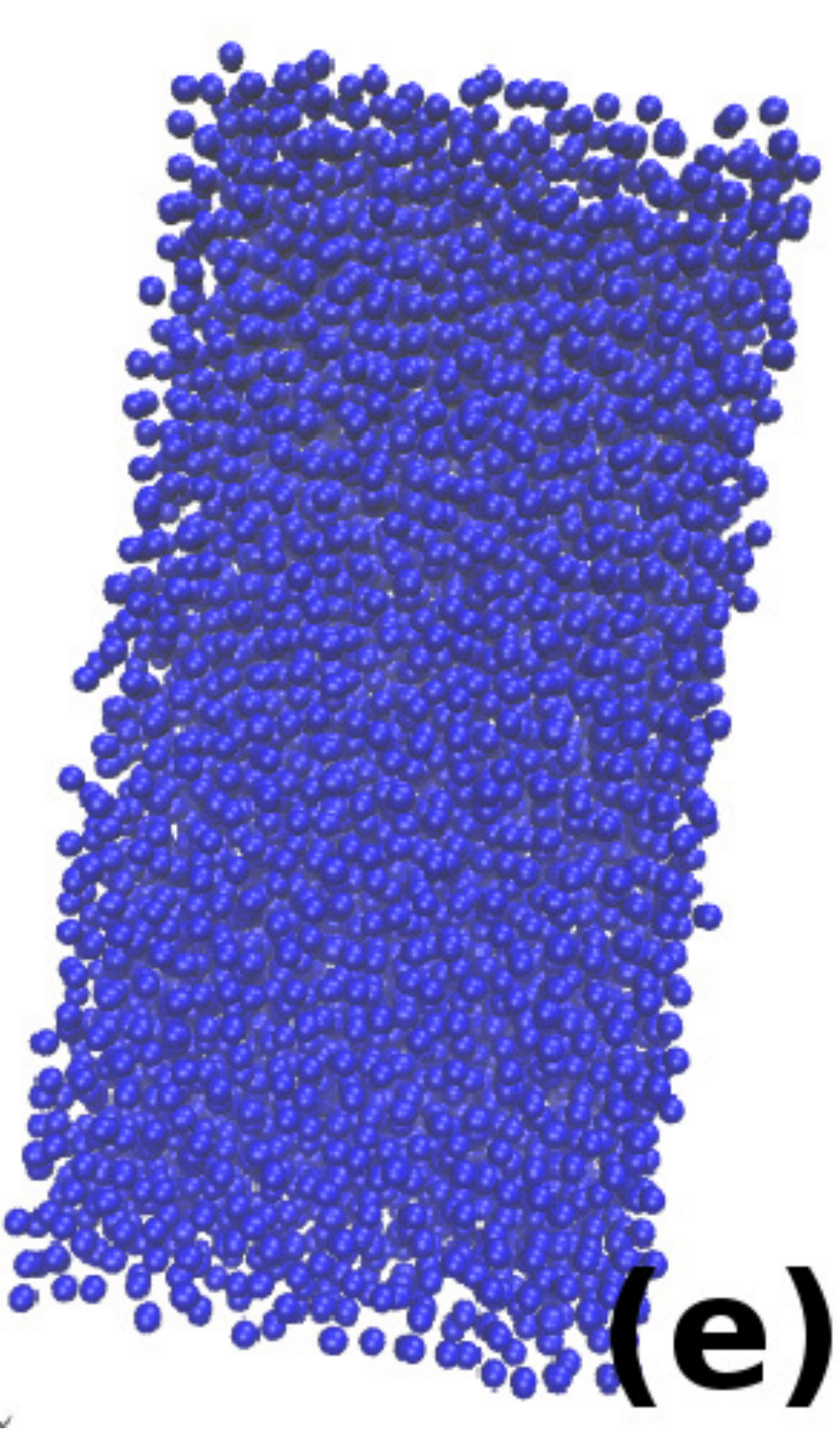}
\hspace{1cm}
\includegraphics[scale=0.2]{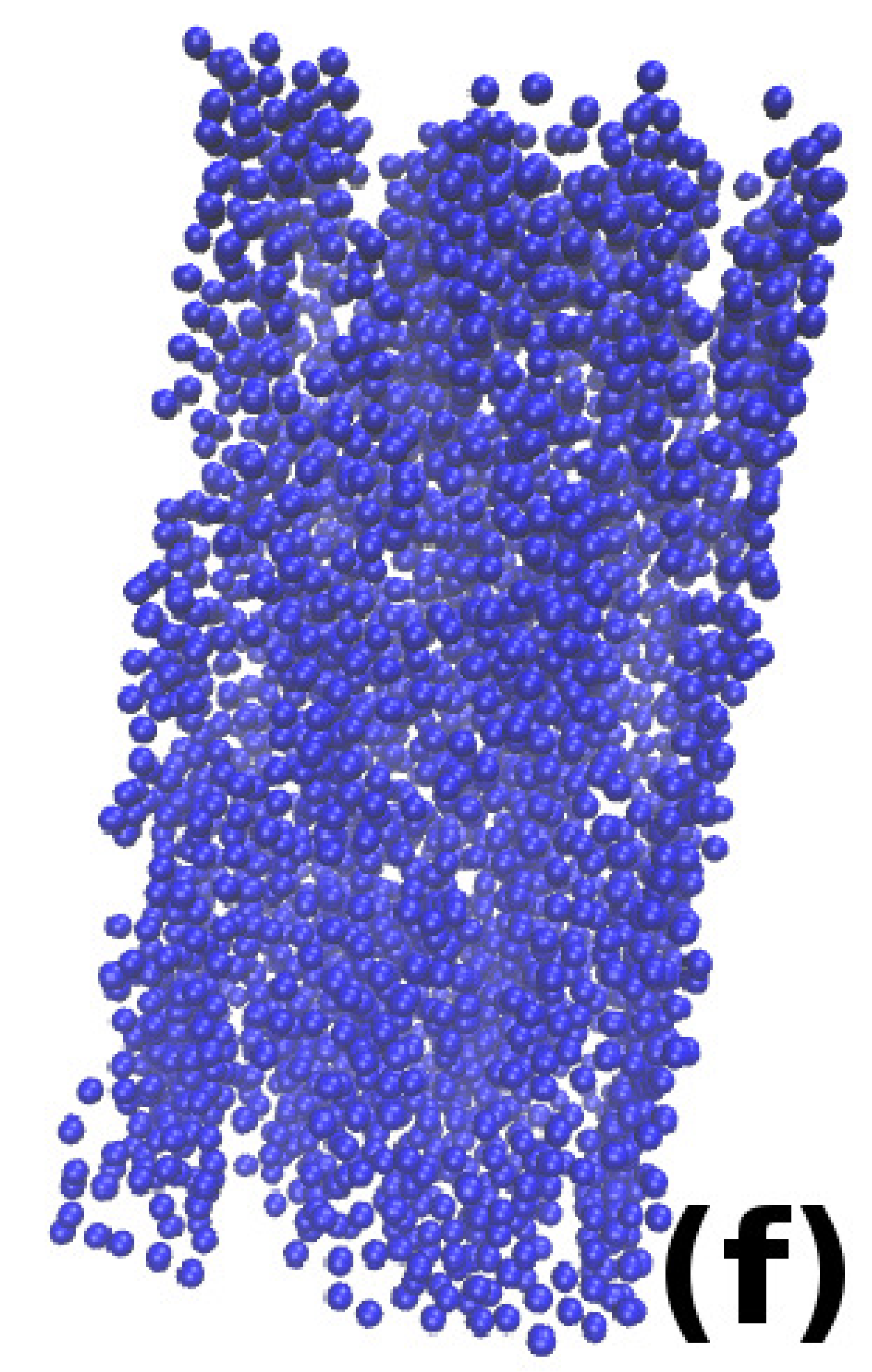}
\hspace{1cm}
\includegraphics[scale=0.2]{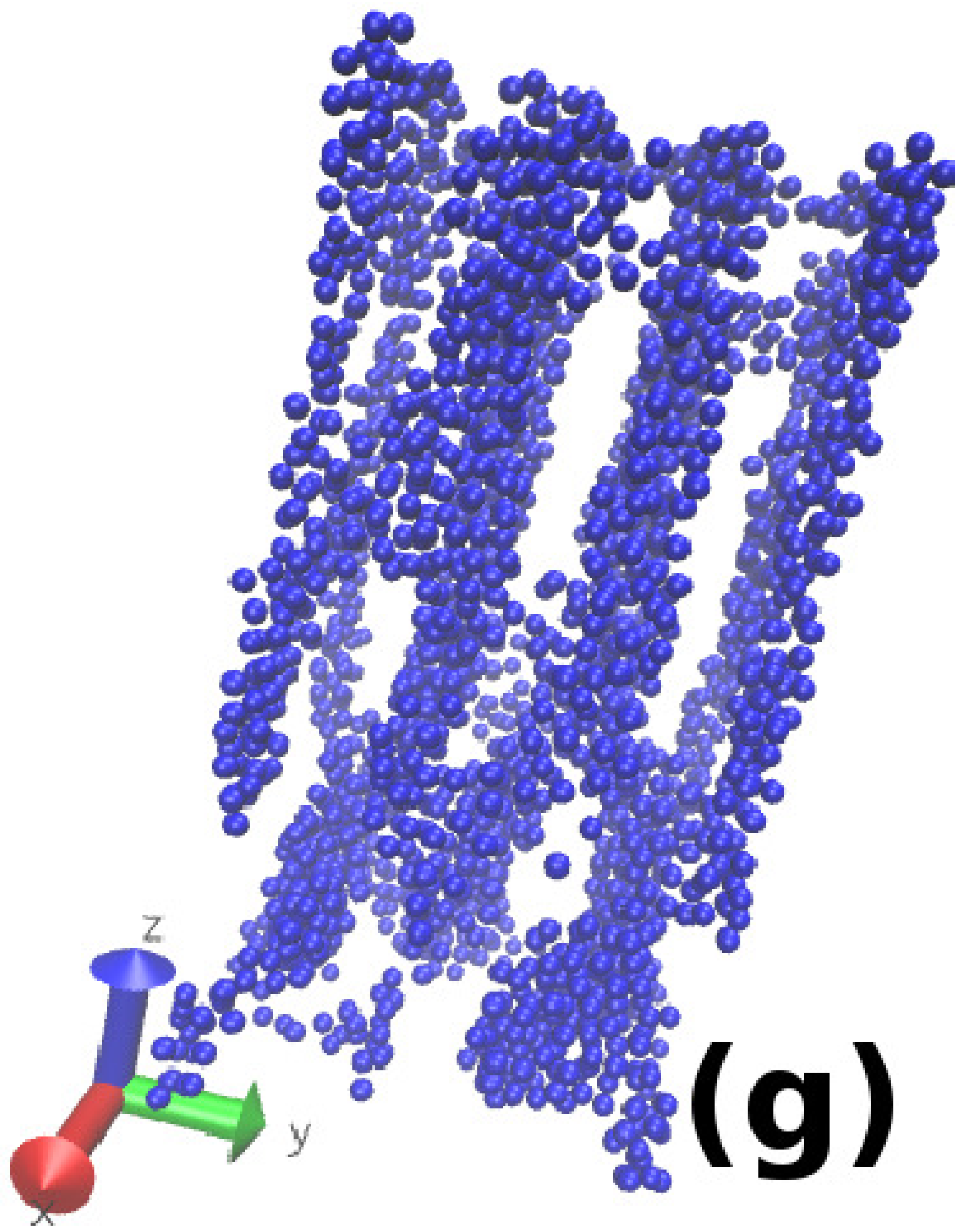}
\hspace{1cm}
\includegraphics[scale=0.2]{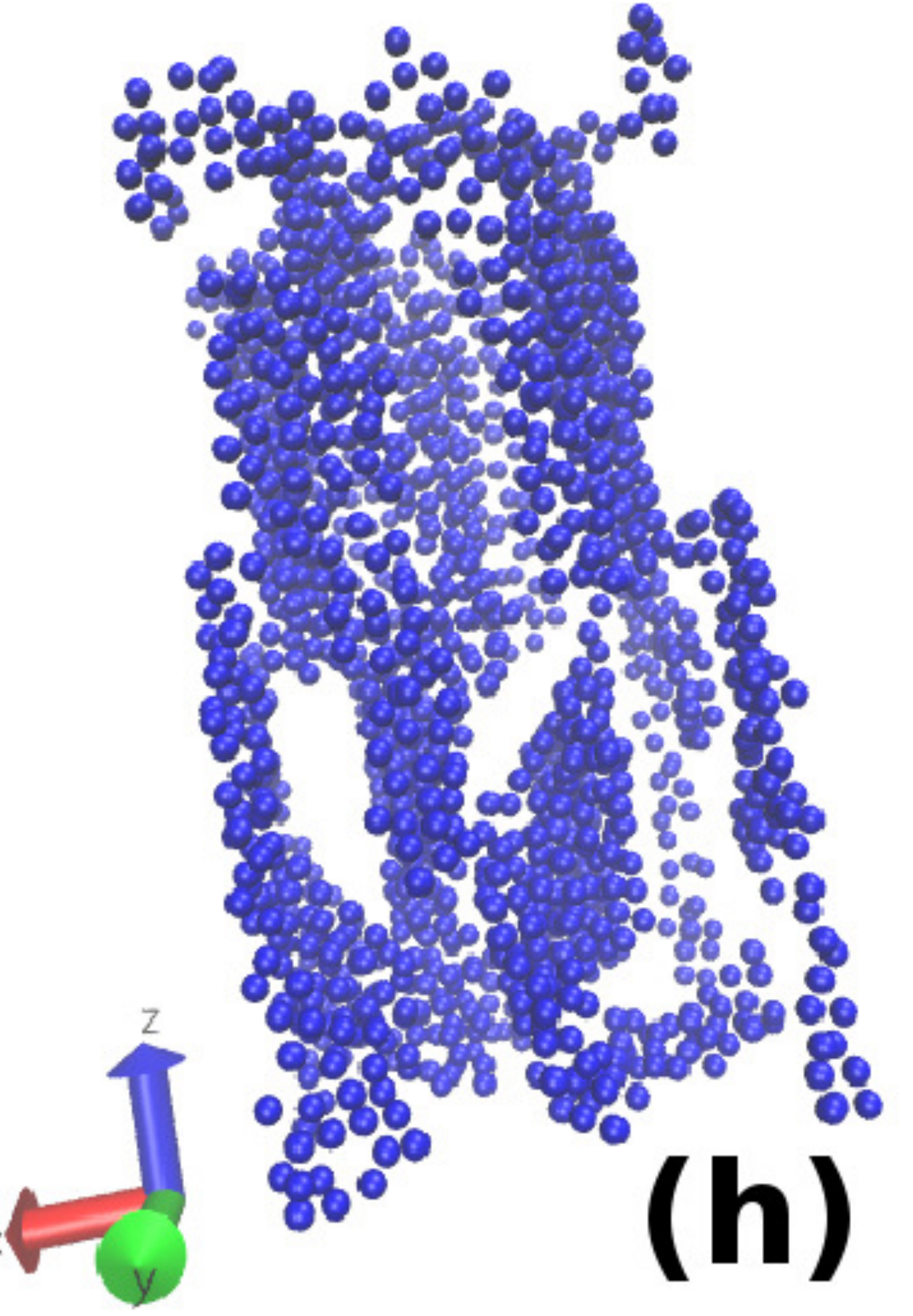}
\caption{ The figure shows snapshots for four different values of $\sigma_{4n}$ from (a)-(d) or (e)-(h), $1.25\sigma, 1.75\sigma, 2.25\sigma$ and $2.5\sigma$, respectively, for $\epsilon_n=0$. The upper row shows both the nanoparticles and monomers, while, the lower row shows only nanoparticles. The micellar chains are in dispersed state for $\sigma_{4n}=1.25\sigma$. For $\sigma_{4n}>1.25\sigma$, the nanoparticles and micellar chains form network-like structures (in (b,c) or (f,g)) which show a structural transition for $\sigma_{4n}=2.5\sigma$ forming individual sheets of nanoparticles (in (h)). }
\label{epsn_0}
\end{figure*}

\begin{figure*}
\centering
\includegraphics[scale=0.2]{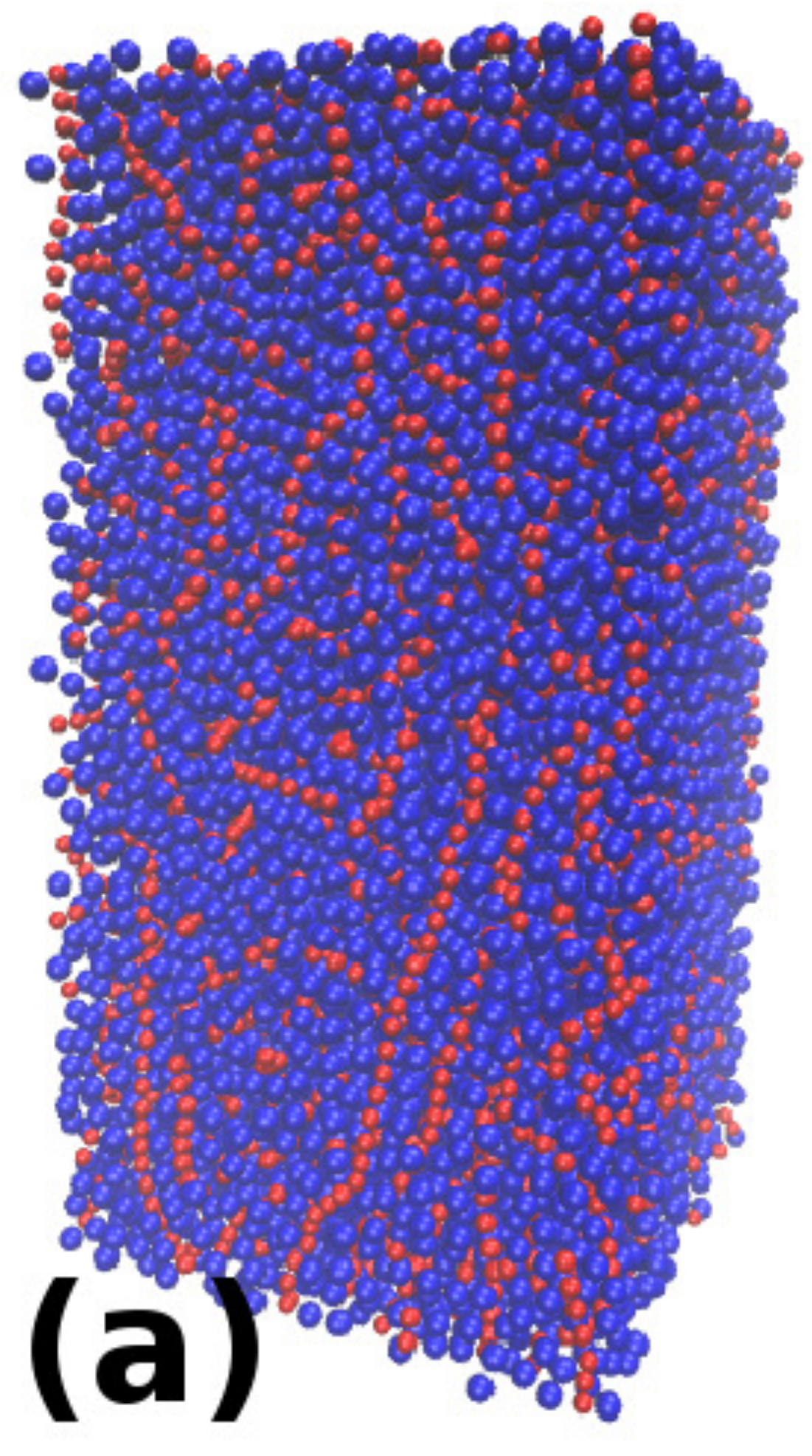}
\hspace{1cm}
\includegraphics[scale=0.2]{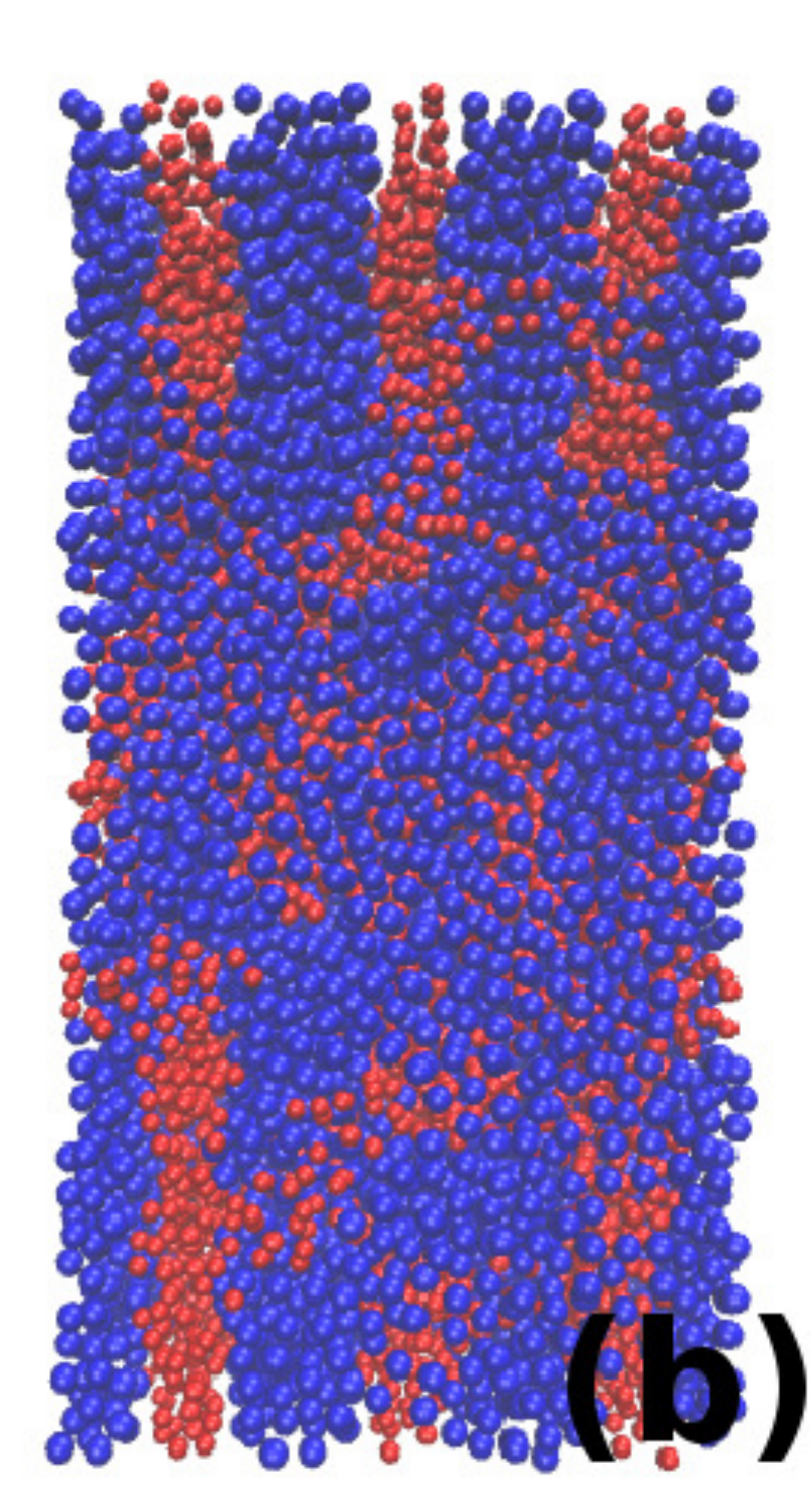}
\hspace{1cm}
\includegraphics[scale=0.2]{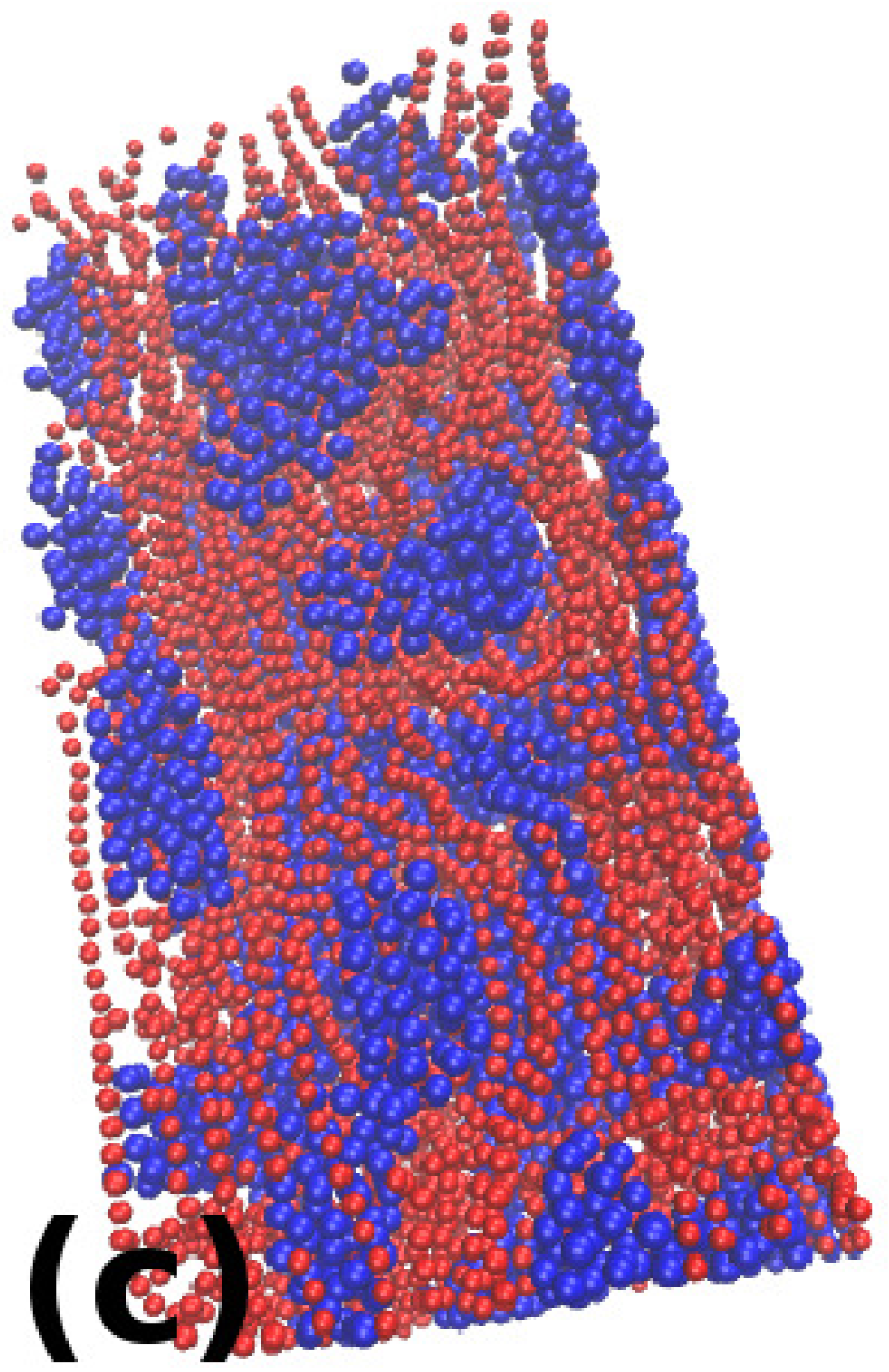}
\hspace{1cm}
\includegraphics[scale=0.2]{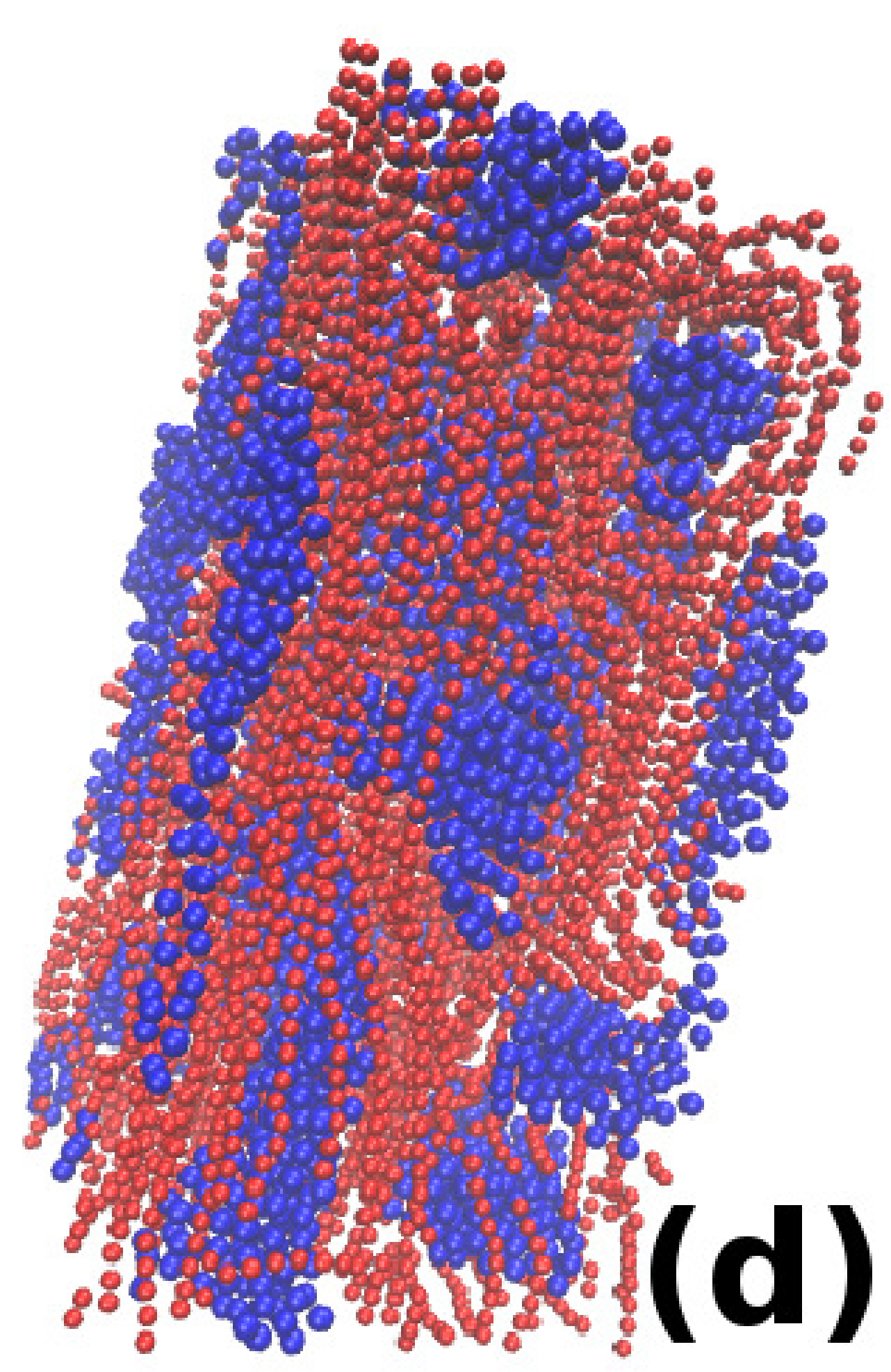} \\
\includegraphics[scale=0.2]{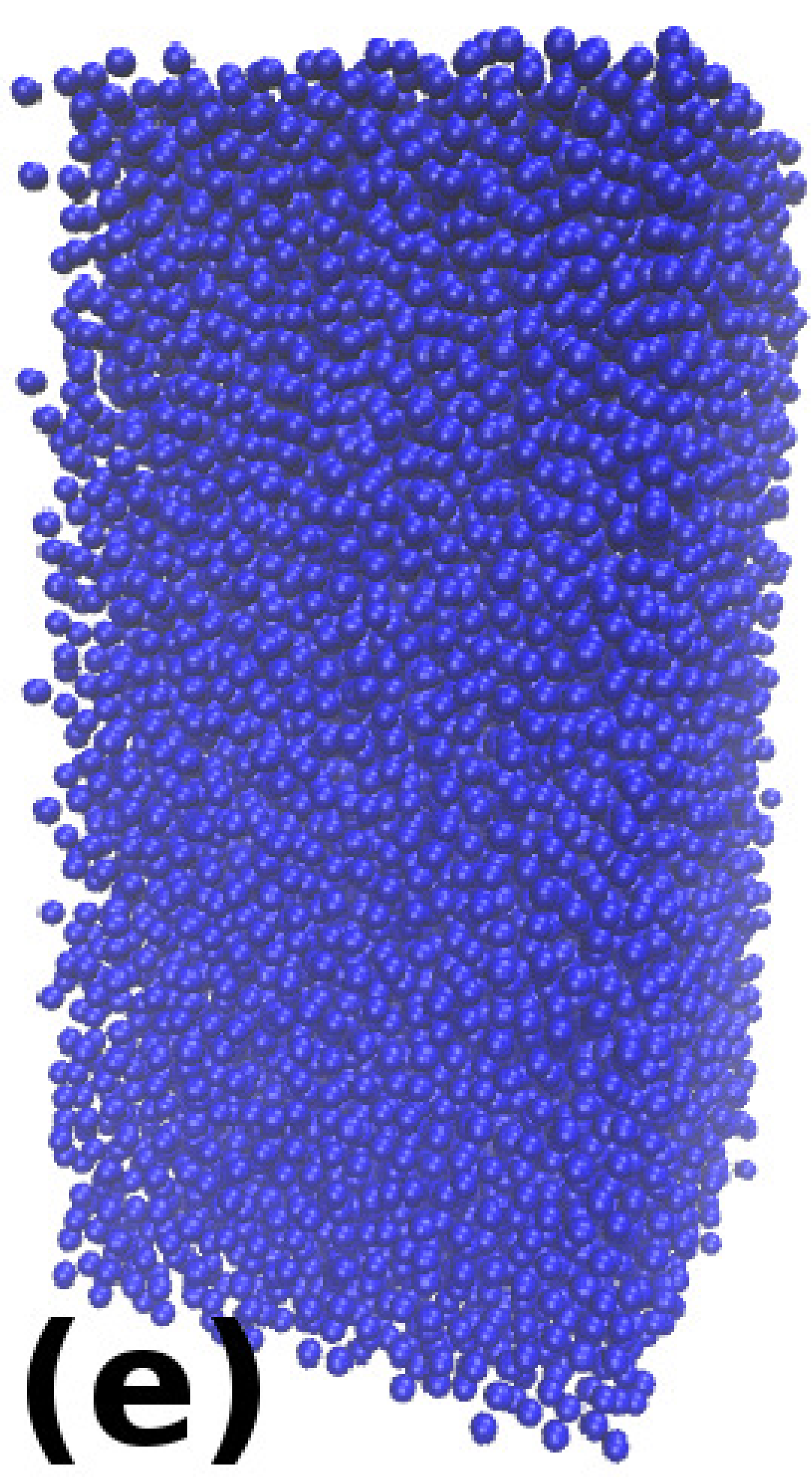}
\hspace{1cm}
\includegraphics[scale=0.2]{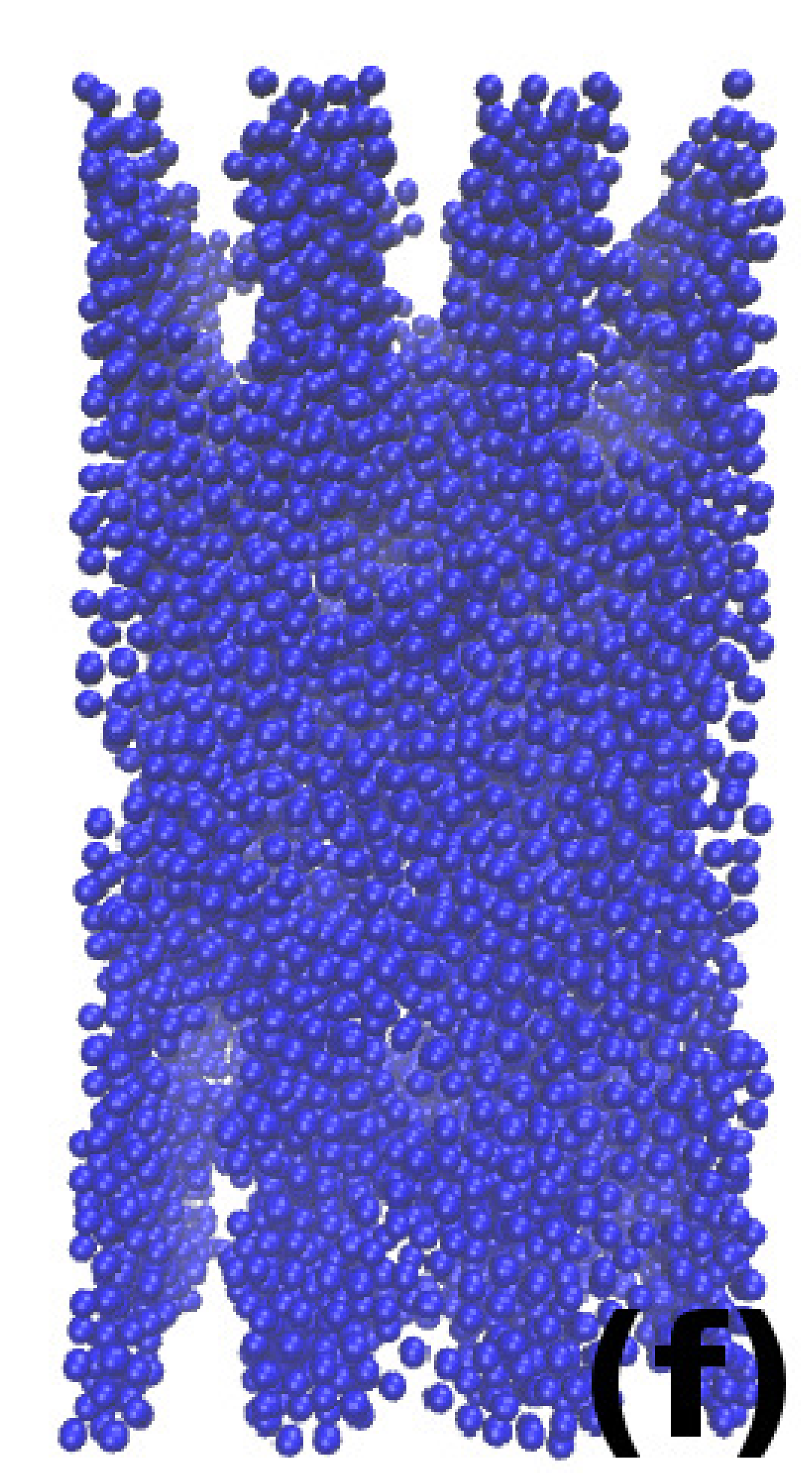}
\hspace{1cm}
\includegraphics[scale=0.2]{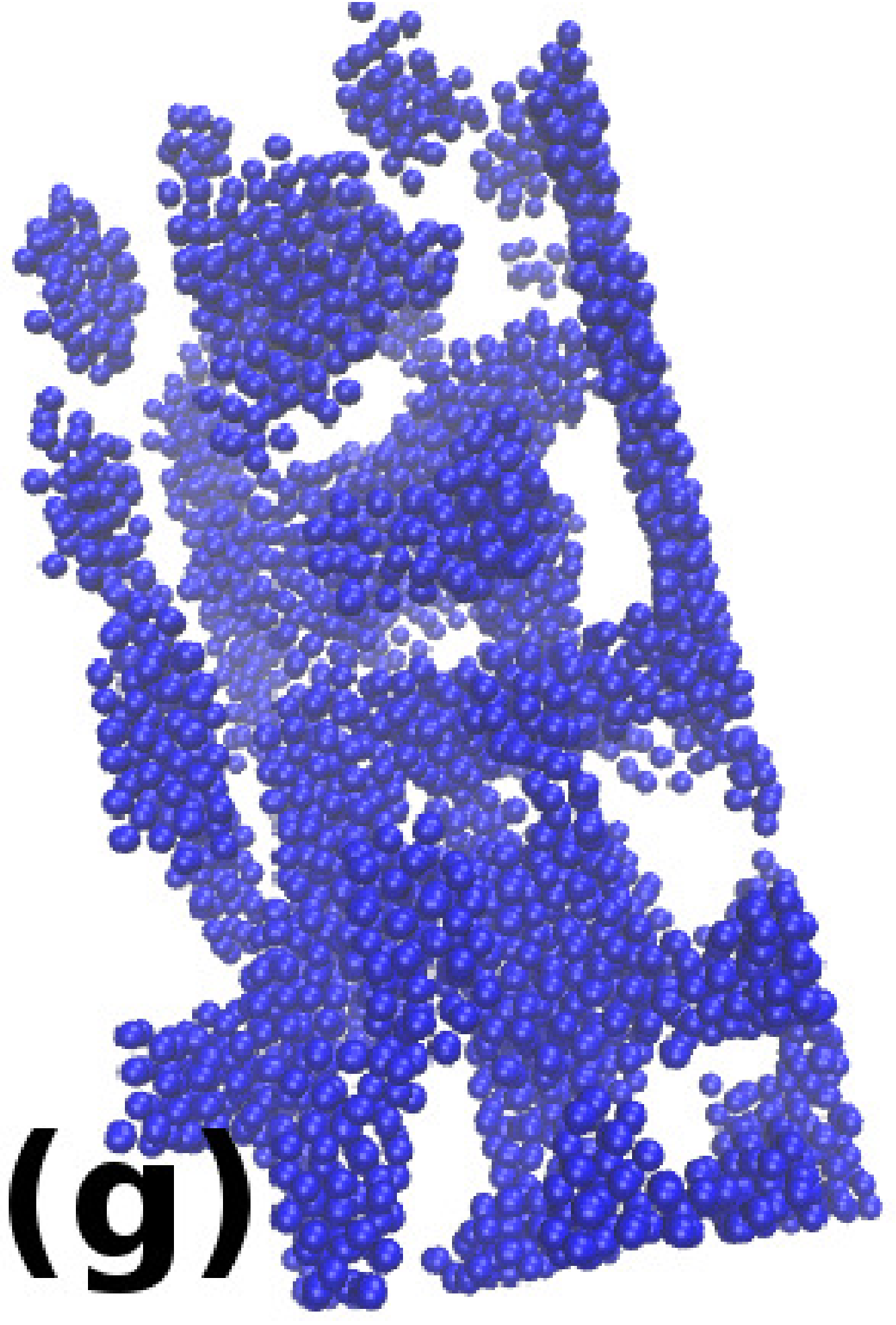}
\hspace{1cm}
\includegraphics[scale=0.2]{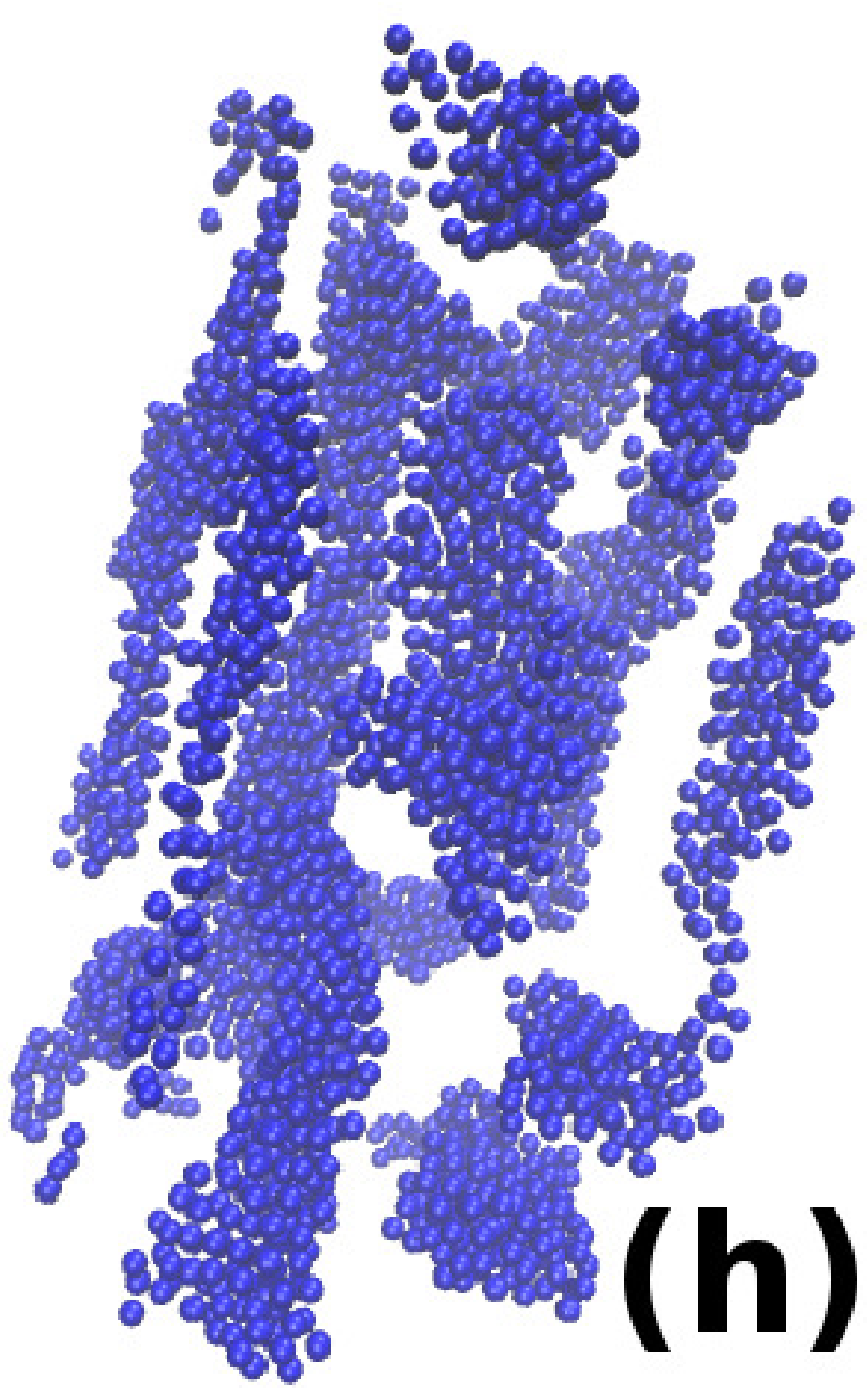}
\caption{The figure shows snapshots for four different values of $\sigma_{4n}$ from left to right, $1.25\sigma, 1.75\sigma, 2.5\sigma$ and $2.75\sigma$, respectively, for $\epsilon_n=5k_BT$. The upper row shows both the nanoparticles and monomers, while the lower row shows only nanoparticles. With increase in $\sigma_{4n}$ from $1.25\sigma$ to $1.5\sigma$, the micellar chains show structural transition from a dispersed state to formation of clusters of micellar chains that joins to form a network-like structure. With further increase in $\sigma_{4n}$, the system form some intermediate structures with nanoparticle network breaking gradually and finally showing a transition for $\sigma_{4n}=2.75\sigma$ forming individual sheets of nanoparticles.}
\label{epsn_5}
\end{figure*}

\begin{figure*}
\centering
\includegraphics[scale=0.2]{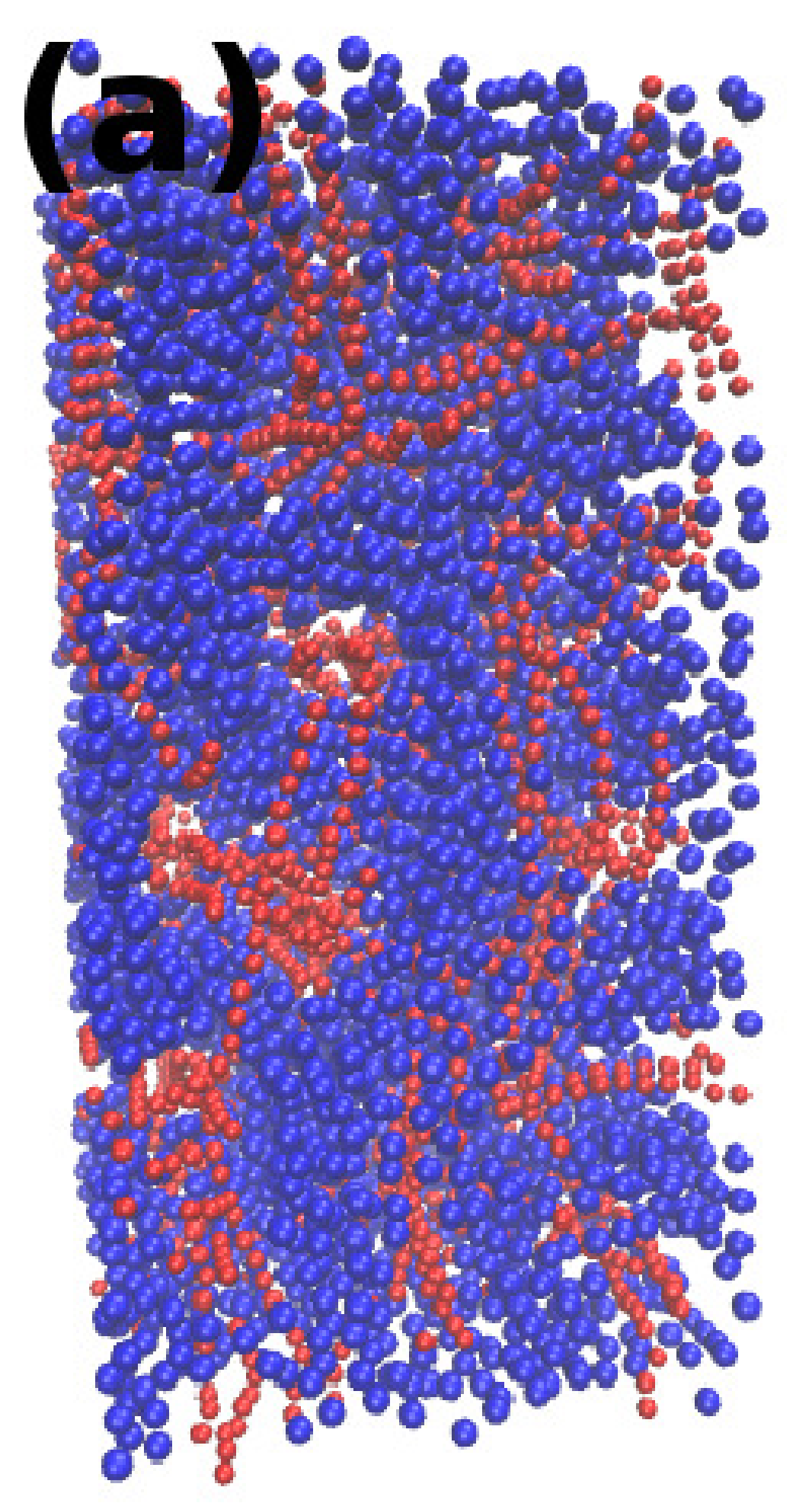}
\hspace{1cm}
\includegraphics[scale=0.2]{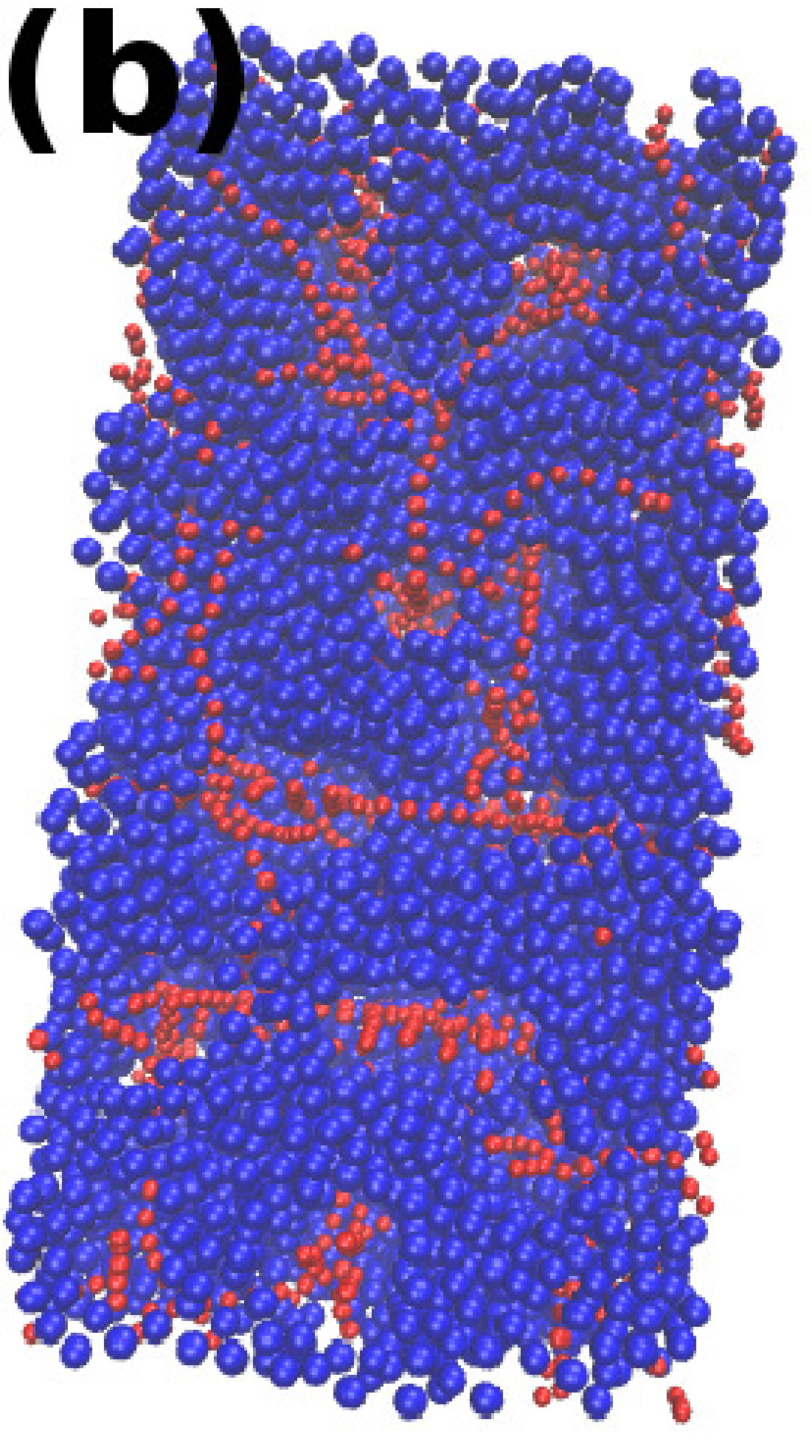}
\hspace{1cm}
\includegraphics[scale=0.2]{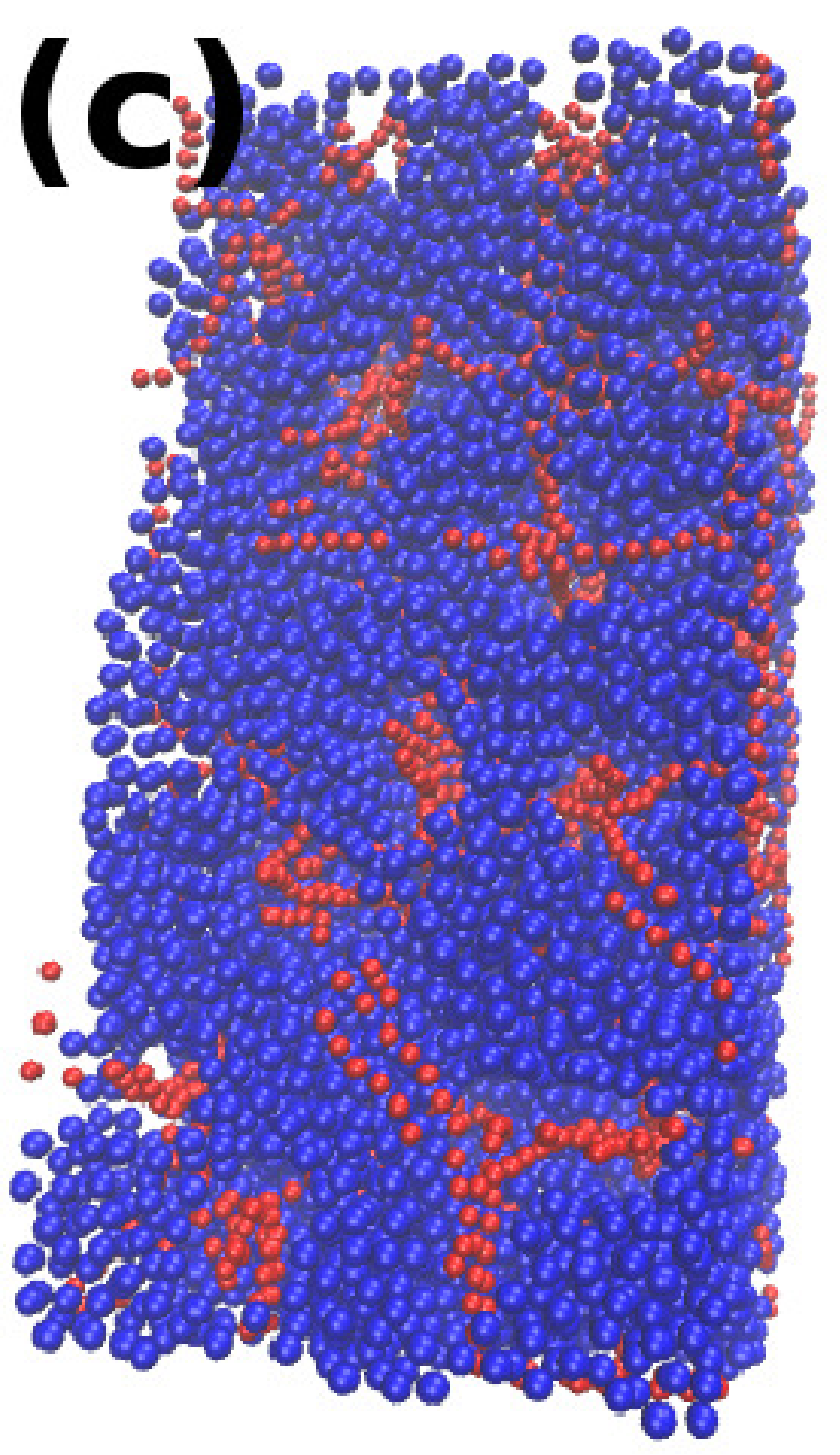}
\hspace{1cm}
\includegraphics[scale=0.2]{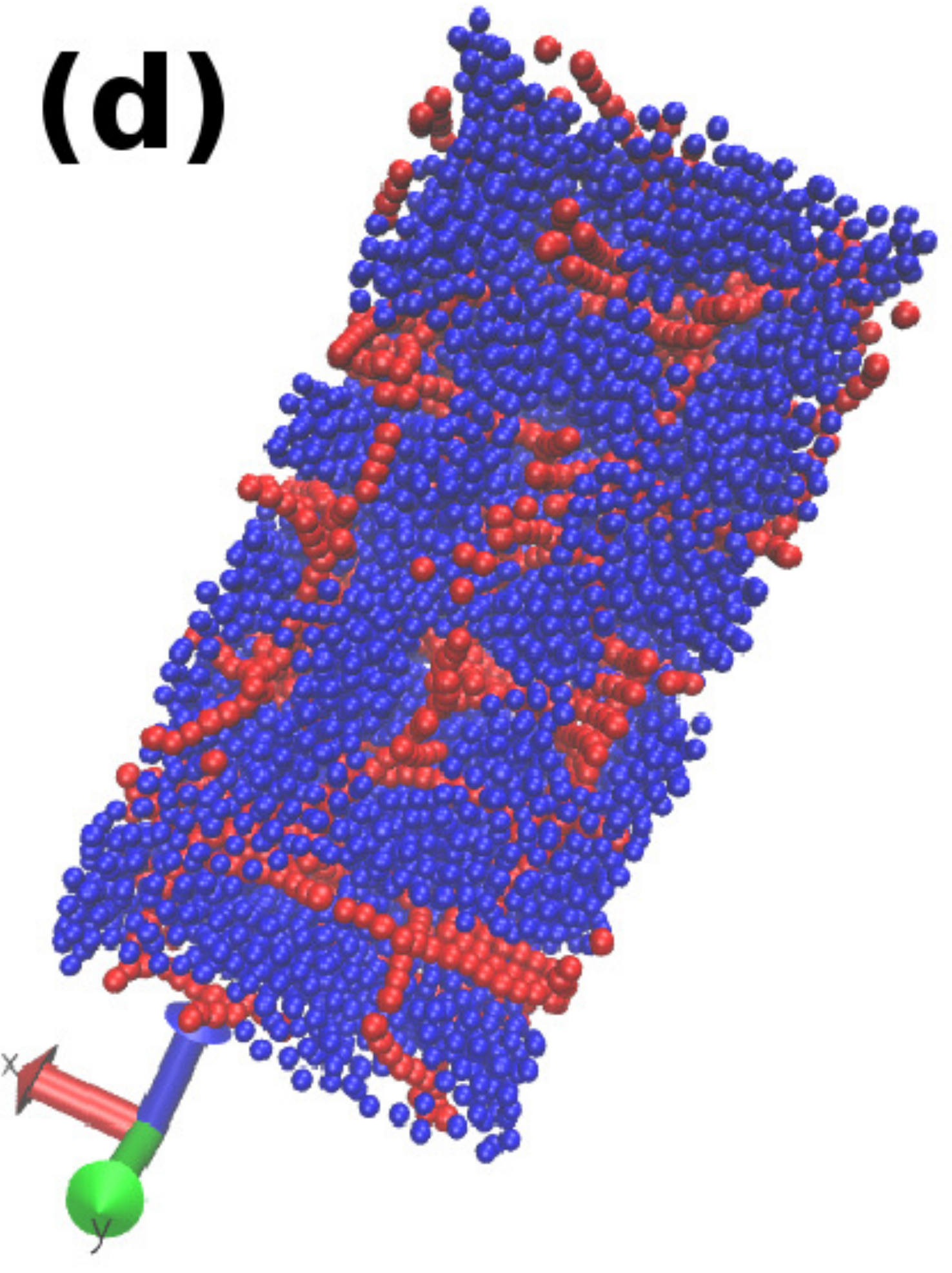} \\
\includegraphics[scale=0.2]{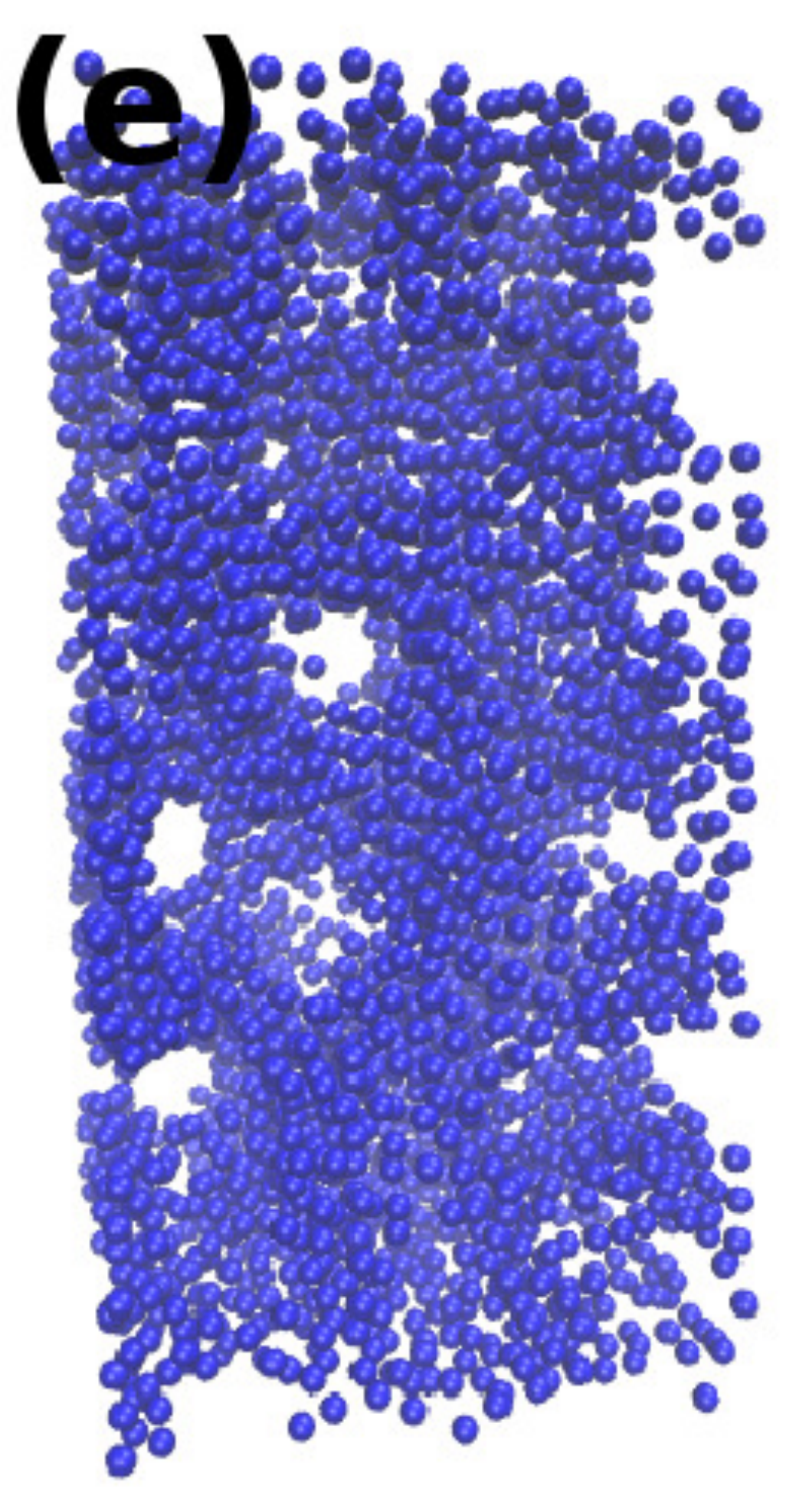}
\hspace{1cm}
\includegraphics[scale=0.2]{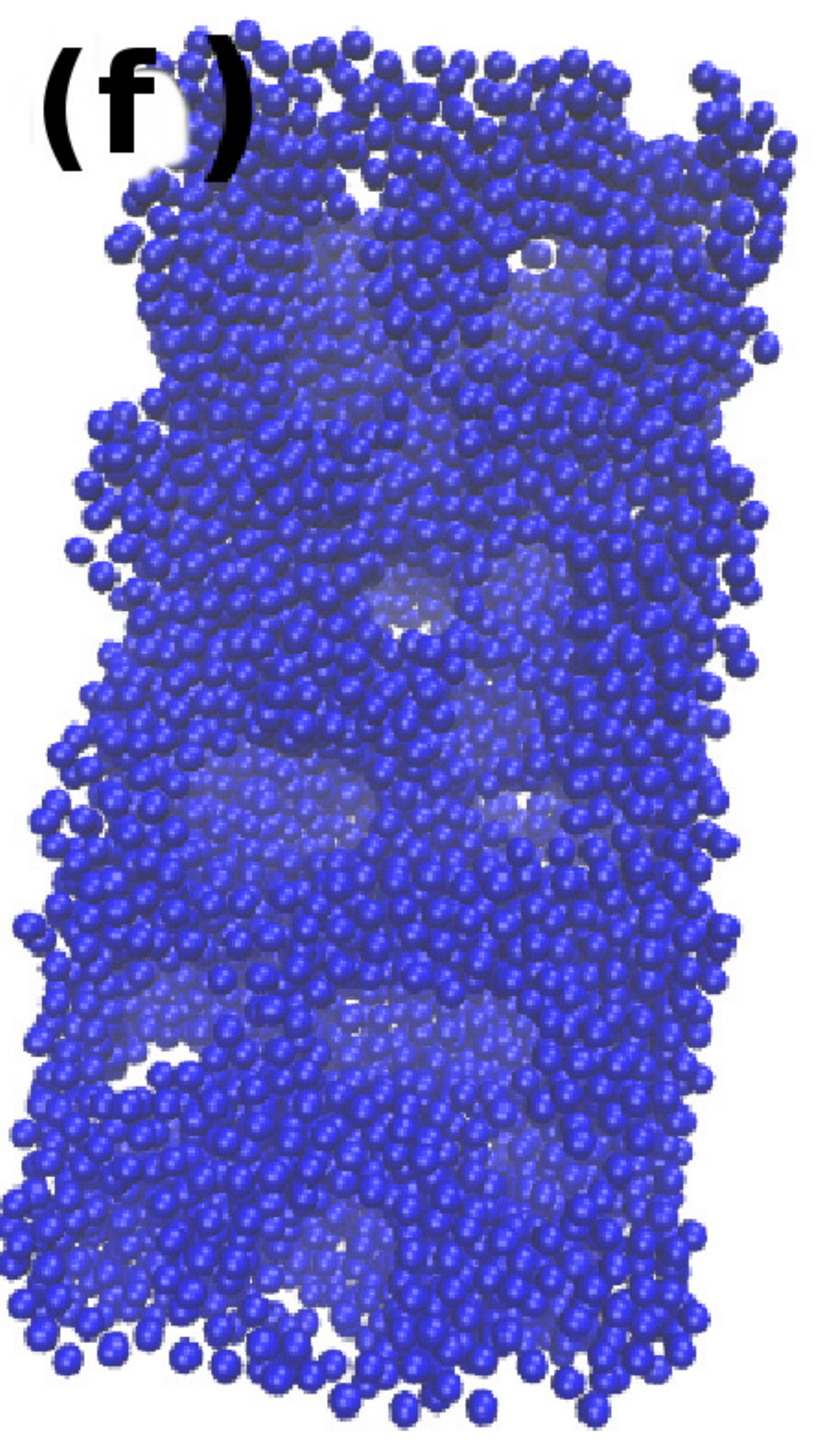}
\hspace{1cm}
\includegraphics[scale=0.2]{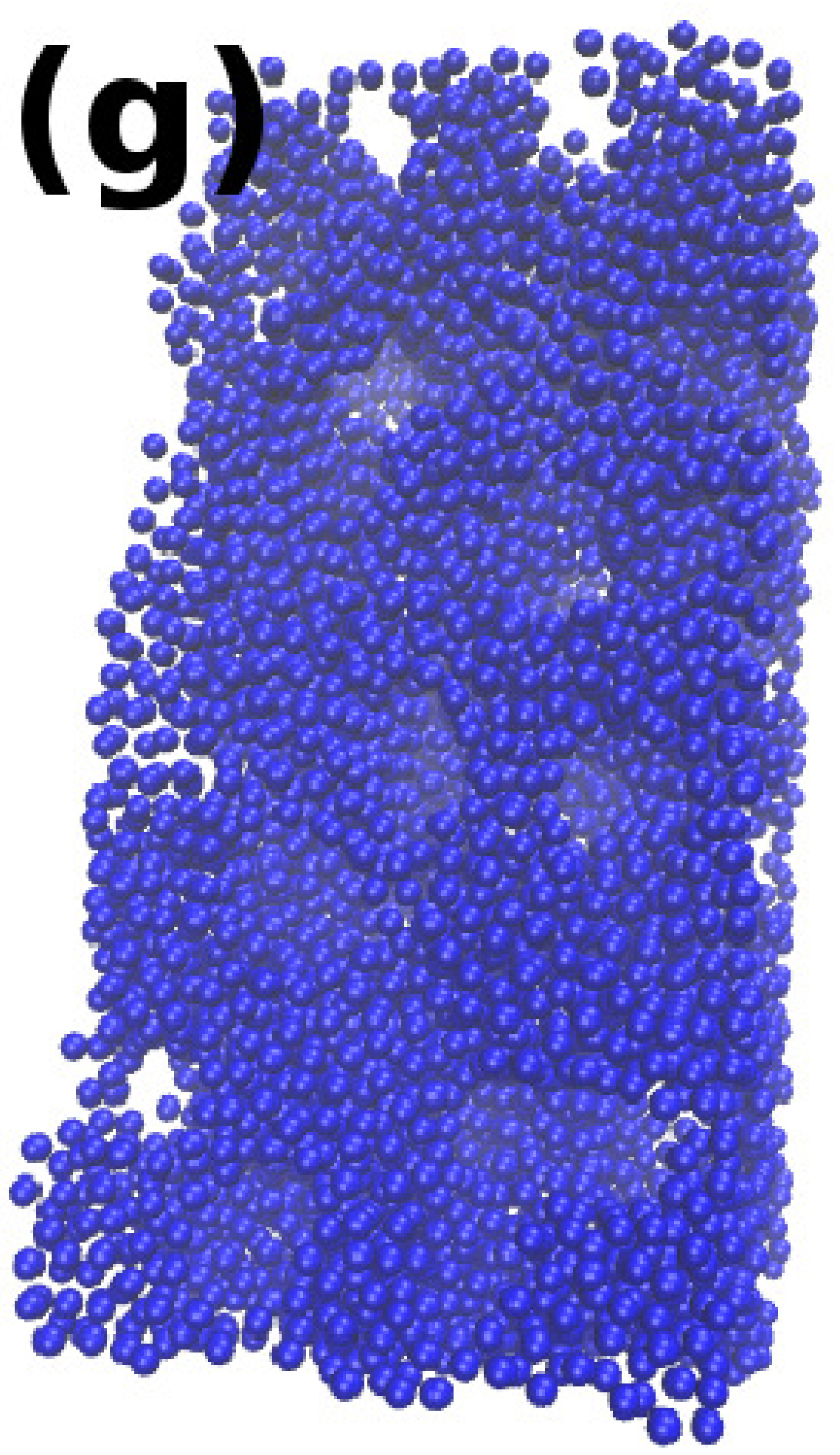}
\hspace{1cm}
\includegraphics[scale=0.2]{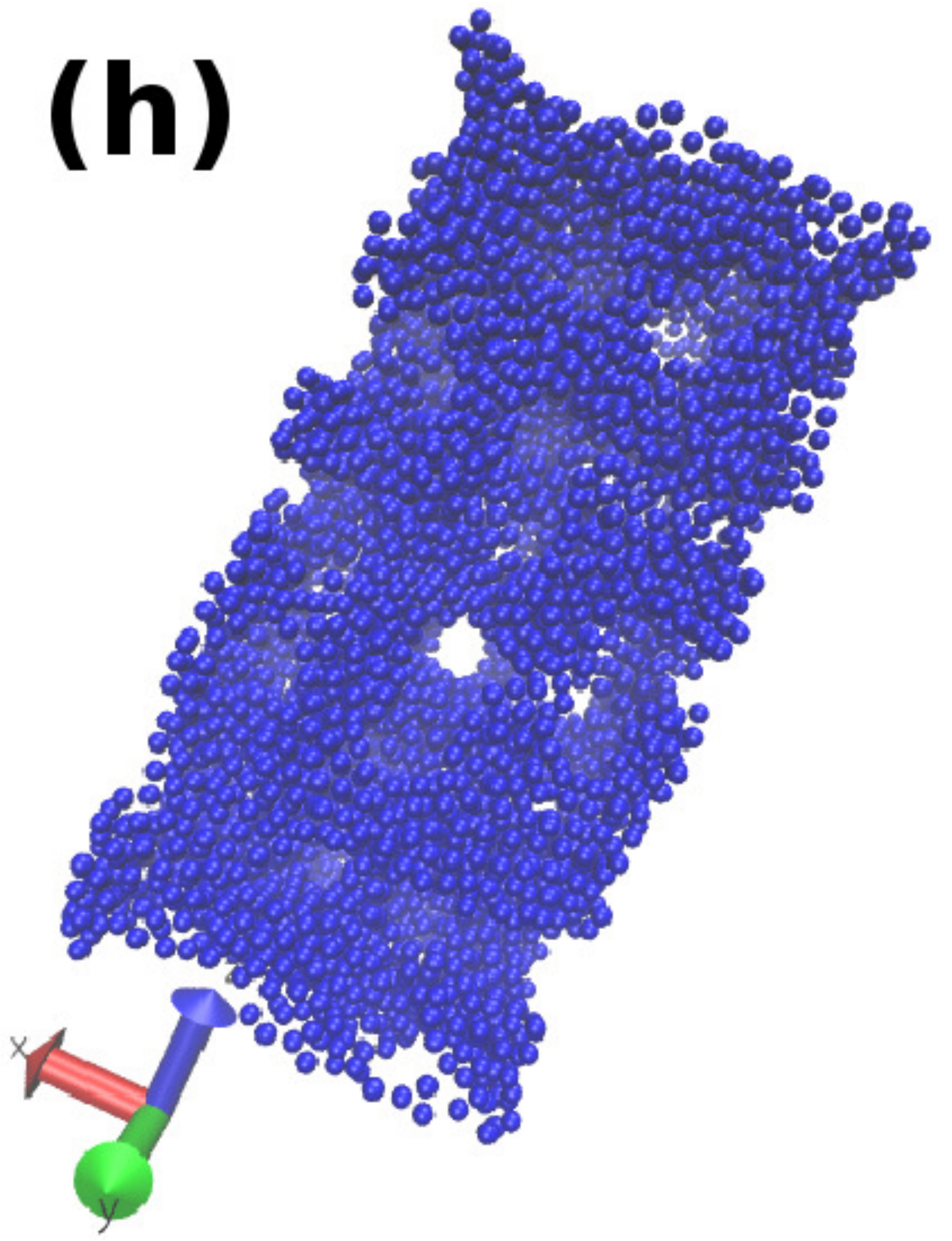}
\caption{ The figure shows the snapshots for monomer no. density $\rho_m=0.037\sigma^{-3}$ and $\sigma_{4n}=2.75\sigma$ for four different values of $\epsilon_n=0,2,5$ and $11 (k_BT)$ from left to right respectively. The upper row shows both the nanoparticles(blue) and monomers(red) while, the lower row shows only nanoparticles. All the snapshots for $\epsilon_n>0$ show similar kind of nanoparticle arrangement but the nanoparticle show lower paking for $\epsilon_n=0$.}
\label{2000}
\end{figure*}

\end{document}